\pdfoutput=1  

\documentclass[aps,prd,superscriptaddress,floatfix,nofootinbib,preprintnumbers,eqsecnum,twocolumn,dvipsnames]{revtex4-1} 

\usepackage{amsmath,float,graphicx,placeins,amssymb,multirow,pgfplots,pgfplotstable}
\usepackage{empheq}
\usepackage{tikz}
\usepackage{tikz-feynman}
\usepackage{subfigure}
\usepackage{comment}
\usepackage{enumerate,slashed,cancel,comment,color,bbm,graphicx,rotating,placeins,mathtools,multirow,hhline}
\usepackage[normalem]{ulem}
\usepackage{array}
\usepackage{tabularray}

\usepackage{lipsum}
\usepackage{hyperref}
\usepackage{color}
\usepackage{xcolor}
 \hypersetup
 {
   colorlinks,
   linkcolor={blue!80!black},
   citecolor={blue!70!black},
   urlcolor={blue!70!black}
 }

\usepackage{lipsum}
\usepackage{diagbox}
\usetikzlibrary{shapes.geometric,arrows}
\graphicspath{}
\usepackage{siunitx}
\usepackage{empheq}
\usepackage{xr}
\def\beq{\begin{equation}}
\def\eeq{\end{equation}}
\def\beqn{\begin{eqnarray}}
\def\eeqn{\end{eqnarray}}
\def\half{\mbox{\small ${\frac{1}{2}}$}}

\def\quarter{\mbox{\small ${\frac{1}{4}}$}}

\newcommand{\newc}{\newcommand}
\def\calZ{{\cal Z}}
\def\calM{{\cal M}}
\def\calV{{\cal V}}
\def\calF{{\cal F}}

\def\bQ{{\bf Q}}
\def\bT{{\bf T}}

\def\Qs{{\bf q}}

\def\KE{{\rm KE}}
\def\CM{{\rm CM}}

\def\barOmega{{\overline{\Omega}}}

\def\barXi{{\overline{\Xi}}}

\def\quarter{{\textstyle{1\over 4}}}
\def\ie{{\it i.e.}\/}
\def\eg{{\it e.g.}\/}
\def\etc{{\it etc}.\/}
\def\inbar{\,\vrule height1.5ex width.4pt depth0pt}
\def\IR{\relax{\rm I\kern-.18em R}}
 \font\cmss=cmss10 \font\cmsss=cmss10 at 7pt
\def\IQ{\relax{\rm I\kern-.18em Q}}
\def\IZ{\relax\ifmmode\mathchoice
 {\hbox{\cmss Z\kern-.4em Z}}{\hbox{\cmss Z\kern-.4em Z}}
 {\lower.9pt\hbox{\cmsss Z\kern-.4em Z}}
 {\lower1.2pt\hbox{\cmsss Z\kern-.4em Z}}\else{\cmss Z\kern-.4em Z}\fi}

\newcommand{\abs}[1]{\left|#1\right|}

\newcommand{\ch}[1]{\check{{#1}}}

\newcommand {\eq}[1]{\mbox{Eq.~#1}}

\newcommand{\pvec}[1]{\vec{#1}\mkern2mu\vphantom{#1}}

\def\be{\begin{equation}}
\def\ee{\end{equation}}
\def\ba{\begin{eqnarray}}
\def\ea{\end{eqnarray}}

\def\W{\Omega}

\def\g{\gamma}

\def\p{\partial}
\def\l{\ell}

\def\d{\delta}

\def\t{\theta}

\def\s{\sigma}

\def\la{\lambda}

\def\hf{\frac{1}{2}}

\def\Ncal{\mathcal{N}}

\def\Lcal{\mathcal{L}}

\renewcommand {\bar}[1]{\overline{#1}}
\nonstopmode

\def\Tmin{T_{\rm min}}
\def\Tmax{T_{\rm max}}
\def\taumin{\tau_{\rm min}}
\def\taumax{\tau_{\rm max}}

\begin{document}

\title{The Pull of Stasis:\\ 
A Study of the Dynamics of the Thermal Stasis Attractor}  

\def\andname{\hspace*{-0.5em}} 

\author{Jonah Barber}
\email[Email address: ]{jbarber2@arizona.edu}
\affiliation{Department of Physics, University of Arizona, Tucson, AZ 85721 USA}

\author{Keith R. Dienes}
\email[Email address: ]{dienes@arizona.edu}
\affiliation{Department of Physics, University of Arizona, Tucson, AZ 85721 USA}
\affiliation{Department of Physics, University of Maryland, College Park, MD 20742 USA}

\author{Brooks Thomas}
\email[Email address: ]{thomasbd@lafayette.edu}
\affiliation{Department of Physics, Lafayette College, Easton, PA  18042 USA}

\begin{abstract}
Cosmological stasis, a surprising phenomenon in which the abundances of different energy 
components in the universe with different equations of state remain constant despite 
cosmological expansion, has been a focus of recent attention.  This behavior emerges as the 
consequence of an attractor that governs the dynamics of the corresponding cosmological 
system and pulls it towards stasis even if the system is not initially in this state. 
However, while some systems actually reach stasis in finite time, it is also possible for 
such systems to spend considerable time under the influence of this attractor, continually 
heading towards stasis without ever quite reaching it.  This too represents behavior that 
is entirely unexpected within standard cosmological scenarios.  In this paper, we present 
an explicit model which realizes both of these behaviors in a thermal context while 
satisfying all relevant phenomenological and cosmological constraints.  Within this model, 
we then examine how the attractor influences the cosmological dynamics and explore the 
potential consequences for the early universe.
\end{abstract}
\maketitle

\tableofcontents

\def\ie{{\it i.e.}\/}
\def\eg{{\it e.g.}\/}
\def\etc{{\it etc}.\/}
\def\taubar{{\overline{\tau}}}
\def\qbar{{\overline{q}}}
\def\kbar{{\overline{k}}}
\def\bQ{{\bf Q}}
\def\calT{{\cal T}}
\def\calN{{\cal N}}
\def\calF{{\cal F}}
\def\calM{{\cal M}}
\def\calZ{{\cal Z}}

\def\beq{\begin{equation}}
\def\eeq{\end{equation}}
\def\beqn{\begin{eqnarray}}
\def\eeqn{\end{eqnarray}}
\def\apo{\mbox{\small ${\frac{\alpha'}{2}}$}}
\def\half{\mbox{\small ${\frac{1}{2}}$}}
\def\sqapo{\mbox{\tiny $\sqrt{\frac{\alpha'}{2}}$}}
\def\sqap{\mbox{\tiny $\sqrt{{\alpha'}}$}}
\def\sqapxtwo{\mbox{\tiny $\sqrt{2{\alpha'}}$}}
\def\aptwo{\mbox{\tiny ${\frac{\alpha'}{2}}$}}
\def\apofour{\mbox{\tiny ${\frac{\alpha'}{4}}$}}
\def\bosqtwo{\mbox{\tiny ${\frac{\beta}{\sqrt{2}}}$}}
\def\btosqtwo{\mbox{\tiny ${\frac{\tilde{\beta}}{\sqrt{2}}}$}}
\def\apofour{\mbox{\tiny ${\frac{\alpha'}{4}}$}}
\def\sqaptwo{\mbox{\tiny $\sqrt{\frac{\alpha'}{2}}$}  }
\def\apoeight{\mbox{\tiny ${\frac{\alpha'}{8}}$}}
\def\sapoeight{\mbox{\tiny ${\frac{\sqrt{\alpha'}}{8}}$}}

\newc{\gsim}{\lower.7ex\hbox{{\mbox{$\;\stackrel{\textstyle>}{\sim}\;$}}}}
\newc{\lsim}{\lower.7ex\hbox{{\mbox{$\;\stackrel{\textstyle<}{\sim}\;$}}}}
\def\calM{{\cal M}}
\def\calV{{\cal V}}
\def\calF{{\cal F}}
\def\bQ{{\bf Q}}
\def\bT{{\bf T}}
\def\Qs{{\bf q}}

\def\half{{\textstyle{1\over 2}}}
\def\quarter{{\textstyle{1\over 4}}}
\def\ie{{\it i.e.}\/}
\def\eg{{\it e.g.}\/}
\def\etc{{\it etc}.\/}
\def\inbar{\,\vrule height1.5ex width.4pt depth0pt}
\def\IR{\relax{\rm I\kern-.18em R}}
 \font\cmss=cmss10 \font\cmsss=cmss10 at 7pt
\def\IQ{\relax{\rm I\kern-.18em Q}}
\def\IZ{\relax\ifmmode\mathchoice
 {\hbox{\cmss Z\kern-.4em Z}}{\hbox{\cmss Z\kern-.4em Z}}
 {\lower.9pt\hbox{\cmsss Z\kern-.4em Z}}
 {\lower1.2pt\hbox{\cmsss Z\kern-.4em Z}}\else{\cmss Z\kern-.4em Z}\fi}



\section{Introduction}


\bigskip

While we know a great deal about the expansion history of the universe, there remains much
that we do not know.  Indeed, this expansion history could have differed significantly from 
that of the standard cosmology in a variety of ways --- especially prior to the 
onset of Big-Bang nucleosynthesis (BBN) --- without tensions arising with current observational 
limits (for reviews, see, \eg, Refs.~\cite{Allahverdi:2020bys,Abdalla:2022yfr,Batell:2024dsi}).  
However, a variety of proposed or planned experiments --- including next-generation CMB observatories 
and gravitational-wave detectors --- offer prospects for better probing and constraining the 
properties of our universe both before and after the beginning of BBN.~ 
It is therefore of interest to consider what modifications to the 
expansion history would arise in different extensions of the Standard Model (SM) 
and what the observational consequences of those modifications might be.

The manner in which the cosmic expansion rate evolves with time in a flat 
Friedmann-Robertson-Walker (FRW) universe depends on  the abundances $\Omega_i$ of the 
individual cosmological energy components present in that universe and on their 
corresponding equation-of-state parameters $w_i$.  Likewise, the manner in which these 
abundances  themselves evolve with time depends in turn on an interplay between the effects 
of cosmic expansion and the effects of additional processes --- often of a particle-physics 
origin --- which induce the transfer of energy density from one energy component to another.  
We shall refer to these latter processes as ``pumps.''

In most cosmological scenarios, the different abundances are driven by the resulting 
dynamics to a configuration in which one abundance is extremely large (essentially unity) 
while the others are extremely small.   It is for this reason that the long-lived cosmological 
epochs within most traditional cosmologies are dominated by a single energy component, and are 
thus either radiation-dominated or matter-dominated or even vacuum-energy-dominated.   However, 
it has recently been shown~\cite{Dienes:2021woi,Dienes:2023ziv} that within many cosmologies 
based on various models of physics beyond the Standard Model (BSM), the corresponding equations 
of motion actually exhibit dynamical attractors which pull the system toward fixed-point 
solutions wherein {\it multiple}\/ different cosmological energy components with different 
equation-of-state parameters nevertheless have fixed, non-zero abundances $\Omega_i$.  
This then gives rise to a new kind of cosmological epoch --- an epoch of 
{\it cosmological stasis}~\cite{Dienes:2021woi, Dienes:2023ziv} --- during which the different 
non-zero abundances $\Omega_i$ remain constant despite cosmological expansion.  Indeed, over 
the past few years this stasis phenomenon has been discovered to exist within a large variety 
of models of BSM physics and their associated cosmologies~\cite{
Dienes:2021woi,Barrow:1991dn,Dienes:2022zgd,Dienes:2023ziv,Dienes:2024wnu,
Halverson:2024oir,Barber:2024vui,Barber:2024izt,Huang:2025odd,Dienes:2025tox,
Long:2025wjw,Dienes:2025qdw,Barenboim:2026zgj,Barenboim:2026txj}.

The rate at which this cosmological attractor pulls our system towards stasis 
can depend on many factors.  These include the different possible abundances 
with which our system starts as well as the properties of the physical pump 
processes that lead to the corresponding energy transfers.   As a result, 
depending on these features, it is possible that a given cosmological system 
might evolve either quickly or slowly towards the fixed-point stasis configuration 
dictated by the corresponding cosmological attractor.   

These different rates of approach towards stasis may have a number of interesting 
phenomenological implications.  However, perhaps the most important is that in 
certain cases these differences in approach rates actually have the potential to 
prevent stasis from occurring altogether!  This possibility arises because the 
physics that gives rise to the stasis-inducing pump terms is itself often subject 
to an intrinsic time-limit --- a so-called ``expiration date'' --- after which the 
dynamics of our system changes completely and the attractor dissolves.  If our system 
has already reached stasis by this time, then the existence of such an expiration 
date can place a fundamental limit on how long the universe can remain in stasis. 
However, if our system has merely been pulled by the attractor 
towards a stasis solution without having yet reached it, our system might begin by 
{\it approaching}\/ a stasis configuration, only to suddenly veer away from this 
stasis solution once the expiration date is reached.  Thus our system would have 
experienced the pull of the stasis attractor --- thereby significantly deforming its 
expected dynamics relative to what would have been expected in more standard 
cosmologies --- without ever actually exhibiting stasis.

In this paper, we shall explore these ideas more fully within the context of a model 
which exhibits all of these features and which makes use of the thermal stasis mechanism 
originally presented  in Ref.~\cite{Barber:2024vui}.  As we shall demonstrate, this model 
gives rise to a thermal 
stasis attractor which contains a wide range of behaviors, including some approaches to stasis which 
proceed relatively quickly and some which approach stasis more slowly.  Equally critically, we shall 
demonstrate that this model also contains a natural expiration date --- a maximum length of time during 
which our system can remain under the influence of the stasis attractor.   As a result, this model 
provides graphic illustration of the rich set of new cosmologies which can emerge as the result of 
stasis attractors in BSM scenarios.

It is important to understand these results in the proper context.  Of course, within 
certain BSM cosmologies, the existence of the stasis  attractor is guaranteed and is completely 
independent of cosmological initial conditions. This is consistent with all previous results 
in the literature concerning the stasis phenomenon.  By contrast, what we are stating here is 
that the duration (and even the existence) of a resulting stasis {\it epoch}\/ can --- and often 
does --- depend on those same initial conditions, often quite sensitively.  Indeed, in extreme 
cases, these initial conditions may even determine whether our stasis {\it attractor}\/ produces 
a stasis {\it epoch}\/.  However, regardless of the choice of initial conditions, the cosmological 
stasis attractor continues to exist and the cosmologies that emerge are wholly new, endowed with 
interesting features in their own right which merit independent exploration.

This paper is organized as follows.  In Sect.~\ref{sec:attractor}, we review the general 
considerations which impact the rates at which the state of a dynamical system with an 
attractive fixed point evolves toward that fixed point along different trajectories.  
In Sect.~\ref{sec:emergence}, we then present our model of the stasis attractor and show that 
this attractor indeed has an expiration date.  In Sect.~\ref{sec:constraints}, we examine the 
consistency conditions and observational bounds which constrain this model.  Although many 
of these considerations significantly restrict that parameter space, we shall nevertheless 
demonstrate in Sect.~\ref{sec:Results} that large regions remain wherein all of these constraints 
are satisfied.  In Sect.~\ref{sec:Dynamics}, we present approximate analytic results for the 
timescales required for the system to evolve from a variety of different initial conditions 
toward the fixed point under the influence of the stasis attractor.  In Sect.~\ref{sec:Results}, 
we present our full numerical results for these timescales.  Indeed, it is here that we
show that our thermal stasis model can give rise to a significant number of $e$-folds of stasis, 
but that the duration of the stasis epoch is highly sensitive to the initial conditions for the 
system.  In Sect.~\ref{sec:conclusions}, we conclude with a summary of our findings and highlight 
possible directions for future work.  We also include two Appendices which provide extra details 
concerning results presented in the main text.


\section{General attractor dynamics:\\ Fast versus slow directions}\label{sec:attractor}


In this section, we shall quickly review the general behavior of dynamical attractors, 
with an eye towards demonstrating that there can generally be both ``fast'' and ``slow'' 
paths along which a given physical system may be pulled by such an attractor.  

Towards this end, let us consider a physical system characterized by a set of dynamical 
variables $u_i(t)$ whose time-evolution is governed by a system of first-order differential 
equations for which there exists an attractive fixed-point solution at which these 
variables take the values $\bar{u}_i$.  We shall assume that sufficiently 
close to this attractor, this system of equations is to a good approximation both linear 
in the $u_i$ and autonomous.  It therefore follows that within the vicinity of the fixed 
point, the equations of motion for the $u_i$ each take the general form
\begin{equation}
  \frac{du_i} {dt} ~\approx~ \sum_j J_{ij} (u_j - \bar{u}_j)~
\label{diffeq} 
\end{equation}
where $J$ is a matrix whose elements $J_{ij}$ are all independent of $t$.  

A system of linear equations of this form may be solved in a straightforward manner via a 
change of basis.  In particular, if $V$ denotes the matrix of eigenvectors of $J$, we have 
\begin{equation}
  J ~=~ V \Lambda V^{-1}~,
  \label{eq:DefOfVMat}
\end{equation}
where $\Lambda_{ij} = \lambda_i \delta_{ij}$ with $\lambda_i$ denoting the eigenvalues 
of $J$.  For all of the physical systems that we shall be considering in this paper, 
these eigenvalues $\lambda_i$ are real-valued, and we shall assume this in what follows.
Inserting Eq.~(\ref{eq:DefOfVMat}) into Eq.~(\ref{diffeq}), we have 
\begin{equation}
  \frac{du_i}{dt} ~\approx~ \sum_{jk} V_{ij} \lambda_j
    (V^{-1})_{jk}  (u_k - \bar{u}_k)~.
\label{eq:dudtWithVSub} 
\end{equation}
Thus, if we define the corresponding dynamical variables
 \begin{equation}
  x_i ~\equiv~ \sum_j (V^{-1})_{ij} (u_j - \bar{u}_j)~,
  \label{eq:DefOfxSubi}
\end{equation}
it is straightforward to show that these evolve as
\begin{equation}
  \frac{dx_i}{dt} ~\approx~ \lambda_ix_i~.
\label{eq:dudtWithVSub} 
\end{equation}
Since the general solution to this differential equation is $x_i(t) = x_i^{(0)}e^{\lambda_it}$
where $x_i^{(0)} \equiv x_i(0)$, it follows from Eq.~(\ref{eq:DefOfxSubi}) that the $u_i(t)$ 
evolve with time in the vicinity of the fixed point according to the relation
\begin{equation}
  u_i(t) ~\approx~ \bar{u}_i + \sum_j V_{ij} x_j^{(0)} e^{\lambda_j t}~.
  \label{genAttractorUt}
\end{equation}

Because the fixed-point solution is by assumption attractive, we know that 
$\lambda_{i} < 0$ for all $i$.  Thus, from Eq.~(\ref{genAttractorUt}),
we see that the time for our system to effectively reach the stasis location 
depends on the eigenvalues $\lambda_i$ as well as its initial coordinates 
$u_i^{(0)}$.  Indeed, we see that $u_i-\overline{u}_i$ approaches zero at a rate 
which is governed by the eigenvalues $\lambda_j$,  with the overall approach rate 
no more rapid than $e^{\lambda_j t}$ where $\lambda_j$ is the most negative 
eigenvalue for which $V_{ij}\not =0$.

More explicitly, we can ask how long it takes the system to ``reach'' the fixed 
point starting from an arbitrary initial location $x_i^{(0)}$.  Looking at 
Eq.~(\ref{genAttractorUt}), we see that no trajectory will ever actually hit the 
fixed point unless it happens to already start there.  Because of this, 
declaring that our system has reached the fixed point requires an inherently 
subjective criterion --- one which is presumably endowed with a cutoff beyond 
which the system is sufficiently close to be considered to have 
reached (and therefore now reside at) the fixed point.  For example, we can adopt 
the criterion that we have reached the fixed point so long as
\begin{equation}
\abs{x_j}~\leq~\delta_j~
\label{stasis_reached} 
\end{equation}
for all $j$ and for some specified values $\delta_j$.  With this condition, the 
time $t_{\rm FP}$ to reach the fixed point would then be given by
\begin{equation}
     t_{\rm FP} ~=~  \max_j \left\{\frac{1}{\abs{\la_j}}
       \log\left|\frac{x_j^{(0)}}{\delta_j}\right|\right\}~.
\label{nevercalledeq} 
\end{equation}
Indeed, any condition of the form $t_{\rm FP}\leq \epsilon$ selects a parallelotope 
within our original parameter space --- \ie, a region around the fixed point in which 
we consider the fixed point to have been reached.   

While such a region is relatively straightforward to delineate in terms of the $x_i$ 
dynamical variables, it becomes far more complex in terms of the $u_i$ variables.  
Moreover, we shall later develop an alternative definition for the point at which we 
may consider our system to have ``arrived'' at the fixed-point location --- one which 
relies on the behaviors of actual physical variables within our system.   Thus, we shall 
not use Eq.~(\ref{nevercalledeq}) in the following.  

That said, we do observe one important point from this analysis:  the rates at which 
our system evolves towards the fixed point can vary significantly between different 
trajectories if the eigenvalues $\lambda_i$ have absolute magnitudes which are sufficiently unequal. 
 Indeed, letting our most-negative (MN) and least-negative (LN) eigenvalues 
be $\lambda_1=\lambda_{\rm MN}$ and $\lambda_2= \lambda_{\rm LN}$, respectively, 
with $\lambda_{\rm MN}\ll \lambda_{\rm LN}$, we find that within the corresponding 
two-dimensional parameter space $(x_1^{(0)}, x_2^{(0)})$ the fastest trajectories for 
reaching the fixed point have any of the starting locations
\beq
(x_1^{(0)},x_2^{(0)})~=~ (A,0) ~~~{\rm for~any}~A~,
\eeq
or equivalently
\beqn
u_1^{(0)} ~&=&~ \bar{u}_1 + V_{11} A\nonumber\\
u_2^{(0)} ~&=&~ \bar{u}_2 + V_{21} A~.
\eeqn 
Note that we are here referring to the fastest {\it trajectories}\/ --- \ie, those 
trajectories governed by the fastest eigenvalues.  Of course, there exist other 
starting locations which might have shorter times to reach the fixed point as the 
result of their initial proximities to that point.   We are likewise imagining that 
$A$ is bounded by our initial assumption that we are focusing on regions of our 
parameter space which are sufficiently close to the fixed point that a linear analysis 
such as we are performing here is valid.  Finally, as discussed above, we are assuming 
that $\lambda_{\rm MN}\ll \lambda_{\rm LN}$ --- \ie, that there exists a hierarchy 
in the absolute sizes of our eigenvalues.  More complex behaviors can emerge if this 
is not the case.

These general considerations can have profound consequences for the cosmologies 
associated with particle-physics models which give rise to a stasis attractor, 
including the model which shall be our focus in this paper.  Indeed, as we shall 
see, the state of the universe can evolve toward this attractor at significantly 
different rates, depending on the initial conditions for the pertinent dynamical 
variables.  


\section{A framework for a thermal stasis attractor\label{sec:emergence}}


In this section, we present an explicit framework in which a stasis attractor 
arises in a thermal context.  It is ultimately within this framework that our 
subsequent analysis will take place.  As we shall demonstrate, this framework 
gives rise to all of the features which will be necessary for our discussion, 
including the existence of fixed-point stasis solution, the existence of
an attractor which pulls our system towards this stasis point and which can do 
so along both ``fast'' and ``slow'' directions, and the existence of an 
``expiration date'' which limits the length of time during which the system 
can be pulled along by the attractor.  Parts of this framework were originally 
presented (for other purposes) in Ref.~\cite{Barber:2024vui}.  Accordingly, in 
this section we shall discuss only the salient details of this framework and 
concentrate on the critical features mentioned above.  We shall then 
further develop this framework into an actual physics model in 
Sect.~\ref{subsec:expiration_date}.

\subsection{Preliminaries}

We begin by considering a flat FRW universe 
containing two primary energy components that will participate in a 
matter/radiation stasis:  one component associated with a non-relativistic 
massive particle $\phi$ of mass $m$ that represents the matter in our model, 
and a second component associated with a massless particle $\chi$ comprising 
the radiation. In general, a universe consisting of both matter and radiation 
inevitably becomes matter-dominated as a result of cosmological expansion 
unless the underlying physics of $\phi$ and $\chi$ includes a process --- a 
so-called ``pump'' --- that converts matter back to radiation.  Indeed, such 
pumps naturally arise in a plethora of BSM models~\cite{Dienes:2021woi,
Dienes:2023ziv, Barrow:1991dn,Dienes:2022zgd,Dienes:2024wnu,Halverson:2024oir,
Barber:2024izt, Huang:2025odd,
Dienes:2025qdw} in which the matter particles naturally decay to radiation.
In our case, however, we shall follow Ref.~\cite{Barber:2024vui} in considering 
a pump arising from the {\it annihilation}\/ of two  $\phi$ particles into 
radiation with a cross-section of the form
\begin{equation}
    \sigma v ~=~ C \left(\frac{\abs{\vec{p}_{\CM}}}{m}\right)^q~ ,
\label{eq:swept_vol_rate}
\end{equation}
where $\sigma v$ denotes the ``swept-volume'' rate (\ie, the product of the 
annihilation cross-section $\sigma$ and the relative velocity $v$ of these 
$\phi$ particles), where $C$ is an overall rate prefactor, where $\vec{p}_{\CM}$ 
is the incoming massive-particle momentum in the center-of-mass frame, and where 
$q$ is an (as yet undetermined) exponent. We will find that only a certain range 
of $q$-values will give rise to a stasis attractor. Because the cross-section 
depends on the momenta of the annihilating particles, the overall annihilation 
rate will end up depending on the temperature of the matter particles.   This 
temperature dependence will ultimately play a decisive role in stasis.  Toward 
this end, we will assume here that the $\phi$ particles form a thermal population 
at a temperature $T$ and are able to scatter elastically off each other in order 
to maintain thermal equilibrium with each other.  Note, in particular, that we 
have no need to invoke the presence of any thermal bath. We will also treat the 
matter as non-relativistic (which amounts to requiring that $T\ll m$).

We now examine the general equations that govern the energy densities and corresponding
abundances of our two energy components in a flat FRW universe. The matter has 
equation-of-state parameter $w_M=0$, whereas the radiation has equation-of-state 
parameter $w_\gamma=1/3$. The energy densities of matter $\rho_M$ and radiation 
$\rho_\gamma$ therefore evolve according to
\begin{eqnarray}
  \frac{d\rho_M}{dt} &=& -3 H \rho_M - P^{(\rho)}_{M,\gamma} ~ \nonumber\\
  \frac{d\rho_\gamma}{dt} &=& -4 H \rho_\gamma + P^{(\rho)}_{M,\gamma}~~,
\label{eq:eoms}
\end{eqnarray}
where $H\equiv \dot{a}/a$ is the Hubble parameter and where  $P_{M,\gamma}^{(\rho)}$ 
(the so-called ``pump'' term) describes the rate at which the annihilation of $\phi$ into 
$\chi$ converts matter into radiation.

Because the rate $P_{M,\gamma}^{(\rho)}$ depends on the temperature $T$, we will also need 
to know how the temperature evolves in this system.  We can determine this by recognizing that 
$T$ is related to a third energy component, the kinetic-energy density $\rho_\KE$ of the $\phi$ 
particles.  Indeed, this energy component is distinct from $\rho_M$, which --- as befitting 
the energy density of non-relativistic matter ---  comprises the rest-mass-energy density of 
these particles alone.  Assuming that the $\phi$ particles form an ideal gas, we then 
have a relation between $T$ and $\rho_\KE$, namely
\begin{equation}
  \rho_\KE ~=~ \frac{3}{2}n_M T ~\approx~ \frac{3}{2} \rho_M \frac{T}{m}~,
  \label{idealgas}
\end{equation}
where $n_M$ denotes the number density of the $\phi$ particles.
The time evolution of $T$ can then be determined from the time evolution of $\rho_\KE$.  
Recognizing that the kinetic energy has an equation-of-state parameter $w_\KE = 2/3$ when 
the matter is non-relativistic, and further recognizing that the annihilation of $\phi$ 
particles produces not only a pump $P^{(\rho)}_{M,\gamma}$ that describes the conversion 
of matter energy (\ie, rest-mass energy) to radiation but also a pump $P^{(\rho)}_{\KE,\gamma}$ 
that describes the corresponding conversion of matter kinetic energy to radiation, we then have
\begin{eqnarray}
  \frac{d\rho_\KE}{dt} ~&=&~ -5 H \rho_\KE - P^{(\rho)}_{\KE,\gamma} ~ .
\label{eq:eomsKE}
\end{eqnarray}

It turns out that $\rho_\KE$ will be vanishingly small during the period of interest.  
Because of this, we can ignore the impact of $\rho_\KE$ on the cosmological evolution 
(essentially treating $\Omega_\KE$ as zero), and only consider the effects that come 
from the time-evolution of the temperature $T$. Under these assumptions, this 
time-evolution is given by
\begin{equation}
  \frac{dT}{dt} ~=~ -2 H T  + T\frac{P^{(\rho)}_{M,\gamma}}{\rho_M} 
    - T\frac{P^{(\rho)}_{\KE,\gamma}}{\rho_\KE}~.
\label{eq:TempDiffEq}
\end{equation}

We can also determine the time-evolution of the corresponding abundances
$\Omega_i \equiv 8\pi G\rho_i/(3H^2)$ for the matter and radiation.  Given the 
above results, we obtain 
\beq 
 \frac{d\Omega_M}{dt} ~=~ 
    H \Omega_M\left( 1 - \Omega_M\right)~ -P_{M,\gamma}~,
\label{eq:convert3} 
\eeq 
where $d\Omega_\gamma/dt = -d\Omega_M/dt$ and where we have defined the pump 
terms for the transfer of abundances rather than of energy densities:
\beq
  P_{ij} ~\equiv~ \frac{8\pi G}{3 H^2} \, P^{(\rho)}_{ij}~.
  \label{eq:abundance_pump}
\eeq
Note that the relation $d\Omega_\gamma/dt = -d\Omega_M/dt$ applies in any 
universe containing only matter and radiation, since such universes necessarily have 
$\Omega_{\rm tot} = \Omega_M+\Omega_\gamma =1$.  This in turn implies that
\begin{equation}
  \frac{d\rho_{\rm tot}}{dt} ~=~ -(4-\Omega_M) H \rho_{\rm tot}~
\label{eq:rhotot_t}
\end{equation}
within any such universe.

In order to proceed further, we require explicit expressions for the pumps 
$P_{M,\gamma}^{(\rho)}$ and $P_{\KE,\gamma}^{(\rho)}$.  Each of these pumps 
depends on a thermal average of the cross-section.  In addition, because any 
$\phi$-annihilation process has two incoming matter $\phi$ particles, each pump 
depends on $\rho_M^2$ rather than $\rho_M$ alone. In general, these pumps are 
given by
\begin{eqnarray}
  P^{(\rho)}_{M,\gamma} &=& \frac{1}{m}\langle \sigma v\rangle \rho_M^2 
     \nonumber \\
  P^{(\rho)}_{\KE,\gamma} &=&
    \frac{1}{2m^2}\Bigl\langle \bigl(\KE_a + \KE_b\bigr)
    \sigma v\Bigr\rangle \rho_M^2~,
  \label{eq:PumpsInTermsofAvgs}
\end{eqnarray}
where $\langle X \rangle$ denotes the thermal average of $X$ and where $\KE_a$ and 
$\KE_b$ denote the kinetic energies of the two annihilating particles $\phi_a$ 
and $\phi_b$. Explicitly evaluating these thermal averages then leads to the 
results~\cite{Barber:2024vui}
\begin{eqnarray}
  P^{(\rho)}_{M,\gamma} &=&
    \frac{ \rho_M^2}{m} \, C \,\left(\frac{T}{m}\right)^{q/2} \, A(q)\nonumber\\
  P^{(\rho)}_{\KE,\gamma} &=&
    \frac{\rho_M^2 T}{2 m^2} \left(\frac{q+6}{2}\right) 
    C \,\left(\frac{T}{m}\right)^{q/2} \, A(q)~~ 
\label{quadratically}
\end{eqnarray}
where 
\begin{equation}
  A(q) ~\equiv~ \frac{2}{\sqrt{\pi}} \, \Gamma\left(\frac{q+3}{2}\right)~.
\label{eq:Adef}
\end{equation}
Note that $A(q)>0$ for all $q> -3$.

\subsection{Stasis fixed point}

At this stage, we can now demonstrate the existence of a thermal-stasis fixed point.
To do this, let us first define a new dynamical variable, the so-called {\it coldness}
\begin{equation}
  \Xi~\equiv~ \frac{T^q \rho_M}{m^{q+4}} ~.
\label{eq:DefOfSParam}
\end{equation}
This name for $\Xi$ reflects the fact that $q$ will be negative within our region 
of interest;  we therefore find that $\Xi$ is larger when $T$ is smaller, and vice 
versa.  Likewise, we can also introduce the quantity $\calN \equiv \log(a/a_0)$, 
which indicates the number of $e$-folds of cosmological expansion which have occurred 
between an early fiducial time (at which the scale factor was $a_0$) and any later 
time (such as the present, with scale factor $a$).  Thus $\calN$ can serve as an 
alternative clock variable, replacing $t$.  In terms of these new variables, the 
dynamical equations governing the evolution of our $(\Omega_M,\Xi)$ system then take 
the relatively simple form
\begin{eqnarray}
  \frac{d\Omega_M}{d\calN} ~&=&~ \Omega_M \left[1-\Omega_M 
    - \widehat{C} A(q) \sqrt{ \Xi\Omega_M} \, \right]\nonumber\\
  \frac{d\Xi}{d\calN} ~&=&~  \Xi\left[ -\left(2q+3\right)   
    -\widehat{C} \left(1+\frac{q^2}{6}\right) A(q) \sqrt{ \Xi \Omega_M}
    \right]~,~\nonumber\\
\label{Sweqs_expanded}
\end{eqnarray}
where we have defined
\begin{equation}
  \widehat C ~\equiv~ \sqrt{\frac{3}{8\pi G}}\, m\, C ~.
\label{Chatdef}
\end{equation}

Given these equations, we can immediately see that there is a fixed-point solution 
given by 
\beqn
  \barOmega_M &=& 1 + \frac{2q+3}{1+q^2/6} \nonumber\\
  \barXi &=& \frac{1}{\barOmega_M} 
    \left[ \frac{1-\barOmega_M}{\widehat C A(q)} \right]^2 ~
\label{eq:stasisvalues}
\eeqn
whenever $q$ is within the range
\begin{equation}
    -6+2\sqrt{3} ~<~ q ~<~ -3/2~~.
\label{eq:q-range}
\end{equation}
Indeed, as long as $q$ is within this range, we find that  $0\leq \barOmega_M\leq 1$.  
The values within Eq.~(\ref{eq:stasisvalues}) then correspond to our stasis 
solution wherein the matter abundance $\Omega_M$ remains constant despite cosmological 
expansion.  During this stasis, both the temperature $T$ and the matter energy density 
$\rho_M$ continue to fall.  However, they each fall in such a way as to keep the 
coldness $\Xi$ constant.   Thus, as explained in more detail in 
Ref.~\cite{Barber:2024izt}, it is the coldness $\Xi$ --- rather than the temperature 
$T$ --- which remains constant during stasis.

\subsection{Stasis attractor}

The next step is to demonstrate that the stasis solution in Eq.~(\ref{eq:stasisvalues}) 
is the end point of a dynamical {\it attractor}, with our system flowing {\it towards}\/ 
(rather than away from) this solution regardless of its initial location 
$(\Omega_M^{(0)}, \Xi^{(0)})$ within the $(\Omega_M,\Xi)$ plane.  To do this, we 
simply approximate our dynamical equations so that they have the form in 
Eq.~(\ref{diffeq}) near the fixed point, specifically
\begin{equation} 
 \begin{pmatrix} 
   d\Omega_M/d\calN \\
   d\Xi/d\calN 
 \end{pmatrix} 
 ~\approx~
 \begin{pmatrix}
   J_{\Omega\Omega} & J_{\Omega \Xi} \\
   J_{\Xi\Omega} & J_{\Xi\Xi} 
 \end{pmatrix}
 \begin{pmatrix}
   \Omega_M - \barOmega_M \\
   \Xi - \barXi 
 \end{pmatrix}~,
\label{simplediffeqs} 
\end{equation} 
where $(\barOmega_M,\barXi)$ are given in Eq.~(\ref{eq:stasisvalues}) and where
\beqn
J_{\Omega\Omega} &=&  -\half \, (1+\barOmega_{M})
        \nonumber\\ 
J_{\Omega\Xi} &=& - \half \left(1 - \barOmega_{M}\right)
\barOmega_M/\barXi \, 
\nonumber\\
J_{\Xi\Omega} &=& - \half\, 
(- 2 q - 3) 
\, \barXi/\barOmega_M 
         \nonumber\\
J_{\Xi\Xi} &=& -\half \,(-2q - 3) ~.
\label{Jelements}
\eeqn
Given this Jacobian matrix, we find the eigenvalues
\beqn
 &&  \lambda_1,\lambda_2 ~=~
 - \frac{\barOmega_{M}}{4} + \frac{q}{2} + \frac{1}{2}  \nonumber\\
   &&~~~~~~\pm  \frac{1}{4} \sqrt{\barOmega_{M}^{2} 
   + 12 \,\barOmega_{M} q + 20 \,\barOmega_{M} + 4 q^{2} + 8 q + 4}
   ~. ~~~ \nonumber\\
\label{eigenvalues}
\eeqn
Within the range for $q$ given in Eq.~(\ref{eq:q-range}), we see that both of these 
eigenvalues are negative.  This verifies that the dynamics of our system actually 
constitutes an attractor, as desired.

\subsection{Fast and slow trajectories} 

As discussed in the Introduction, another important required ingredient is that 
our attractor exhibit both ``fast'' and ``slow'' trajectories.   However, it is 
straightforward to verify that the attractor above also exhibits this feature.
Indeed, given the results in Eq.~(\ref{eigenvalues}), we can immediately see 
that these Jacobian eigenvalues vary within the approximate numerical ranges
\begin{eqnarray}
    0 &\lesssim& \abs{\lambda_1} ~\lesssim~ 0.2 \nonumber\\
    0.9 &\lesssim& \abs{\lambda_2} ~\lesssim~ 1.6~.
\end{eqnarray}
Thus, our eigenvalues exhibit a moderate hierarchy between them.
In fact, 
for the $q= -2$ special case, the eigenvalues are both rational, \ie, 
\begin{equation}
    (\la_1,~\la_2) ~=~ \left(-\frac{1}{5},~ -1\right)~,
    \label{lambdasForQEqNeg2}
\end{equation}
with a relative factor of five between them.

This in turn implies that there will exist both fast and slow trajectories along 
which our system can evolve towards stasis.  Indeed, whether our system is pulled 
towards stasis along a fast or slow trajectory ultimately depends on the starting 
location of our system within the $(\Omega_M, \Xi)$ plane.   

There are several ways in which we might quantify this.  One measure of the 
``speed'' with which our system approaches the stasis fixed point is the length 
of time it takes our system to arrive there from a given starting location.
In Eq.~(\ref{stasis_reached}) we  provided one possible definition for when 
stasis has been reached, but it turns out that using such a definition is 
somewhat cumbersome and not always easy to implement without prior knowledge 
of the location of the stasis point.  We shall therefore adopt an alternative 
approach which relies on the fundamental property of stasis itself, namely
that relevant abundances such as $\Omega_M$  remain unchanged despite cosmological 
expansion.  Accordingly, we shall define the ``arrival'' of our system into stasis 
along a given trajectory according to a criterion which involves the rate of change 
of a given abundance.  For example, we could associate this arrival with the time 
at which this rate of change falls below a given cutoff.   However, such a criterion 
might also depend on features intrinsic to the  cosmological expansion which are 
completely independent of the existence of the pump and which are therefore independent 
of the emergence of the stasis.  In order to establish a convention for the arrival 
of our system which is independent of these pump-independent details, we should 
therefore normalize the time evolution of the relevant abundance by the time 
evolution of the relevant abundance that would have emerged even if the pump had 
been absent.

We shall therefore adopt the convention that our system has arrived at the stasis 
location as soon as 
\beq 
 \abs{\left\langle\frac{d\Omega_i/d \Ncal}
   {\left. \left(d\Omega_i/d\Ncal\right) \right\vert_{P_{ij}=0}}
   \right\rangle} ~<~\delta~,
\label{trueDeltaDefinition}    
\eeq
where $\delta$ is an arbitrary cutoff.  Of course, it may happen that this condition 
is satisfied instantaneously during an early period far from the stasis fixed 
point if the abundances happen to behave non-monotonically.  However, we are 
focused here on the late-time evolution of our system as it approaches the stasis 
fixed point and therefore exhibits increasingly small values of $|d\Omega_i/d\calN|$.

\begin{figure}[t]
  \centering
  \includegraphics[width=0.55\textwidth]{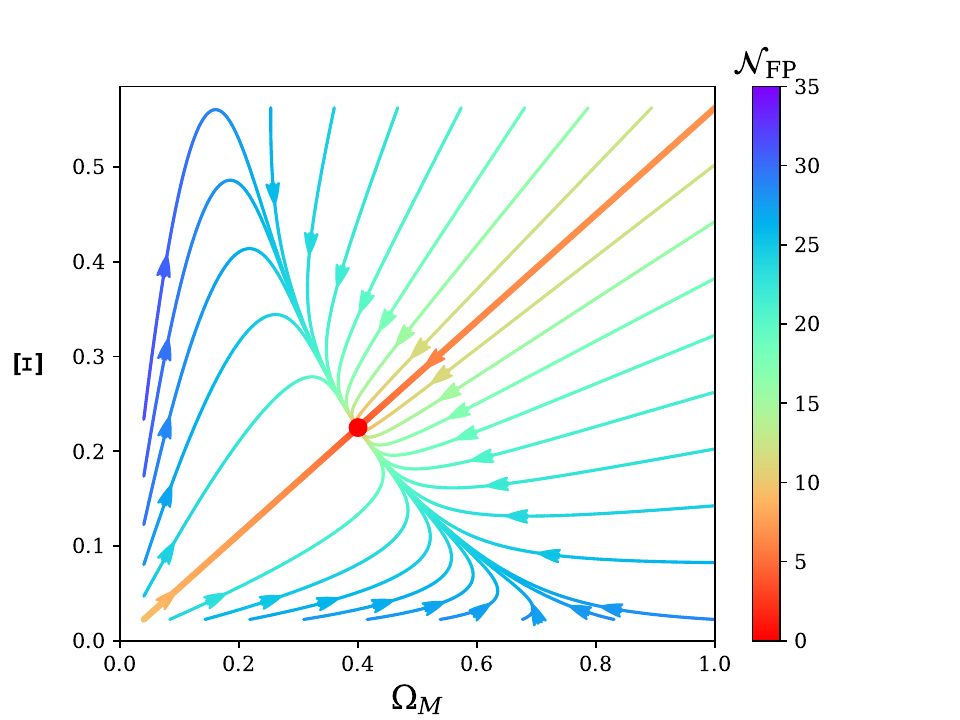}
  \caption{The number of $e$-folds $\calN_{\rm FP}$ that are required to reach stasis 
    along any attractor trajectory, starting from any point in the $(\Omega_M,\Xi)$ plane.   
    The fixed-point stasis solution is indicated with a solid red dot, while the values 
    of $\calN_{\rm FP}$ along any trajectory are indicated by the corresponding colors 
    and evaluated with $q= -2$ taken as a reference value, utilizing the convention for 
    arrival at the stasis location given in Eq.~(\ref{trueDeltaDefinition}) with 
    $\delta= 0.001$.  We observe the existence of a wide range of trajectories exhibiting 
    a wide range of times required to reach stasis, including a ``fastest'' straight-line 
    diagonal trajectory indicated with a thicker  red/orange line. 
\label{fig:heatmap_attractor_wait_time}}
\end{figure}

Adopting Eq.~(\ref{trueDeltaDefinition}) as our definition for arrival at the 
stasis fixed point, we can now evaluate the number of $e$-folds $\calN_{\rm FP}$ 
that are required to reach stasis along any trajectory, starting from any point 
in the $(\Omega_M,\Xi)$ plane.  The results are shown in 
Fig.~\ref{fig:heatmap_attractor_wait_time} for $q= -2$ and $\delta=0.001$.

As evident in this figure, our system exhibits a large variety of attractor 
trajectories which collectively exhibit a broad range of times required for
the system to reach stasis.  Interestingly, 
we also see within this figure the existence of a trajectory which proceeds in a 
straight line through the $(\Omega_M,\Xi)$ plane and which reaches the stasis 
point in the shortest time possible.  It turns out that for $q=-2$ the velocity 
with which our system proceeds along this ``fastest'' trajectory can be obtained 
analytically.  To see this, we recognize that this fastest trajectory 
is described by the equation
\begin{equation}
  (\Omega_M, \Xi) ~=~ s\,(\barOmega_M, \barXi)~,
\label{fastest}
\end{equation}
where $s$ indicates the instantaneous (time-dependent) position of our system 
along this trajectory.  Indeed, for $q= -2$, substituting this solution into 
Eq.~(\ref{Sweqs_expanded}) we obtain a differential equation for $s$, namely 
the so-called ``logistic'' equation
\begin{equation}
   \frac{ds}{d\calN}~=~ s(1-s)~,
\end{equation}
with solutions of the general form
\begin{equation}
  s(\calN) ~=~ \frac{1}{1+ A \,e^{-\calN}}~.
\label{gensoln}
\end{equation}
Here $A$ depends on the initial conditions of our system.   Clearly initial conditions 
with $s<1$ correspond to solutions with $A>0$, while those with $s>1$ correspond to 
solutions with $A<0$.  For $A>0$, the solution $s(t)$ is a sigmoid function which 
evolves from $0$ to $1$ over the range $\calN \in (-\infty,\infty)$.  In such cases, 
our system starts at a location below and to the left of the stasis fixed point in
Fig.~\ref{fig:heatmap_attractor_wait_time} and begins to accelerate towards the fixed 
point along the fastest trajectory before reaching a maximum velocity and then slowing 
down again, ultimately asymptoting to the stasis fixed point exponentially slowly. 
By contrast, systems which have initial values $s>1$ have $A<1$.  They therefore 
begin above and to the right of the stasis fixed point within 
Fig.~\ref{fig:heatmap_attractor_wait_time}.  Such systems immediately evolve toward 
the fixed point with velocities that continually decrease as the stasis fixed point 
is reached.  Indeed, in each case, the stasis fixed point is approached only 
asymptotically, with ever-decreasing velocities.

Given this result, one might wonder if it is a general principle
that the fastest approach to stasis is always governed by a logistic equation.   
Indeed, for the stasis attractor discussed in Ref.~\cite{Dienes:2023ziv},
which is associated with the overdamped/underdamped transitions of a tower of scalar-field
zero-modes, the fastest approach to stasis turns out also to be governed by a logistic 
equation.  However, this is not a general feature, and there exist many examples of 
stasis attractors  for which the fastest trajectories do not correspond to logistic 
equations.

\subsection{Explicit model and expiration dates}\label{subsec:expiration_date}

Our final step is to demonstrate that our scenario comes with a built-in mechanism 
that limits the time during which our cosmological system can experience the pull of 
our stasis attractor.  In other words, as we shall now demonstrate, our scenario gives 
rise to an automatic {\it expiration date}\/ for the attractor beyond which it no longer 
functions.

Thus far, we have merely {\it asserted}\/ the existence of an annihilation cross-section 
of the form in Eq.~(\ref{eq:swept_vol_rate}), with $q$ within the range in 
Eq.~(\ref{eq:q-range}).  Indeed, as we have seen, it is necessary to have a pump term 
of this form with $q$ within this range in order to obtain our fixed-point stasis 
solution which serves as a cosmological attractor.  However, it still remains to 
determine whether it possible to build a realistic scenario within which such an
annihilation cross-section actually arises.

Towards this end, we now quickly review a particle-physics mechanism which can 
accomplish this feat~\cite{Barber:2024vui}.  This will thereby not only furnish us 
with an explicit realization of our pump, but in so doing also lead to a natural 
understanding of various constraints that ultimately limit the time during which our 
cosmological attractor can exist.

As discussed above, our pump involves the annihilation of two matter fields $\phi$ 
into radiation (with a corresponding radiation field denoted $\chi$).  To be more 
explicit about this annihilation process, and to demonstrate that this process can 
have an annihilation cross-section of the desired form, we shall consider a model 
consisting of {\it three}\/ real scalar fields $\phi$, $X$, and $\chi$ which are all 
singlets with respect to all Standard-Model (SM) symmetries and which do not couple to 
the scalar sector of the SM in any way.  We shall also assume that the Lagrangian for 
these three fields is invariant under two independent $\IZ_2$ symmetry 
transformations: one under which $\phi$ is odd while $\chi$ and $X$ are both even, 
and another under which $\chi$ is odd while $\phi$ and $X$ are both even.  With 
these restrictions, this Lagrangian takes the form 
\begin{equation}
  \Lcal~=~ \hf (\p \phi)^2 +\hf (\p X)^2+\hf (\p \chi)^2 - U~,
  \label{eq:Lagrangian}
\end{equation}
where the scalar potential $U$ contains the terms
\begin{eqnarray}
  U &=& \hf m^2 \phi^2 + \hf m_X^2 X^2 + \frac{1}{2} g_\phi m\,\phi^2 X 
    + \frac{1}{2} g_\chi m X \chi^2 \nonumber\\
    &&  
      + \frac{1}{4!}\la_\phi \phi^4 
      + \frac{1}{4!}\la_\chi \chi^4 + \frac{1}{4}\la_{\phi\chi} \phi^2\chi^2
      + \ldots~~~
  \label{eq:ScalarPot}
\end{eqnarray}
In principle, we can also include within this scalar potential $U$ a number of 
additional trilinear and quartic interaction terms involving two or more factors 
of $X$.  However, such additional terms will not have a significant impact on the 
dynamics of our model within our eventual regime of interest, and we shall not 
consider them further.

Two additional comments about this Lagrangian are in order.  First, we have 
elected to parametrize the super-renormalizable interactions appearing in the first 
line Eq.~(\ref{eq:ScalarPot}) in terms of the overall mass scale $m$
and a pair of dimensionless couplings $g_\phi$ and $g_\chi$.  This choice is merely 
a convention and does not constitute a loss of generality.  Indeed, for reasons 
that will shortly become apparent, within this framework we will require that 
$g_\chi \ll g_\phi \lsim 1$.

Second, the quartic interaction terms which appear in the second line of 
Eq.~(\ref{eq:ScalarPot}) --- while consistent with the symmetries of our 
model --- do not play an essential role in the feature that will ultimately 
interest us most, namely the emergence of stasis.  Moreover, as we shall 
demonstrate in detail in Appendix~\ref{sec:Quartics}, there exist broad regions 
of our model-parameter space wherein these interactions contribute negligibly to 
the scattering and annihilation rates which impact the stasis dynamics, but 
wherein $\lambda_\phi$, $\lambda_\chi$, and $\lambda_{\phi\chi}$ are also not 
unnaturally small or finely tuned.  Thus, in what follows, we shall simply
assume that the values of $\lambda_\phi$, $\lambda_\chi$, and $\lambda_{\phi\chi}$ 
are such that they can be ignored in our analysis.  

\begin{figure}[H]
    \centering
    \begin{tikzpicture}
    \begin{feynman}
        \node[ blob,draw=gray,pattern color=gray,very thick] (loop);
        \vertex [left=of loop] (mx);
        \vertex [right=of loop] (gx);
        \vertex [below left=of mx] (i1) {\(\phi\)};
        \vertex [above left=of mx] (i2) {\(\phi\)};
        \vertex [above right=of gx] (f1) {\(\chi\)};
        \vertex [below right=of gx] (f2) {\(\chi\)};
        \diagram* {
        (mx) -- [red, plain, very thick, edge label=\(\textcolor{black}{X}\)] (loop) 
          -- [red, plain, very thick, edge label=\(\textcolor{black}{X}\)] (gx),
        (i1) -- [blue, plain, very thick] (mx) -- [blue, plain, very thick] (i2),
        (f1) -- [green, plain, very thick] (gx) -- [green, plain, very thick] (f2)
        };
    \end{feynman}
 \end{tikzpicture}
    \caption{
    Diagrammatic depiction of the $s$-channel 
    annihilation process which underlies our stasis pump.
    The one-loop correction for the $X$ propagator (shown in gray) includes 
    contributions from both virtual $\phi$-particle and $\chi$-particle pairs.  
    Figure taken from  Ref.~\cite{Barber:2024vui}. 
    \label{fig:pump_diagram}}
\end{figure}
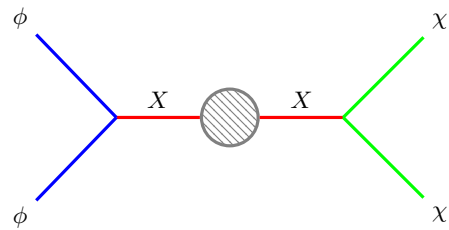

With these simplifications, the parameter space of our model becomes effectively 
four-dimensional, comprising the couplings $g_\phi$ and $g_\chi$ and the mass 
scales $m$ and $m_X$.  Given the three fields $\phi$, $\chi$, and $X$, we shall 
then realize our pump as resulting from $\phi$-annihilation of the form 
$\phi\phi\to X \to \chi\chi$ through an $s$-channel process with $X$ as the 
mediator~\cite{Barber:2024vui}. This process is depicted diagrammatically in 
Fig.~\ref{fig:pump_diagram}.  
Evaluating the propagator of the intermediary $X$ particle, we obtain 
\begin{equation}
\Delta(p_X) ~\sim~
   \frac{i}{m^2} 
  \left( \widetilde{\alpha}+i\beta
  \frac{|\vec p_{\CM}|}{m} + 4
   \frac{|\vec p_{\CM}|^2}{m^2}\,\right)^{-1},
 \label{eq:propag}
\end{equation}
where the coefficients are given by
\begin{eqnarray}
  \widetilde{\alpha}  &=& 
    -\frac{\mu^2}{m^2} + \frac{ig_\chi^2}{32\pi} \nonumber\\
  \beta &=& \frac{g_\phi^2}{32\pi}~
  \label{eq:alpha_beta_model}
\end{eqnarray}
and where we have defined the parameter
\begin{equation}
   \mu^2 ~\equiv~ m_X^2 - 4m^2~.
\label{eq:mudef}
\end{equation}

The fact that a term linear in $|\vec p_\CM|$ appears in the denominator of 
$\Delta(p_X)$ in Eq.~(\ref{eq:propag}) is crucial for the emergence of the 
stasis attractor.  Indeed, so long as the couplings associated with the 
interaction vertices in Fig.~\ref{fig:pump_diagram} are independent of 
$|\vec p_\CM|$, we see that the swept-volume rate will scale as 
$\sigma v \sim |\Delta(p_X)|^{2}$.  Thus, if the term linear in $|\vec p_\CM|$ 
within the denominator of the propagator $\Delta(p_X^2)$ in Eq.~(\ref{eq:propag})
dominates over the other two terms, the swept-volume rate will have the form 
Eq.~(\ref{eq:swept_vol_rate}) with $q = -2$, a value which falls within the 
range in Eq.~(\ref{eq:q-range}).  Indeed, regardless of the complex phase of 
$\widetilde{\alpha}$, we find~\cite{Barber:2024vui} that this linear term dominates 
whenever the our system satisfies the conditions
\begin{eqnarray}
   4\alpha &\ll& \beta^2 
  ~~~~~~~~~~~~~~ \nonumber \\
  \frac{\alpha}{\beta}m &\ll& |\vec p_\CM |
    ~\ll~ \frac{\beta}{4} m~,
  \label{eq:allowedp}
\end{eqnarray} 
where $\alpha \equiv |\widetilde{\alpha}|$.  The first of these conditions stipulates 
that
\begin{equation}
  g_\phi ~\gg~ 2\left(g_\chi^4 + 1024 \pi^2\frac{\mu^4}{m^4}\right)^{1/4}~,      
\end{equation}
which is tantamount to requiring that
\begin{equation}
  g_\phi~\gg~ g_\chi~,~~~~
  \frac{g_\phi}{\sqrt{32\pi}} ~\gg~ \frac{|\mu|}{m}~.
\label{conditions}
\end{equation}
Likewise, the second condition in Eq.~(\ref{eq:allowedp}) implies that
\begin{equation}
  \left[\frac{g_\chi^4}{g_\phi^4} + 
    \left(\frac{32\pi\mu^4}{g_\phi^2 m^4}\right)^2\,\right]^{1/2} \!\! m
    ~\,\ll~ |\vec p_\CM| ~\ll~ \frac{g_\phi^2 }{128\pi}\,m~.
  \label{eq:allowedp_model}
\end{equation}
In what follows, we shall focus our attention on regions of our parameter space 
within which these conditions are satisfied. In other words, we shall neglect 
subleading terms in all expressions which are proportional to $g_\chi/g_\phi$ or to 
$\mu^2/m^2$.

The first of the conditions in Eq.~(\ref{eq:allowedp}) simply constitutes a 
constraint on the parameters which characterize a given model realization of the 
stasis attractor.  By contrast, the second implies that even in situations in 
which this first condition is satisfied, this attractor can be
realized only while the temperature of the $\phi$-particle gas lies within
a particular window.  Indeed, since the momentum distribution for 
particles within this gas is Maxwellian by assumption, the characteristic 
scale for $|\vec p_\CM|$ is  
\begin{equation}
  T ~\sim~ |\vec p_{\CM}|^2/m~.
  \label{eq:emp_sim_from_pcm}
\end{equation} 
Thus, the rough temperature window which corresponds to the allowed range of 
$|p_{\rm CM}|$ in Eq.~(\ref{eq:allowedp}) is given by
\begin{equation}
  \Tmin ~\lesssim ~ T ~\lesssim~ \Tmax~,
  \label{Trange}
\end{equation}
where 
\begin{eqnarray} 
  \Tmax ~&\equiv &~ \left(\frac{g_\phi^2}{128 \pi}\right)^2 m
    \nonumber\\
  \Tmin ~&\equiv &~ \left[\frac{g_\chi^4}{g_\phi^4} +
    \left(\frac{32\pi\mu^2}{g_\phi^2m^2}\right)^2 \right]m~.
  \label{eq:T_bounds_propagator}
\end{eqnarray}

In most regimes of phenomenological interest, the temperature drops with time 
during stasis.  This lower limit on $T$ therefore provides us with an ``expiration 
date'' for our attractor --- \ie, a time beyond which our attractor no longer 
functions in the manner needed in order to provide stasis. Indeed, for times 
beyond that at which the temperature is given by $T_{\rm min}\/$, the 
annihilation process we have been studying no longer has an effective scaling 
exponent $q= -2$, but instead has $q= 0  $. This is clearly outside the range 
given in Eq.~(\ref{eq:q-range}).  Thus, for $T<T_{\rm min}$, this is no longer 
an attractor that gives rise to stasis --- in fact, even the attractor ceases 
to exist.

But this is not all.   This analysis also demonstrates that there is not only a 
{\it minimum}\/ allowed temperature for our stasis attractor, but also a 
{\it maximum}\/ allowed temperature.  Indeed, if the existence of a minimum 
allowed temperature $T_{\rm min}$ can be viewed as providing an ``expiration'' 
date for our stasis attractor, then the existence of a maximum allowed 
temperature $T_{\rm max}$ can likewise be viewed as providing an early time 
before which our stasis attractor also fails to exist.  We may therefore 
refer to this as a ``manufacture'' date for the attractor.   Thus, our 
attractor exists and can give rise to stasis only between its manufacture date 
and its expiration date.  Indeed, like many grocery items, such attractors have a 
finite shelf life.

For some choices of the couplings, we find that $\Tmin \ll \Tmax$.  This then leads to a 
sizable range of temperatures across which the attractor is active. However, for some 
choices of couplings, it is possible to have $\Tmax < \Tmin$.  In such cases, our 
attractor has already expired at the time of manufacture.  Of course, if a physical effect 
subsequently raises the temperature, this could act to ``revive'' the attractor
once again.

Note that achieving the desired scaling behavior for $\sigma v$ is also 
contingent on the annihilating $\phi$ particles being non-relativistic --- \ie, 
on the assumption that $|\vec p_\CM|\ll m$.  However, given that 
$\beta/4 = g_\phi^2/(128\pi)\ll 1$ in our model for all $g_\phi$ within the 
$g_\phi\lsim 4\pi$ perturbative regime, the upper bound on $|\vec p_\CM|$ in 
Eq.~(\ref{eq:allowedp_model}) subsumes this additional constraint.  Thus, within 
the context of this model, we need not impose $|\vec p_\CM|\ll m$ as an independent 
constraint.

At times when the temperature of our $\phi$-particle gas satisfies the 
condition in Eq.~(\ref{Trange}), we may obtain an explicit expression for $\sigma v$.
Since the linear term within denominator of the propagator in 
Eq.~(\ref{eq:propag}) dominates for typical $\phi$ particles at such values of 
$T$, we can approximate 
\begin{equation}
  \Delta(p_X) ~\approx ~ \frac{1}{\beta m |\vec p_\CM|}
     ~=~ \frac{32\pi}{g_\phi^2} 
     \frac{1}{m |\vec p_\CM|}~.~
  \label{eq:DeltaInAttractorLand}
\end{equation}
This in turn leads to a swept-volume rate $\sigma v$ which takes the form in
Eq.~(\ref{eq:swept_vol_rate})
with $q= -2$ and with
\begin{equation} 
  C ~=~ \frac{16\pi g_\chi^2}{m^2 g_\phi^2}~.
  \label{formula_for_C}
\end{equation}

For later purposes, it will also prove useful to calculate the proper decay width 
$\Gamma_X$ of our mediator particle.  An elementary QFT calculation yields
\begin{eqnarray}
  \Gamma_X ~&=&~\Gamma_{X\to\chi\chi} + \Gamma_{X\to\phi\phi}\nonumber\\
   ~&=&~ \frac{g_\chi^2}{64\pi} \frac{2m^2}{m_X} 
    + \frac{g_\phi^2}{64\pi}\frac{m}{m_X}\abs{\mu} \Theta(\mu^2/m^2)~,\nonumber\\
\end{eqnarray}
where $\Theta(x)$ is the Heaviside theta-function.

\subsection{Summary}

To summarize, we see that our framework successfully exhibits all of the required 
key features outlined in the Introduction, specifically:
\begin{itemize}
  \item  A fixed-point stasis solution given in Eq.~(\ref{eq:stasisvalues});  
  \item Jacobian eigenvalues $\lambda_{1,2}$ which are both negative in the 
    vicinity of this stasis solution, as shown in Eq.~(\ref{lambdasForQEqNeg2}) 
    and in Fig.~\ref{fig:heatmap_attractor_wait_time}, thereby demonstrating 
    that our fixed-point stasis solution is actually a stasis {\it attractor}\/;  
  \item ``Fast'' and ``slow'' attractor trajectories, with the fast trajectory 
    having an exact solution given in Eqs.~(\ref{fastest}) and (\ref{gensoln}).  
    Indeed, the regions of the two-dimensional space of initial dynamical 
    variables ($\Omega_M,\Xi$) which lead to fast/slow trajectories are shown 
    in Fig.~\ref{fig:heatmap_attractor_wait_time}.  The $q= -2$ ``fastest'' 
    trajectory is also shown in this figure;   and 
  \item  Manufacture and expiration times, defined as in 
    Eq.~(\ref{eq:T_bounds_propagator}), which respectively indicate the times 
    at which our attractor first begins and later ceases to operate.    
\end{itemize}


\section{Model constraints and consistency conditions\label{sec:constraints}}


In the previous section, we presented a model which gives rise to a $2\to 2$ annihilation 
cross-section for our $\phi$ particles that scales with $|\vec p_{\rm CM}|$ according to
Eq.~(\ref{eq:swept_vol_rate}) with $q=-2$.  Indeed, this is within the range for $q$  that 
allows the emergence of a stasis epoch.  However, models such as this are also subject to a 
number of internal self-consistency conditions and phenomenological constraints.  Therefore, 
before proceeding further, it is necessary for us to assess whether there exist regions 
within the parameter space of this model wherein all of these consistency conditions and 
constraints are satisfied.  Furthermore, within such regions of that parameter space, we 
must also assess how long the corresponding stasis epoch can possibly last.  Indeed, as 
we saw in Sect.~{\ref{subsec:expiration_date}}, stasis can only be achieved while the 
stasis attractor is active, \eg, while the temperature $T$ of the $\phi$-particle gas lies 
within the range specified in Eq.~(\ref{Trange}).  This consideration imposes an upper 
limit on the number $\Ncal_s$ of $e$-folds of expansion that the universe can possibly 
undergo during stasis for any given combination of our model parameters.  Moreover, the 
actual value of $\Ncal_s$ obtained for an particular set of initial conditions for our 
cosmological energy components may fall below smaller --- and potentially significantly 
below --- this upper limit.

In performing this analysis of the constraints on our model, we shall often find it 
convenient to parametrize the mass scales $m$ and $\mu$ which characterize our model in 
terms of dimensionless quantities.  In order to do this, we shall parametrize $m$ in
terms of the quantity
\begin{eqnarray}
  g_G ~\equiv~ \frac{m}{M_P} ~=~ \sqrt{G}m~,
\end{eqnarray}
where $M_P$ is the Planck mass.  With this definition, the parameter space of our model 
can effectively be characterized by four dimensionless parameters: the coupling constants 
$g_\phi$ and $g_\chi$ and the dimensionless ratios $g_G$ and $\mu/m$. We find that the fourth 
quantity $\mu/m$ is of lesser importance as it will have essentially no effect on the model 
dynamics at all if it below a certain threshold given by
\begin{equation}
  \frac{\abs{\mu}}{m} ~\ll~ \frac{g_{\gamma}}{2^{5/2}\sqrt{\pi}}~,
  \label{mumaximumeff}
\end{equation}
which means the other three quantities will typically be the important parameters of the model.
Moreover, we shall also find it convenient to recast our dynamical variables $\rho_M$ 
and $T$ in terms of the dimensionless quantities
\beq
\varrho_M ~\equiv~ \frac{\rho_M}{m^4} ~,~~~~
  \tau ~\equiv~ \frac{T}{m}~.~~~
\eeq
We note that our coldness parameter $\Xi$, which is inherently dimensionless, 
is completely specified by $\varrho_M$ and $\tau$.  For $q = -2$, the 
relationship between these quantities is
\begin{equation}
  \Xi ~=~ \frac{\varrho_M}{\tau^2}~.
  \label{eq:xi_def}
\end{equation}

The constant value $\barXi$ of $\Xi$ during stasis is entirely determined by the 
values of $g_\phi$, $g_\chi$, and $g_G$.  Indeed, substituting our expression for 
$C$ in \eq(\ref{formula_for_C}) into \eq(\ref{eq:stasisvalues}) and taking $q=-2$, 
we have 
\begin{equation}
  \barXi ~=~ \frac{3}{5\cdot 2^{8}\pi} \frac{g_\phi^4 g_G^2}{g_\chi^{4}}~.
  \label{eq:xibar_from_couplings}
\end{equation}

For any combination of our model parameters $g_\phi$, $g_\chi$, $g_G$, and $\mu/m$, 
stasis can only be realized while $T$ lies within the range specified in 
\eq(\ref{Trange}) --- or, equivalently, when $\tau$ lies within the range 
\begin{equation}
  \taumin ~\lsim ~ \tau ~\lsim~ \taumax~,
  \label{tau_constrained_range}
\end{equation}
where $\taumax$ and $\taumin$ are the values of 
$\tau$ which correspond to $\Tmax$ and $\Tmin$, 
respectively.  In practice, since one or the other of the two terms in 
Eq.~(\ref{eq:T_bounds_propagator}) which contribute to $\Tmin$ dominates 
throughout the vast majority of our parameter space, we shall frequently 
approximate $\Tmin$ at any point within that parameter space simply 
as the larger of these two terms.

Consistency with observation also requires that the energy density of universe 
be dominated by the contribution from the visible-sector radiation bath by the time
that BBN begins.  The duration of the stasis epoch depends 
not only on the values of the model parameters $g_\phi$, $g_\chi$, $g_G$, and $\mu/m$, 
but also on the initial conditions for our cosmological energy components.  
In particular, in order to determine $\Ncal_s$, we must specify the initial values 
$\Omega_M^{(0)}$, $T^{(0)}$, and $H^{(0)}$ of the variables $\Omega_M$, $T$, and $H$, 
respectively, at some appropriately 
chosen fiducial time $t^{(0)}$.  Since stasis can only develop in this model once our 
population of $\phi$ particles has already begun to behave like massive matter, we 
choose $t^{(0)}$ to be sufficiently late that this population of particles is already 
non-relativistic, but sufficiently early that the universe is not already in stasis.  
The appropriate range of values for $t^{(0)}$ depends on the values of our model parameters.
For simplicity in what follows, we shall focus on the case in which the initial abundance
of our radiation field $\chi$ is negligible, and thus default to assuming that 
$\Omega_M^{(0)} = 1$ (essentially assuming there is a sizable amount of initial matter 
such that $\Omega_M^{(0)} \sim 1$).  By contrast, we shall take $T^{(0)}$
and $H^{(0)}$ to be free parameters.  However, we shall find it more convenient to specify
these initial conditions in terms of the initial values $\varrho_M^{(0)}$ and $\tau^{(0)}$ 
of the dimensionless matter-energy-density and temperature variables $\varrho_M$ and $\tau$,
respectively. 

We now examine, in turn, each additional consideration that constrains our 
thermal-stasis model.  The conditions on $\varrho_M$ and $\tau$ which follow from
these consideration (in addition to the conditions which we have already 
discussed) will ultimately be compiled for reference in Table~\ref{tab:constraints}.

\subsection{Maintaining kinetic equilibrium among the matter particles
\label{sec:sub:matterSelfEquilibrium}}

The dynamics which give rise to the stasis attractor in our thermal-stasis model 
is predicated on the assumption that the momentum distribution for our population of 
$\phi$ particles takes the form of a Maxwell-Boltzmann distribution throughout the 
stasis epoch and can therefore be completely characterized at any particular time $t$ 
by a corresponding temperature $T$.  This assumption is in turn predicated on the 
presence of rapid interactions which serve to maintain kinetic equilibrium among 
these particles.  The thermalization rate associated with these interactions 
must be sufficiently large throughout the stasis epoch that it exceeds the rate 
at which abundance is transferred from matter to radiation via the 
annihilation process $\phi\phi \to \chi\chi$.  We note that during stasis, this 
annihilation rate, which is simply the pump term $P_{M,\gamma}$ in Eq.~(\ref{eq:convert3}).
is necessarily equal to the rate of change of $\Omega_M$ due to cosmic expansion.

Within the context of our model, the principal process which serves to 
redistribute kinetic energy and momentum among our population of $\phi$ particles 
is the scattering process $\phi \phi \to \phi \phi$.  As a result, requiring that 
the thermalization rate exceed the annihilation rate associated with the process 
$\phi\phi\to \chi\chi$ is tantamount to requiring that 
\begin{equation}
  \frac{\langle\sigma v\rangle_{\phi\phi\to \phi\phi}}
    {\langle\sigma v \rangle_{\phi\phi\to \chi\chi}} ~\gg~ 1~,
  \label{eq:therm_cond_raw}
\end{equation}
where $(\sigma v )_{\phi\phi\to \phi\phi}$ and $(\sigma v )_{\phi\phi\to \chi\chi}$ 
denotes the swept-volume rates for the thermalization and annihilation processes, 
respectively.  

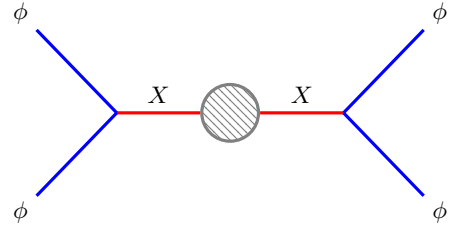
\begin{figure}[H]
    \centering
    \begin{tikzpicture}
    \begin{feynman}
        \node[ blob,draw=gray,pattern color=gray,very thick] (loop);
        \vertex [left=of loop] (mx);
        \vertex [right=of loop] (gx);
        \vertex [below left=of mx] (i1) {\(\phi\)};
        \vertex [above left=of mx] (i2) {\(\phi\)};
        \vertex [above right=of gx] (f1) {\(\phi\)};
        \vertex [below right=of gx] (f2) {\(\phi\)};
        \diagram* {
        (mx) -- [red, plain, very thick, edge label=\(\textcolor{black}{X}\)] (loop) 
          -- [red, plain, very thick, edge label=\(\textcolor{black}{X}\)] (gx),
        (i1) -- [blue, plain, very thick] (mx) -- [blue, plain, very thick] (i2),
        (f1) -- [blue, plain, very thick] (gx) -- [blue, plain, very thick] (f2)
        };
    \end{feynman}
 \end{tikzpicture}
    \caption{Feynman diagram for the $s$-channel matter-matter scattering process.} 
    \label{fig:elastic_scattering_diagram}
\end{figure}

While a variety of Feynman diagrams contribute to $(\sigma v )_{\phi\phi\to \phi\phi}$, 
the dominant contribution within our regime of interest for stasis is the one 
associated with the $s$-channel diagram depicted in 
Fig.~\ref{fig:elastic_scattering_diagram}.  Other contributions --- such as that 
associated with the corresponding $t$-channel diagram --- do not feature the same 
resonant enhancement for $|p_{\rm CM}| \approx 2m_X$ and are therefore subleading.
Thus, for simplicity in what follows, we shall focus exclusively on the contribution to 
$(\sigma v)_{\phi\phi\to\phi\phi}$ associated with the $s$-channel
diagram in Fig.~\ref{fig:elastic_scattering_diagram}.  

Given this approximation, and given the similarity between the diagram in
Fig.~\ref{fig:elastic_scattering_diagram} and the corresponding diagram for 
$\phi\phi \to \chi\chi$ annihilation in Fig.~\ref{fig:pump_diagram}, we note that
the ratio of swept-volume rates for the thermalization and annihilation process is
\begin{equation}
  \frac{(\sigma v)_{\phi\phi\to \phi\phi}}{(\sigma v )_{\phi\phi\to \chi\chi}} 
    ~=~ \frac{g_\phi^2}{g_\chi^2} \frac{\abs{\vec{p}_{\text{CM}}}}{m}~.
\end{equation}
Thus, since $(\sigma v )_{\phi\phi\to \chi\chi} \propto \abs{\vec{p}_{\text{CM}}}^{-2}$ 
for $q = -2$, we find that the ratio of corresponding thermal averages
of these swept-volume rates takes the form
\begin{equation}
  \frac{\langle\s v\rangle_{\phi\phi \to \phi\phi}}
      {\langle\s v\rangle_{\phi\phi\to \chi\chi}} 
    ~=~ \frac{g_\phi^2}{g_\chi^2}  
      \frac{\langle |p_{\rm CM}|^{-1}\rangle}{\langle |p_{\rm CM}|^{-2}\rangle m}
    ~=~ \frac{g_\phi^2}{g_\chi^2}\frac{1}{\sqrt{\pi}} 
      \left(\frac{T}{m}\right)^{1/2}~,
  \label{eqLtherm_rat_raw}
\end{equation}
where in going from the first to the second equality we have used
the fact that $\bigl\langle |\vec p_\CM |^q \bigr\rangle ~=~ (mT)^{q/2} \, A(q)$,
which follows from Eqs.~(\ref{eq:PumpsInTermsofAvgs}) and (\ref{quadratically}), along
with Eq.~(\ref{eq:Adef}).  We can also express Eq.~(\ref{eqLtherm_rat_raw}) in 
the more revealing form
\begin{eqnarray}
  \frac{\langle\s v\rangle_{\phi\phi\to\phi\phi}}{\langle\s v\rangle_{\phi\phi\to\chi\chi}} 
    ~=~ \left(\frac{\tau}{\tau_{\rm therm}}\right)^{1/2}
  \label{eq:therm_rat}
\end{eqnarray}
by defining the dimensionless thermalization temperature 
\begin{equation}
  \tau_{\rm therm} ~\equiv~ \pi \left(\frac{g_\chi}{g_\phi}\right)^4~.
  \label{eq:tau_therm}
\end{equation}
The result in Eq.~(\ref{eq:therm_rat}) implies that the condition in 
Eq.~(\ref{eq:therm_cond_raw}) will be satisfied so long as
\begin{equation}
  \tau ~\gg~ \tau_{\rm therm}~.
  \label{eq:therm_cond}
\end{equation}

We can assess impact of this requirement on the duration of stasis by   
comparing the expression for $\tau_{\rm therm}$ in Eq.~(\ref{eq:tau_therm}) to 
the expression to the expression for $\Tmin$ in 
Eq.~(\ref{eq:T_bounds_propagator}).  Within regions of our 
parameter space wherein the first term in brackets in the latter expression 
dominates, $\tau_{\rm therm}$ differs from $\taumin$ by only a factor 
of $\pi$.  By contrast, within regions wherein the second term dominates, 
we have $\taumin \gg \tau_{\rm therm}$.  Thus, even in the worst-case 
scenario, kinetic equilibrium is maintained among our population of $\phi$ 
particles until only shortly before the stasis epoch would have ended anyhow.
Thus, we conclude that the consistency condition in Eq.~(\ref{eq:therm_cond})
does not lead to a significant reduction in $\calN_s$ at any point within our
parameter space. 

We emphasize that the constraint on $\tau$ in Eq.~(\ref{eq:therm_cond}) applies
only once the universe is already in stasis, not while it is evolving toward stasis. 
Indeed, as we shall see, there exist certain regions of the $(\tau,\varrho_M)$ plane 
wherein additional scattering processes distort the phase-space distribution of the $\phi$ 
particles away from thermality at early times, but in such a way that the universe 
nevertheless evolves toward stasis.  These processes and their ramifications for 
cosmological dynamics shall be discussed extensively in Sect.~\ref{sec:Exothermic}.

Finally, we note that in addition to $\phi\phi\to \phi\phi$ scattering, there is yet 
another scattering process which arises in our thermal stasis model and which can 
potentially affect the cosmological dynamics associated with the stasis attractor.  
This is the scattering process $\phi\chi \to \phi \chi$, which facilitates the exchange 
of energy and momentum between the populations of $\phi$ and $\chi$ particles present 
in the universe.  Since this process drives these two particle populations toward 
thermal equilibrium with each other, we must ensure that the the swept-volume rate for 
this process is sufficiently small that $\phi\chi \to \phi \chi$ scattering does not 
significantly affect the cosmological dynamics.  In Appendix~\ref{sec:compton}, we analyze 
the impact of this scattering process and demonstrate that it has negligible impact on 
evolution of $\Omega_M$ and $T$ within our regime of interest.  Moreover, we also show 
that the overall effect of this process on the stasis dynamics is simply a shift in the 
value of $\barOmega_M$.

\subsection{General-relativistic effects on the $X$ propagator}\label{subsec:GravQFT}

The structure of the $X$ propagator $\Delta(p)$ plays a crucial role in 
the stasis dynamics of our model and is ultimately what leads to a 
swept-volume rate for $\phi$-particle annihilation of the form in 
Eq.~(\ref{eq:swept_vol_rate}) with $q= -2$.  However, in an expanding FRW 
universe, the Minkowski-space form of the propagator in Eq.~(\ref{eq:propag})
is modified by general-relativistic effects.  We must therefore ensure that
this modification is negligible within our parameter-space region of interest. 

In Minkowski space, the classical equation of motion for our mediator 
field is simply the Klein-Gordon equation:
\begin{equation}
  (-\partial^2 - m_X^2) X ~=~ 0~.
\end{equation}
By contrast, the corresponding equation of motion in an FRW universe (in the 
background frame) is
\begin{equation}
  (-\partial^2 -3H \partial_t - m_X^2) X ~=~ 0~,
  \label{eq:FRW_EOM_X}
\end{equation}
where the additional term proportional to $H$ accounts for the effect of
cosmic expansion.  We note that this additional term, which includes a 
partial derivative with respect to the coordinate time $t$ in this frame, 
explicitly breaks Lorentz invariance. 
The Green's function for the operator which acts on $X$ in Eq.~(\ref{eq:FRW_EOM_X}) --- 
which we may to a good approximation associate with the Feynman propagator 
$\Delta_{\rm FRW}(x,x')$ in position space --- is 
\begin{equation}
  (-\partial^2  - 3H\partial_t - m_X^2) \Delta_{\rm FRW} (x,x') ~=~ 
    i\delta^4(x-x')~.
\end{equation}
The corresponding momentum-space propagator $\Delta_{\rm FRW}(p)$, which is 
simply the Fourier transform of $\Delta_{\rm FRW}(x,x')$, is therefore
\begin{equation}
  \Delta_{\rm FRW}(p) ~=~ \frac{i}{p^2 - m_X^2 + 3iH E_X }~,
  \label{eq:Classical_FRW_Prop}
\end{equation}
where $E_X$ denotes the energy of the mediator field.

At the quantum level, this classical result for $\Delta_{\rm FRW}(p)$ is 
modified.  At one loop, one might expect the modified propagator to take
the form
\begin{equation}
  \Delta_{\rm FRW} (p) \,=\,
    \frac{i}{p^2 - m_X^2 + 3iH E_X + \Pi_X (p^2,H)}~,
\end{equation}
where the one-loop radiative correction $\Pi_X (p^2,H)$ depends on $H$ as
well as on $p^2$.  Without loss of generality, we may expand 
$\Pi_X(p^2,H) = \Pi_X(p^2) + b^{(1)}_X(p^2)E_X H + \mathcal{O}(H^2)$
as a power series in $H$, where $b^{(1)}_X(p^2)$ is a dimensionless coefficient 
whose value depends on $p^2$.  Thus, to leading order in $H$, we find that 
$\Delta_{\rm FRW}(p)$ is related to the Minkowski-space propagator $\Delta(p)$ 
in Eq.~(\ref{eq:propag}) by 
\begin{equation}
  \frac{1}{\Delta_{\rm FRW}(p)} ~\approx~ \frac{1}{\Delta(p)} 
    +\left[3+ b_X^{(1)}(p^2)\right]E_X H~.
  \label{eq:DeltaFRWp_Deltap_rel}
\end{equation}
It therefore follows that the condition under which general-relativistic
effects on the $X$ propagator can be ignored is 
\begin{equation}
  \left[3+ b_X^{(1)}(p^2)\right] E_X H  ~\ll~ \frac{1}{\Delta(p)}~.
  \label{eq:Minkowki_approx_condit_raw}
\end{equation}

The precise form of the coefficient $b_X^{(1)}(p^2)$ within our parameter-space regime 
of interest is not well known, and deriving it is beyond the scope of 
this paper.  Thus, in what follows, we shall obtain a conservative bound within  
the $(\tau,\varrho_M)$ plane from the condition in Eq.~(\ref{eq:Minkowki_approx_condit_raw}) 
by assuming that $b_X^{(1)}(p^2)$ is $\mathcal{O}(1)$ or smaller and can therefore 
be neglected.  We emphasize that this assumption does not have a significant
impact on our overall results.  Indeed, as we shall see in Sect.~\ref{sec:Results}, this 
bound is subleading in comparison with other constraints on our model and would have 
no appreciable impact on our results even if this coefficient were 
$b_X^{(1)}(p^2) \sim \mathcal{O}(10^6)$ for relevant values of $p^2$.
 
Within our regime of interest for stasis, wherein the scattering and 
annihilation processes mediated by virtual $X$ particles are $s$-channel
processes which involve highly non-relativistic initial-state $\phi$ 
particles and wherein $|\mu|/m \ll 1$, we have $E_X \approx m_X \approx 2m$.  
Thus, since Eq.~(\ref{eq:emp_sim_from_pcm}) implies that the characteristic 
three-momentum magnitude of one of these initial-state $\phi$ particles in
the CM frame is $|\vec{p}_{\rm CM}| \sim m\sqrt{\tau}$ within this regime of 
interest, we find that the condition in Eq.~(\ref{eq:Minkowki_approx_condit_raw}) 
can be expressed as an upper bound on $H$ of the form
\begin{equation}
  H ~\ll~ \frac{1}{6 m \Delta(m\sqrt{\tau})}~.
\end{equation}
Alternatively, since the relationship between $H$ and the critical density
implies that
\begin{equation}
  H^2 ~=~ \frac{8\pi g_G^2 m^2\varrho_M}{3\Omega_M}~,
  \label{eq:H2_rel_with_varrho}
\end{equation}
we may also express this condition as a constraint on the 
relationship between $\varrho_M$ and $\tau$.  In particular, 
for $\tau$ within the range specified in 
Eq.~(\ref{tau_constrained_range}), we find that this constraint
takes the form 
\begin{equation}
  \frac{\varrho_{M}}{\tau} ~\ll~ 
    \frac{ g_{M}^{4}\Omega_M}{3 \cdot 2^{15} \pi^{3} g_{G}^{2}}~.
\end{equation}

In principle, other general-relativistic effects, such as the non-adiabatic
production of particles from the vacuum, can also have a non-negligible 
effect on the phase-space distributions of particles in the early
universe.  However, we find that the corresponding bounds on the parameter 
space of our model are subleading in comparison with other constraints and
therefore need not be considered further.

\subsection{Particle destruction via $4\to 2$ annihilation\label{sec:Exothermic}}

In addition to $2\to 2$ elastic-scattering processes of the sort illustrated in
Fig.~\ref{fig:elastic_scattering_diagram}, which simply redistribute kinetic
energy across our population of $\phi$ particles, the interaction Lagrangian in 
Eq.~(\ref{eq:Lagrangian}) also gives rise to processes in which
the initial and final states comprise $\phi$ particles alone, but in which 
the final state comprises fewer such particles than the initial state.
While these latter processes arise at higher order in $g_\phi^2$, they can 
nevertheless have a significant impact on the cosmological dynamics due to 
resonance effects.  Exothermic processes of this sort convert a portion of the 
rest-mass energy of the initial-state particles into kinetic energy, thereby 
heating the $\phi$-particle gas.  In other words, they give rise to an 
additional energy-density pump $P_{M,{\rm KE}}^{(\rho)}$ in 
Eq.~(\ref{eq:eoms}) which serves as a sink term in the evolution equation
for $\rho_M$ and a source term in the evolution equation for $\rho_{\rm KE}$.  
If this pump converts rest-mass energy to kinetic energy energy at a
significant rate, the dynamics which give rise to stasis can consequently be 
disrupted.  Thus, we must demand that $P_{M,{\rm KE}}^{(\rho)}$ be negligible
during stasis.

\begin{figure}[H]
    \centering
    \begin{tikzpicture}
    \begin{feynman}
        \vertex (vf1);
        \vertex [below= 1.25 cm of vf1] (vf2);
        
        \node[blob,draw=gray,pattern color=gray,very thick, 
           above left = 0.1cm and 1.15cm of vf1] (loop1) ;
        \vertex [above left = 0.25cm and 1.25cm of loop1] (vi1) ;
         \node[blob,draw=gray,pattern color=gray,very thick,
           below left = 0.1cm and 1.15cm of vf2] (loop2) ;
        \vertex [below left = 0.25cm and 1.25cm of loop2] (vi2) ;
        
        \vertex [above left = 0.45cm and 1.25cm of vi1] (i11)  {\(\phi_1\)} ;
        \vertex [below left = 0.45cm and 1.25cm of vi1] (i12)  {\(\phi_2\)} ;
        \vertex [above left = 0.45cm and 1.25cm of vi2] (i21)  {\(\phi_3\)} ;
        \vertex [below left = 0.45cm and 1.25cm of vi2] (i22)  {\(\phi_4\)} ;
        
        \vertex [above right=of vf1] (f1) {\(\phi_5\)};
        \vertex [below right=of vf2] (f2) {\(\phi_6\)};

        \diagram* {
        (i11) -- [blue, plain, very thick] (vi1) -- [blue, plain, very thick] (i12),
        (i21) -- [blue, plain, very thick] (vi2) -- [blue, plain, very thick] (i22),
        
        (vi1) -- [red, plain, very thick, pos=0.35, edge label=\(\textcolor{black}{X}\)] (loop1) 
          -- [red, plain, very thick, pos=0.3, edge label=\(\textcolor{black}{X}\)] (vf1),
        (vi2) -- [red, plain, very thick, pos=0.8, edge label=\(\textcolor{black}{X}\)] (loop2)  
          -- [red, plain, very thick, pos=0.65, edge label=\(\textcolor{black}{X}\)] (vf2),
        
        (f1) -- [blue, plain, very thick] (vf1) 
          -- [blue, plain, very thick, edge label=\(\textcolor{black}{\phi}\)] (vf2) 
          -- [blue, plain, very thick] (f2)
        };
    \end{feynman}
    \end{tikzpicture}
    \caption{One of the Feynman diagrams which contributes to the cross-section for 
      $4\phi\to 2\phi$ scattering.  Five additional diagrams associated with distinct 
      ways of contracting the momenta of the external $\phi$ particles into the 
      interaction vertices also contribute to this cross-section.}
    \label{fig:fusion_diagram}
\end{figure}
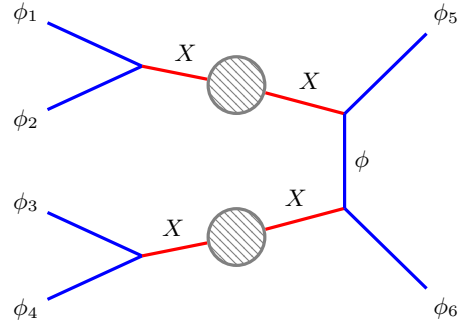

The leading contribution to the interaction rate for exothermic processes of 
this sort is the contribution from $4\to 2$ processes of the sort illustrated
in Fig.~\ref{fig:fusion_diagram} --- processes in which two pairs of initial-state 
$\phi$ particles annihilate pairwise through virtual mediator particles. 
The overall amplitude $\mathcal{M}_{4\to2}$ for this process, accounting for all 
six distinct ways in which the four-momenta of the external particles may be 
contracted into the diagram, can be written in the form
\begin{equation}
  i\mathcal{M}_{4\to 2} ~ = ~ g_\phi^4 m^4 \!\sum_{i,j,k,\ell = 1}^4
    \frac{i\Delta(p_i+p_j)\Delta(p_k+p_\ell)}{(p_i+p_j-p_5)^2-m^2} 
    S_{ijk\ell}\,,
  \label{eq:general_M42}
\end{equation}
where the indices $i$, $j$, $k$, and $\ell$ run over the values $1$, $2$, $3$, and 
$4$ which label the $\phi$ particles in the initial state; where the indices $5$ and $6$ 
label the $\phi$ particles in the final state; where $\Delta(p)$ is defined as in 
Eq.~(\ref{eq:propag}); and where
\begin{equation}
  S_{ijk\ell} ~\equiv~ 
    \begin{cases}
    1/4 & \mbox{all indices different} \\ 
    0 & \mbox{otherwise}
    \end{cases}
\end{equation}
is a combinatorial factor.

\FloatBarrier
\FloatBarrier

In the non-relativistic regime in which $|\vec{p}_i| \ll m$ for all $i$;
we have $(p_i + p_j)^2 \approx 4m^2 + |\vec{p}_i + \vec{p}_j|^2$ to quadratic order
in the three-momenta of the initial-state particles.  Thus, we find that in any 
reference frame in which this approximation holds, each $X$ propagator in 
Eq.~(\ref{eq:general_M42}) takes the approximate form    
\begin{equation}
  \Delta(p_i+p_j) ~\approx~ \frac{i}{m^2} 
  \left[ \widetilde{\alpha} + i\beta
  \frac{|\vec{p}_{ij}|}{m} + 4
   \frac{|\vec{p}_{ij}|^2}{m^2}\,\right]^{-1}\,,
  \label{eq:Delta_pij}
\end{equation}
where $\widetilde{\alpha}$ and $\beta$ are given in Eq.~(\ref{eq:alpha_beta_model}) 
and where we have defined $\vec{p}_{ij} \equiv (\vec{p}_i - \vec{p}_j)/2$.  In other 
words, $\Delta(p_i+p_j)$ has the same functional form as in Eq.~(\ref{eq:propag}), but 
with $|\vec{p}_{\rm CM}|$ replaced by $|\vec{p}_{ij}|$.  By contrast, the final-state 
$\phi$ particles produced in this exothermic reaction have far larger three-momenta in
the background frame within this same regime.  These particles are produced
effectively back to back, with energies $E_5 \approx E_6 \approx 2m$ and 
three-momentum magnitudes $|\vec{p}_5|\approx|\vec{p}_6|\approx \sqrt{3}m$, up to
corrections of $\mathcal{O}(|\vec{p}_i|)$.  The denominator in the expression in
Eq.~(\ref{eq:general_M42}) therefore reduces to
\begin{equation}
  (p_i+p_j - p_5)^2 - m^2 ~\approx~ -4m^2~.
\end{equation}
Thus, we find that $\mathcal{M}_{4\to3}$ in independent of $\vec{p}_5$ and 
$\vec{p}_6$ in this regime.

Since $\rho_M \approx m n_M$, the energy-density pump $P^{(\rho)}_{M,{\rm KE}}$ 
associated with this $4\to 2$ process may be obtained from the corresponding 
collision term in the Boltzmann equation for $n_M$.  In particular, since the 
inverse process is highly Boltzmann-suppressed for a gas of non-relativistic $\phi$ 
particles in kinetic equilibrium, this energy-density pump takes the form
\begin{eqnarray}
  P^{(\rho)}_{M,{\rm KE}} &\,=\,&
  m\big[(2\pi)^{3} n_M\big]^4 \int \Pi_1 \Pi_2 \Pi_3 \Pi_4 \Pi_5 \Pi_6 
    \bigg[|\mathcal{M}_{4\to2}|^2 \nonumber \\
    & & ~~~\times \, 
      (2\pi)^4\delta^4 \left(p_1 + p_2 + p_3 + p_4 - p_5 - p_6\right) 
      \nonumber \\
    & & ~~~\times \, f_\phi(\vec{p}_1)f_\phi(\vec{p}_2)
      f_\phi(\vec{p}_3)f_\phi(\vec{p}_4)\bigg]~,
  \label{eq:Pump_from_collision_term}
\end{eqnarray}
where the normalized phase-space density $f_\phi(\vec{p}_i)$ for each initial-state 
particle takes the form
\beq
f_\phi(\vec{p}\,) ~\equiv~ \frac{1}{(2\pi m T)^{3/2}}\,
       \exp\left ( - \frac{|\vec{p}\,|^2}{2mT}\right)~,
\label{eq:thermalf}
\eeq
and where we have defined
\begin{equation}
  \Pi_i ~\equiv~ \frac{1}{(2\pi)^3}\frac{d^3p_i}{2E_i}~.
\end{equation}  
Within the regime in which $|\vec{p}_i| \ll m$ for all the incoming momenta, we may approximate 
$E_i \approx m$ and $\vec{p}_i \approx 0$ inside the four-dimensional Dirac 
delta function in Eq.~(\ref{eq:Pump_from_collision_term}).  Thus, since 
$|\mathcal{M}_{4\to 2}|^2$ is likewise approximately independent of $\vec{p}_5$ and 
$\vec{p}_6$ in this regime, as discussed above, we find that the 
integral in this may be factored as a product of two 
independent phase-space integrals, one over initial-state momenta and one over 
final-state momenta.  In particular, we find that $P^{(\rho)}_{M,{\rm KE}}$ 
may be written in the form
\begin{equation}
  P^{(\rho)}_{M,{\rm KE}} ~=~ \frac{ Z \rho_M^4 }{256\pi^2m^7}
    \langle|\mathcal{M}_{4\to2}|^2\rangle~,
\end{equation}
where 
\begin{eqnarray}
  \langle |\mathcal{M}_{4\to2}|^2\rangle &\equiv& 
    \int d^3p_1 d^3p_2 d^3p_3 d^3p_4 \bigg[
    |\mathcal{M}_{4\to2}|^2 \nonumber \\ 
    & & \!\!\times \, f_\phi(\vec{p}_1)f_\phi(\vec{p}_2)f_\phi(\vec{p}_3)f_\phi(\vec{p}_4)
    \bigg]~~~~~
  \label{eq:Amplitude_avg_4to2}
\end{eqnarray}
represents the thermal average of $|\mathcal{M}_{4\to2}|^2$ and where  
\begin{eqnarray}
    Z &\,\equiv\,& \int \frac{d^3p_5}{E_5} \frac{d^3p_6}{E_6} 
      \delta(E_5 + E_6 - 4m)\delta^3(p_5 + p_6) \nonumber \\
    &\, =\, & \pi \int \frac{dE_5}{E_5}(E_5-2m^2)^{1/2} 
      \delta(E_5-2m) \nonumber \\
    &\,=\,& \frac{\sqrt{3} \pi}{2}
  \label{eq:Z_int_def}
\end{eqnarray}
is simply a numerical factor.  We note that in going from the first to the 
second line of Eq.~(\ref{eq:Z_int_def}), we have accounted for the fact that 
the two final-state $\phi$ particles are identical particles in evaluating the 
integrals over the angular components of $\vec{p}_5$.

In order to determine how $P^{(\rho)}_{M,{\rm KE}}$ behaves as a function of
$\tau$ for any choice of our model parameters, it remains for us 
to evaluate the expression for $\langle|\mathcal{M}_{4\to2}|^2\rangle$ in 
Eq.~(\ref{eq:Amplitude_avg_4to2}).
Given the manner in which $|\mathcal{M}_{4\to2}|^2$ depends on $|\vec{p}_{ij}|$, 
there are three regimes we need to consider.  The first is the ``cold'' regime wherein 
$\tau \lesssim \taumin$.  Within this regime, the $f_\phi(\vec{p}_i)$ only 
receive non-negligible support at values of $|\vec{p}_i|$ at which the constant terms 
in the denominator of $\Delta(p_i+p_j)$ dominate.  The second is the ``temperate'' 
regime in which $\taumin \lesssim \tau \lesssim \taumax$ --- a
regime which is of particular interest, given that it is also the regime wherein 
$q$ satisfies the stasis condition in Eq.~(\ref{eq:q-range}).
Within this regime, the $f_\phi(\vec{p}_i)$ receive support across a significant region of 
phase space wherein the denominator of $\Delta(p_i+p_j)$ is dominated by terms linear 
in $|\vec{p}_{ij}|$.  The third is the ``hot'' regime wherein 
$\tau \gtrsim \taumax$.  Within this regime, the $f_\phi(\vec{p}_i)$ receive 
substantial support even within regions of phase space wherein the denominator of
$\Delta(p_i+p_j)$ is dominated by terms quadratic in $|\vec{p}_{ij}|$.  We shall
examine the form that $\langle|\mathcal{M}_{4\to2}|^2\rangle$ takes within each of 
these regimes in turn. 

While a detailed analysis of the effect of exothermic processes on the 
phase-space distribution of our $\phi$-particle gas is beyond the scope
of this paper, we can nevertheless asses the circumstances under which 
they will have a non-negligible effect.  The presence of the additional terms 
associated with $P^{(\rho)}_{M,{\rm KE}}$ in the evolution equations for 
$\rho_M$ and $\rho_{\rm KE}$ in Eq.~(\ref{eq:eoms}) and (\ref{eq:eomsKE}) 
leads to the presence of an additional term in the evolution equation for $T$ 
in Eq.~(\ref{eq:TempDiffEq}), which is modified to
\begin{eqnarray}
  \frac{dT}{dt} &\,=\,&
    -2 H T  - \frac{2m}{3\Omega_M} \left( P_{\KE,\gamma} 
    - \frac{\Omega_\KE}{\Omega_M} \,P_{M,\gamma} \right) \nonumber \\ 
  & & ~~~~~+ \frac{2m}{3\Omega_M} 
    \left( 1 + \frac{\Omega_\KE}{\Omega_M}\right) P_{M,\KE} ~,
  \label{eq:dTdtequation_mod_pump}
\end{eqnarray}
where $P_{M,\KE}$ is defined relative to $P^{(\rho)}_{M,{\rm KE}}$ according
to Eq.~(\ref{eq:abundance_pump}).  Recasting this equation as an expression for
the rate of change of $\log\tau$, we have 
\begin{equation}
  \frac{d\log\tau}{dt} ~=~
    -2 H  - \frac{2}{3m^4\varrho_M\tau} \left( P_{\KE,\gamma}^{(\rho)} 
    - \frac{\Omega_\KE}{\Omega_M} \,P_{M,\gamma}^{(\rho)} \right)~.   
\end{equation}
Since $\Omega_{\rm KE} \ll \Omega_M$, this relation implies that in order for 
$P^{(\rho)}_{M,{\rm KE}}$ to have a non-negligible effect on $d\log\tau/dt$, it must 
be the case that   
\begin{equation}
  P^{(\rho)}_{M,{\rm KE}} ~\gtrsim~ \max \Bigg\{ 
   \varrho_M\tau m^4 H,
  \,\abs{P_{\KE,\gamma}^{(\rho)}
    - \frac{\Omega_\KE}{\Omega_M} \,P_{M,\gamma}^{(\rho)} }
    \Bigg\}~.
  \label{eq:PMKE_rapid_condit_T}
\end{equation}

However, if this condition is not satisfied, this does not necessarily mean
that $4\phi\to 2\phi$ scattering has no appreciable impact on the cosmological 
dynamics which give rise to stasis.  Indeed, this process can also have an impact 
on the evolution of $\rho_M$.  Indeed, one finds that the first equation in
Eq.~(\ref{eq:eoms}) is modified in the presence of $4\phi\to 2\phi$ scattering to 
\begin{eqnarray}
  \frac{d\rho_M}{dt} ~&=&~ -3 H \rho_M - P^{(\rho)}_{M,\gamma} 
    - P^{(\rho)}_{M,\KE}  ~ .
\label{eq:eoms_with_PumpMKE}
\end{eqnarray}
Recasting this equation as an expression for the rate of change of $\log\varrho_M$, 
we have
\begin{equation}
  \frac{d\log\varrho_M}{dt} ~=~ 
    -3 H - \frac{P^{(\rho)}_{M,\gamma}}{m^4\varrho_M} 
    - \frac{P^{(\rho)}_{M,\KE}}{m^4\varrho_M}~.    
\end{equation}
This implies that the regime within which $P^{(\rho)}_{M,{\rm KE}}$ has a significant 
impact on $d\varrho_M/dt$ is that within which 
\begin{equation}
  P^{(\rho)}_{M,{\rm KE}} ~\gtrsim~ \max \bigg\{ 
   P^{(\rho)}_{M,\gamma},\, 
   \varrho_M m^4 H\bigg\}~.
  \label{eq:PMKE_rapid_condit_rhoM}
\end{equation}

The conditions in Eqs.~(\ref{eq:PMKE_rapid_condit_T}) and~(\ref{eq:PMKE_rapid_condit_rhoM})
whether $4\to 2$ processes have an impact on $\tau$ and $\varrho_M$, respectively.
Ultimately, however, it is not their impact have on $\tau$ and $\varrho_M$
individually that determines whether these processes have an non-negligible effect on the 
cosmological dynamics, but rather their impact on the manner in which the system evolves within 
the $(\tau,\varrho_M)$ plane.  For example, in situations in which $4\to 2$ processes
have a non-negligible impact the evolution of $\tau$ but not $\varrho_M$, but in which 
$|\log\tau/dt| \ll |\log\varrho_M|$, these processes don't significantly impact the trajectory 
of the system within the $(\tau,\varrho_M)$ plane and thus won't significantly disturb stasis.
Thus, we need to account for these subtleties and derive an upper bound on $P^{(\rho)}_{M,{\rm KE}}$ 
which truly reflects its impact on the evolution of our cosmological system within the 
$(\tau,\varrho_M)$ plane.

As a first step in this direction, we note that $P_{M,\KE}^{(\rho)}$ typically has a far more 
significant impact on the manner in which $\log\tau$ evolves than it does on 
the manner in which $\log\varrho_M$ evolves.  Indeed, since $\tau \ll 1$ and 
$\Omega_{\KE} \ll \Omega_M$ for a non-relativistic population of $\phi$-particles, in the limit 
where $P_{M,\KE}^{(\rho)}$ is the dominant contribution to both terms we would have
\begin{equation}
  \frac{d\ln \varrho_M}{d\ln\tau} ~\approx~ -\frac{3}{2}\tau~.
  \label{eq:Exothermic_rho_tanking_eqn}
\end{equation}
Nevertheless, if $P_{M,\KE}^{(\rho)}$ is sufficiently
small that $d\log\rho_M/dt \gg d\log\tau/dt$, the impact of $4\phi\to 2\phi$ annihilation
on the trajectory of the system within the $(\tau,\varrho_M)$ plane is insignificant even 
if it is the dominant contribution to $d\log\tau/dt$.  Indeed, within this regime, 
$P_{M,\KE}^{(\rho)}$ only has a significant impact on this trajectory if 
$d\log\rho_M/dt \lesssim d\log\tau/dt$ --- \ie, when 
\begin{equation}
  P_{M,\KE}^{(\rho)} ~\gtrsim~ \max\left\{
    \tau \varrho_M m^4 H,\, \tau P_{M,\gamma}^{(\rho)} \right\}~.
  \label{eq:PMKE_rapid_condit_traject}
\end{equation}
Since $\tau \ll 1$, this condition is always satisfied whenever 
Eq.~(\ref{eq:PMKE_rapid_condit_rhoM}) is satisfied. 

Thus, combining the results in Eqs.~(\ref{eq:PMKE_rapid_condit_T}),
(\ref{eq:PMKE_rapid_condit_rhoM}), and (\ref{eq:PMKE_rapid_condit_traject}), 
we find that the condition under which the effects of $4\phi\to 2\phi$ 
scattering on the evolution of our cosmological system {\it can}\/ safely be 
neglected is 
\begin{eqnarray}
  \!\! P^{(\rho)}_{M,{\rm KE}} &~\lesssim~& \max \Bigg\{ 
  \tau P^{(\rho)}_{M,\gamma},\, \varrho_M\tau m^4 H, \nonumber \\ 
  && ~~~~~~~~~~~\abs{P_{\KE,\gamma}^{(\rho)}
    - \frac{\Omega_\KE}{\Omega_M} \,P_{M,\gamma}^{(\rho)} }
    \Bigg\}~.~~~~~
  \label{eq:PMKE_rapid_condit_T_with_strong_pump}
\end{eqnarray}
While the first of the three quantities that appears within the curly braces in 
this expression might at first glance seem unnecessary to include, given that 
$P_{\KE,\gamma}^{(\rho)}$ and $\Omega_\KE\, P_{M,\gamma}^{(\rho)}/\Omega_M$
are both comparable to $\tau P^{(\rho)}_{M,\gamma}$ --- at least within an order 
of magnitude or so --- this first quantity in fact plays a crucial role in 
determining this upper limit on $P^{(\rho)}_{M,{\rm KE}}$.  This is because 
$P_{\KE,\gamma}^{(\rho)}$ and $\Omega_\KE\,P_{M,\gamma}^{(\rho)}/\Omega_M$
are almost identical within the regime wherein $\tau \ll \tau_{\rm min}$, and as a
result the last quantity which appears within the curly braces in 
Eq.~(\ref{eq:PMKE_rapid_condit_T_with_strong_pump}) effectively vanishes.

In order to determine how the constraint in Eq.~(\ref{eq:PMKE_rapid_condit_T_with_strong_pump})
impacts the parameter space of our stasis model, we must first recast this constraint 
as a condition relating the dynamical variables $\varrho_M$ and $\tau$ within each of our three 
temperature regimes.  Within any region of the $(\tau,\varrho_M)$ plane wherein the condition 
in Eq.~(\ref{eq:PMKE_rapid_condit_T_with_strong_pump}) is violated, the impact of exothermic
processes generically tends to disrupt the cosmological dynamics which lead to the stasis
attractor, provided of course that this attractor would otherwise be realized within that 
region of the plane.  Nevertheless, as we shall see, it turns out that within regions of the
$(\tau,\varrho_M)$ plane wherein the stasis attractor is {\it not}\/ realized, these processes 
can often serve to propel the system toward a region within that plane from which stasis 
{\it can}\/ be realized. 

We begin by considering the regime wherein $\taumin \lesssim \tau \lesssim\taumax$, as this 
is the regime within which $q = -2$ and the stasis attractor is active.  Within 
this ``temperate'' regime, as discussed above, the $f_\phi(\vec{p}_i)$ receive support 
across a significant region of phase space wherein the denominator of $\Delta(p_i+p_j)$ 
is dominated by terms linear in $|\vec{p}_{ij}|$.  Within such regions of phase space, we 
may approximate
\begin{equation}
  |\mathcal{M}_{4\to2}|^2 \,\approx\, 2^{18}\pi^4 \bigg(
    \frac{1}{|\vec{p}_{12}| |\vec{p}_{34}|} +
    \frac{1}{|\vec{p}_{13}| |\vec{p}_{24}|}
    +\frac{1}{|\vec{p}_{14}| |\vec{p}_{23}|}\bigg)^2.
  \label{eq:MatElSq_4to2_temperate_raw}
\end{equation}
In order to derive an estimate for $P^{(\rho)}_{M,{\rm KE}}$ valid within this regime, 
we begin by noting that triangle inequalities of the form
\begin{equation}
  \frac{2}{|\vec{p}_{ij}||\vec{p}_{k\ell}||\vec{p}_{ik}||\vec{p}_{j\ell}|} 
    ~\leq~ \frac{1}{|\vec{p}_{ij}|^2|\vec{p}_{k\ell}|^2} + 
    \frac{1}{|\vec{p}_{ik}|^2|\vec{p}_{j\ell}|^2} 
\end{equation}
imply that the expression for $|\mathcal{M}_{4\to2}|^2$ in 
Eq.~(\ref{eq:MatElSq_4to2_temperate_raw}) is bounded from above by 
\begin{eqnarray}
  |\mathcal{M}_{4\to2}|^2 &\,\lesssim\,& 3\cdot 2^{18}\pi^4 \bigg(
    \frac{1}{|\vec{p}_{12}|^2 |\vec{p}_{34}|^2} +
    \frac{1}{|\vec{p}_{13}|^2 |\vec{p}_{24}|^2} ~~\nonumber \\
    &&~~~~~~~~~~~~~~~~~+
    \frac{1}{|\vec{p}_{14}|^2 |\vec{p}_{23}|^2}\bigg)~.
  \label{eq:MatElSq_4to2_temperate_bound}
\end{eqnarray}

We observe that if $|\mathcal{M}_{4\to2}|^2$ is replaced in 
Eq.~(\ref{eq:Amplitude_avg_4to2}) by this upper bound, each of the 
terms in the resulting expression reduces to a product of 
two identical integrals, each involving only two of the four $\vec{p}_i$.
Thus, we find that the corresponding upper bound on 
$\langle |\mathcal{M}_{4\to2}|^2\rangle$ is
\begin{equation}
  \langle |\mathcal{M}_{4\to2}|^2\rangle ~\lesssim~ 
    9\cdot 2^{18}\pi^4
    \left[\int d^3p_a d^3p_b 
    \frac{f_\phi(\vec{p}_a)f_\phi(\vec{p}_b)}{|\vec{p}_{ab}|^2}
       \right]^2\,.
\end{equation}
The integral appearing in this expression has the same general form as the thermal 
average $\langle \abs{\vec{p}_{\rm CM}}^q\rangle$ with $q = -2$.  For this value of $q$,
the integral converges and the upper bound on 
$\langle|\mathcal{M}_{4\to2}|^2\rangle$ is   
\begin{equation}
  \langle |\mathcal{M}_{4\to2}|^2 \rangle ~\lesssim~
    \frac{9\cdot 2^{20}\pi^4}{(mT)^2}\,.
\end{equation}

The corresponding upper bound on $P^{(\rho)}_{M,{\rm KE}}$, which 
captures the parametric dependence of this energy-density pump on $\varrho_M$ 
and $\tau$, is 
\begin{equation}
  P^{(\rho)}_{M,{\rm KE}} ~\lesssim~ 
    9\sqrt{3}\cdot 2^{11} \pi^3 m^5 
    \left(\frac{\varrho_M^4}{\tau^2}\right)~.
  \label{eq:Pump_4to2_temperate_upper}  
\end{equation}
Moreover, since we expect the true value of $P^{(\rho)}_{M,{\rm KE}}$ to come 
reasonably close to saturating this upper bound, we may also regard the value 
of the numerical coefficient in this expression as an order-of-magnitude estimate 
for its true value.  A more precise estimate of this coefficient may be obtained 
via Monte-Carlo integration, using the full expression for $|\mathcal{M}_{4\to2}|^2$ 
in Eq.~(\ref{eq:MatElSq_4to2_temperate_raw}) rather than the approximation in 
Eq.~(\ref{eq:MatElSq_4to2_temperate_bound}).  Proceeding in this manner, we find that
\begin{equation}
  P^{(\rho)}_{M,{\rm KE}} ~\approx~ 
    9\sqrt{3}\cdot 2^{11} \pi^3 \epsilon\, m^5 
    \left(\frac{\varrho_M^4}{\tau^2}\right)~,
  \label{eq:Pump_4to2_temperate_final}  
\end{equation}
where $\epsilon \approx 0.669$.\\

In order to assess the form that the condition in
Eq.~(\ref{eq:PMKE_rapid_condit_T_with_strong_pump}) takes within the 
``temperate'' regime, we begin by noting that 
Eqs.~(\ref{quadratically}) and~(\ref{formula_for_C}) together imply that
$P^{(\rho)}_{M,\gamma}$ takes the form
\begin{equation}
  P^{(\rho)}_{M,\gamma}  ~=~ \frac{32\pi g_\chi^2 m^5}{g_\phi^2}
    \frac{\varrho_M^2}{\tau}
  \label{eq:PMgamma_cold}
\end{equation}
within this regime.  It therefore also follows from
Eq.~(\ref{idealgas}) and~(\ref{quadratically}) that
\begin{eqnarray}
    \abs{P_{\KE,\gamma}^{(\rho)}
    - \frac{\Omega_\KE}{\Omega_M} \,P_{M,\gamma}^{(\rho)}} &~=~& 
    \frac{1}{2} \tau P_{M,\gamma}^{(\rho)} \nonumber \\ 
    &=& \frac{16\pi g_\chi^2 m^5}{g_\phi^2}
    \varrho_M^2~.~~~~~
  \label{eq:abs_of_pumps_equiv_temperate}
\end{eqnarray}
Alternatively, up to an $\mathcal{O}(1)$ prefactor, this quantity may 
be expressed in the form\footnote{It has not escaped the attention 
of the authors that the final quantity on the second line of 
Eq.~(\ref{ninelines}) --- including the overhanging ``roof'' of the square-root 
sign --- consists of {\it nine}\/ horizontal lines.}
\begin{eqnarray}
    \abs{P_{\KE,\gamma}^{(\rho)}
    - \frac{\Omega_\KE}{\Omega_M} \,P_{M,\gamma}^{(\rho)}} &~\sim~& 
    H m^4\varrho_M\tau \sqrt{\frac{\Xi\,\Omega_M}{\barXi\,\barOmega_M}}\nonumber\\
    &~\sim~& H m^4\varrho_M\tau \Omega_M^{1/2}\sqrt{\frac{\Xi}{\barXi}}~.~~~\nonumber\\
\label{ninelines}
\end{eqnarray}
Thus, we find that up to $\mathcal{O}(1)$ prefactors,  the first and third terms
on the right side of Eq.~(\ref{eq:PMKE_rapid_condit_T_with_strong_pump}) are 
identical within the ``temperate'' regime, while the second and third terms differ 
by a factor of $\sqrt{\Omega_M\Xi/\bar{\Xi}}$.

Given this, it is straightforward to determine the constraint contour in the $(\tau, \varrho_M)$ 
plane which follows from demanding that $P_{M,\gamma}^{(\rho)}$ satisfy the condition in
Eq.~(\ref{eq:PMKE_rapid_condit_T_with_strong_pump}).  Within the regime wherein 
$\Xi\, \Omega_M \lesssim \barXi$, the second term on the right side of 
Eq.~(\ref{eq:PMKE_rapid_condit_T_with_strong_pump}) dominates and this condition reduces to
\begin{equation}
  \frac{\varrho_M^{5/2}}{\tau^3} ~\lesssim~ 
    \frac{g_G}{27\cdot 2^{19/2}\pi^{5/2}\epsilon\,\Omega_M^{1/2}}~.
  \label{eq:Exotherm_varrho_tau_condit_temperate}
\end{equation}
By contrast, within the regime wherein $\Xi\,\Omega_M \gtrsim \barXi$, the first and third 
terms dominate and the condition reduces to 
\begin{equation}
   \frac{\varrho_M^{2}}{ \tau^{2}} ~\lesssim ~\frac{5^{1/2}  
    g_{\chi}^{2}}{ 2^{11/2} 3^{7/2} \pi^{2} \epsilon g_{\phi}^{2}}~.
\end{equation}
Thus, within the ``temperate'' regime, we find that 
Eq.~(\ref{eq:PMKE_rapid_condit_T_with_strong_pump}) reduces to a condition within 
the $(\tau,\varrho_M)$ plane of the form 
\begin{eqnarray}
    \varrho_M^{5/2} ~&\lesssim&~ B\,\tau^3\max\left\{\Omega_M^{-1/2},
    \frac{\varrho_M^{1/2}}{\tau\,\barXi^{1/2}} \right\}~,
    \label{fusion_condition_full_temperate}
\end{eqnarray}
where we have defined
\begin{equation}
  B ~ \equiv~ \frac{g_G}{27\cdot 2^{19/2}\pi^{5/2}\epsilon}~.
  \label{eq:Bdef}
\end{equation}

We now turn to the ``cold'' regime, wherein $\tau \lesssim \taumin$. 
Within this regime, the constant terms in $|\mathcal{M}_{4\to2}|^2$ dominate.  
As a result, the thermal average of this amplitude is simply
\begin{equation}
  \langle |\mathcal{M}_{4\to2}|^2 \rangle ~\approx~ |\mathcal{M}_{4\to2}|^2
  ~\approx~
    \frac{9\cdot 2^{18}\pi^4g_\phi^8m^4}{(g_\chi^4 m^4 + 1024\pi^2\mu^4)^2}~.    
\end{equation}
We note that this result does not depend on the form that $f_\phi(\vec{p}_i)$ takes,
and therefore holds regardless of whether or not the condition for kinetic 
equilibrium in Eq.~(\ref{eq:therm_cond}) is satisfied.  The corresponding form
for the energy-density pump $P^{(\rho)}_{M,{\rm KE}}$ is
\begin{equation}
  P^{(\rho)}_{M,{\rm KE}} ~\approx~ 
    \frac{9\sqrt{3}\cdot 2^9\pi^3 g_\phi^8 m^5\varrho_M^4}
    {[g_\chi^4 + 1024\pi^2(\mu/m)^4]^2}~.
  \label{eq:Pump_4to2_cold_final}
\end{equation}

The result in Eq.~(\ref{eq:Pump_4to2_cold_final}) applies within the 
regime in which the average kinetic energy of our $\phi$-particle gas lies 
below the threshold for stasis and the $2\to 2$ processes which typically 
serve to maintain kinetic equilibrium among the $\phi$ particles are 
inefficient.  However, since the exothermic processes which give rise to 
$P^{(\rho)}_{M,{\rm KE}}$ serve to increase this average kinetic energy, they 
can potentially re-establish kinetic equilibrium and propel the system from a 
configuration wherein $\tau$ is too low for the system to achieve stasis 
before the attractor reaches its expiration date toward a configuration from which 
stasis can in fact be achieved.  As a result, stasis can in principle emerge for 
certain combinations of $\varrho_M^{(0)}$ and $\tau^{(0)}$ for which this 
phenomenon could otherwise never have arisen.  

In order to assess the form that the condition in 
Eq.~(\ref{eq:PMKE_rapid_condit_T_with_strong_pump}) takes within the 
``cold'' regime, we begin by noting that the pump $P^{(\rho)}_{M,\gamma}$ 
takes the form
\begin{equation}
  P^{(\rho)}_{M,\gamma}  ~\approx~ m^7 \varrho_M^2\, \sigma v~,
  \label{eq:PMgamma_cold}
\end{equation}
within this regime, where $\sigma v$ is to a good approximation momentum-independent.
Thus, given the expressions for $P_{M,\KE}^{(\rho)}$ and $P_{M,\gamma}^{(\rho)}$ in
Eqs.~(\ref{eq:Pump_4to2_cold_final}) and~(\ref{eq:PMgamma_cold}), we find that when
the first of the three terms within the curly brackets on the right side of 
Eq.~(\ref{eq:PMKE_rapid_condit_T_with_strong_pump}) is largest, the 
corresponding condition on $\varrho_M$ and $\tau$ is 
\begin{equation}
  \frac{\varrho_M^{5/2}}{\tau} ~\lesssim~ 
    \frac{[g_\chi^4 + 1024\pi^2(\mu/m)^4]^2 g_G}
    {27\cdot 2^{15/2}\pi^{5/2} g_\phi^8 \Omega_M^{1/2}}~.
  \label{eq:Exotherm_varrho_tau_condit_cold_vs_expansion}
\end{equation}
Likewise, when the second term is largest, the corresponding
condition on $\varrho_M$ and $\tau$ is 
\begin{equation}
    \frac{\varrho_M^2}{\tau} ~\lesssim~ \frac{\sqrt{3} g_{G}^{2} 
    \left[g_{\chi}^{4} + 1024 \pi^{2} (\mu/m)^{4}\right]}
    {27\cdot 2^5 \pi^{2} g_{\phi}^{6}}~.
  \label{eq:Exotherm_varrho_tau_condit_cold_vs_pump}
\end{equation}
We note that up to an $\mathcal{O}(1)$ numerical factor, the two conditions
in Eqs.~(\ref{eq:Exotherm_varrho_tau_condit_cold_vs_expansion}) 
and~(\ref{eq:Exotherm_varrho_tau_condit_cold_vs_pump}) are equivalent to
\begin{eqnarray}
  \frac{\varrho_M^{5/2}}{\tau} &\lesssim& 
    \frac{B\ \taumin^{2}}{\Omega_M^{1/2}}\nonumber\\
  \frac{\varrho_M^{2}}{\tau} &\lesssim& 
    \frac{B\ \taumin}{\barXi^{1/2}}~,
\end{eqnarray}
respectively, where $B$ was defined in Eq.~(\ref{eq:Bdef}).  Combining these 
conditions into a single relation, we obtain
\begin{equation}
  \varrho_M^{5/2} ~\lesssim~  B\, \tau\,\taumin^{2}
    \max\left\{\Omega_M^{-1/2},
      \frac{\varrho_M^{1/2}}{\barXi^{1/2} \taumin}\right\}~.
  \label{fusion_condition_full_cold}
\end{equation}
We note that this relation has the same form as the corresponding condition 
in Eq.~(\ref{fusion_condition_full_temperate}), which holds within the 
``temperate'' regime.  We also note that the second term within the curly brackets 
in Eq.~(\ref{fusion_condition_full_cold}) always dominates within any region 
of our model-parameter space wherein stasis can be realized. 

Finally, we consider the ``hot'' regime, wherein $\tau \gtrsim \taumax$.  
Within this regime, the $f_\phi(\vec{p}_i)$ receive support across a a significant 
region of phase space wherein the denominator of $\Delta(p_i+p_j)$ is dominated by 
terms quadratic in $|\vec{p}_{ij}|$.  Within such regions of phase space, we may 
approximate
\begin{eqnarray}
  |\mathcal{M}_{4\to2}|^2 &\,\approx\,& \frac{g_\phi^8 m^4}{1024} \bigg(
    \frac{1}{|\vec{p}_{12}|^2 |\vec{p}_{34}|^2} +
    \frac{1}{|\vec{p}_{13}|^2 |\vec{p}_{24}|^2} \nonumber \\ 
    & &~~~~~~~~~~~~~~~+
    \frac{1}{|\vec{p}_{14}|^2 |\vec{p}_{23}|^2}\bigg)^2~.
  \label{eq:MatElSq_4to2_hot_raw}
\end{eqnarray}
Once again, we note that triangle inequalities imply that this expression
is bounded from above by
\begin{eqnarray}
  |\mathcal{M}_{4\to2}|^2 &\,\lesssim\,& \frac{3g_\phi^8 m^4}{1024} \bigg(
    \frac{1}{|\vec{p}_{12}|^4 |\vec{p}_{34}|^4} +
    \frac{1}{|\vec{p}_{13}|^4 |\vec{p}_{24}|^4} \nonumber \\ 
    & &~~~~~~~~~~~~~~~+
    \frac{1}{|\vec{p}_{14}|^4 |\vec{p}_{23}|^4}\bigg)~.  
\end{eqnarray}
If $|\mathcal{M}_{4\to2}|^2$ is replaced in Eq.~(\ref{eq:Amplitude_avg_4to2}) 
by this upper bound, we once observe that each term in the resulting 
expression reduces to a product of two identical integrals, each involving only 
two of the four $\vec{p}_i$.  Thus, we find that the corresponding upper bound on 
$\langle |\mathcal{M}_{4\to2}|^2\rangle$ is given by
\begin{equation}
  \langle |\mathcal{M}_{4\to2}|^2\rangle \,\lesssim\, 
    \frac{9g_\phi^8 m^4}{1024}
    \left[\int d^3p_a d^3p_b \frac{1}{|\vec{p}_{ab}|^4}
      f_\phi(\vec{p}_a)f_\phi(\vec{p}_b) \right]^2\!.
  \label{eq:MalElSq_4to2_hot_diverge}
\end{equation}

While the integral appearing in this expression nominally has the same form as the 
thermal average $\langle \abs{\vec{p}_{\rm CM}}^q\rangle$ with $q = -4$, this integral 
diverges for all $q \leq -3$.  However, this is merely a reflection of the fact that 
$|\mathcal{M}_{4\to2}|^2$ is only well approximated by 
Eq.~(\ref{eq:MatElSq_4to2_hot_raw}) within regions of phase space wherein
$|p_{ij}| \gtrsim \beta m/4 = \sqrt{m\Tmax}$ for all $i$ and $j$.  We shall
regulate the ``infrared divergences'' in Eq.~(\ref{eq:MalElSq_4to2_hot_diverge}) 
after changing integration variables from $\vec{p}_a$ and $\vec{p}_b$ to  
the combinations $\vec{p}_+$ and $\vec{p}_-$, where 
$\vec{p}_\pm\equiv \vec{p}_a\pm \vec{p}_b$, by taking the lower limit of integration 
for the magnitude $|\vec{p}_-| = 2|\vec{p}_{ab}|$ of $\vec{p}_-$ to be $\beta/2$.
Doing so, we obtain
\begin{eqnarray}
  \langle |\mathcal{M}_{4\to2}|^2\rangle &\,\lesssim\, & 
    \frac{9 g_\phi^8}{64\pi^2 m^2 T^6} 
    \left( \int_0^\infty d|\vec{p}_+| |\vec{p}_+|^2
    e^{-\frac{|p_+|^2}{4mT}}\right)^2 \nonumber \\
  & &\!\!\!\! \!\!\!
  \times \bigg(\int_{\beta/2}^\infty d|\vec{p}_-| 
    \frac{1}{|\vec{p}_-|^2} e^{\frac{-|p_-|^2}{4mT}}\bigg)^2 
    + \Delta \langle|\mathcal{M}|^2\rangle ~, \nonumber \\
\end{eqnarray}
where $\Delta \langle|\mathcal{M}|^2\rangle$ represents the contribution to
$\langle|\mathcal{M}_{4\to2}|^2\rangle$ from the low-momentum region of phase 
space in which $|\vec{p}_{ij}|< \sqrt{m\Tmax}$ for one or more combinations 
of $i$ and $j$.  Evaluating these integrals, we obtain  
\begin{eqnarray}
\langle |\mathcal{M}_{4\to2}|^2\rangle &\,\lesssim\, &
    \frac{9 g_\phi^8}{64\pi T^4} 
    \Bigg[\frac{T^{1/2}}{\Tmax^{1/2}} 
    e^{-\Tmax/T} \nonumber\\
    & & \!\!\!\!
    -\sqrt{\pi}\,{\rm erfc}
    \left(\frac{\Tmax^{1/2}}{T^{1/2}}\right)\Bigg]^2
    + \Delta \langle|\mathcal{M}|^2\rangle~,~~~~~~~~
  \label{eq:MatEl_4to2_hot_int_eval_step}
\end{eqnarray}
where ${\rm erfc}(z) \equiv 1-{\rm erf}(z)$ denotes the complementary error 
function.  For $T \gg \Tmax$, the contribution
$\Delta \langle|\mathcal{M}|^2\rangle$ is comparatively negligible and
this expression reduces to
\begin{equation}
  \langle |\mathcal{M}_{4\to2}|^2\rangle ~\lesssim~ 
    \frac{9\cdot 2^8 g_\phi^4}{mT^3}~. 
\end{equation}
The corresponding upper bound on the energy-density pump, which captures the 
parametric dependence of this pump on $\varrho_M$ and $\tau$, is therefore 
\begin{equation}
  P^{(\rho)}_{M,{\rm KE}} ~\lesssim~ 
    \frac{9\sqrt{3} g_\phi^4 m^5 }{2\pi}
      \left(\frac{\varrho_M^4}{\tau^3}\right)~.
  \label{eq:Pump_4to2_hot_final}
\end{equation}
Since we expect the true value of $P^{(\rho)}_{M,{\rm KE}}$ to come 
reasonably close to saturating this upper bound, we may once again regard the value 
of the numerical coefficient in this expression as an order-of-magnitude estimate 
for its true value.

Approximating $P^{(\rho)}_{M,{\rm KE}}$ with 
the upper bound in this equation, we can determine the condition on $\varrho_M$ 
and $\tau$ which follows from Eq.~(\ref{eq:PMKE_rapid_condit_T_with_strong_pump}) 
within the ``hot'' regime.  The form which this upper bound takes depends on whether  
the first and third terms in Eq.~(\ref{eq:PMKE_rapid_condit_T_with_strong_pump})
dominate over the second term.  We first consider the case in which the second term ---
the term associated with Hubble expansion --- dominates, which occurs whenever
\begin{equation}
    \Omega_M ~\lesssim~ \frac{\barXi\tau}{\Xi\tau_{\rm max}}~.
\end{equation}
In this case, the resulting condition on $\varrho_M$ and $\tau$ takes the form
\begin{equation}
  \frac{\varrho_M^{5/2}}{ \tau^4} ~\lesssim~ 
    \frac{2^{5/2}\pi^{3/2} g_G}{27 g_\phi^4 \Omega_M^{1/2}}~.
  \label{eq:Exotherm_varrho_tau_condit_hot}
\end{equation}
By contrast, in the case wherein the first and third terms in 
Eq.~(\ref{eq:PMKE_rapid_condit_T_with_strong_pump}) --- the terms associated with 
annihilation processes --- dominate over the expansion term, we must first evaluate 
$P^{(\rho)}_{M,\gamma}$ in order to obtain the corresponding 
condition on $\varrho_M$ and $\tau$.  The thermal average of the squared matrix element 
for the $\phi\phi\to\chi\chi$ scattering process which gives rise to $P^{(\rho)}_{M,\gamma}$ 
may be evaluated within the ``hot'' regime via the imposition of a momentum cutoff analogous 
to the cutoff which we imposed in evaluating $\langle |\mathcal{M}_{4\to2}|^2\rangle$ in 
Eq.~(\ref{eq:MatEl_4to2_hot_int_eval_step}).
After some algebra, we find that the stasis pump is given in this regime by
\begin{equation}
  P_{M,\gamma}^{(\rho)} ~=~ 
    \frac{m^5\varrho_{M}^{2} g_{\chi}^{2}}{2 \sqrt{\pi} \tau^{\frac{3}{2}}}~.
\end{equation}
Thus, in such situations, we find that the $\phi\phi\to\chi\chi$ scattering rate exceeds the 
expansion rate when the condition
\begin{equation}
  \frac{\varrho^2_M}{\tau^{\frac{5}{2}}} ~\lesssim~ 
  \frac{ \sqrt{\pi}  g_{\g}^{2}}{9\sqrt{3} g_{M}^{4}}
\label{eq:Exotherm_varrho_tau_condit_hot_AnnDom}
\end{equation}
is satisfied.

We note that up to an $\mathcal{O}(1)$ numerical factor, the two conditions in
Eqs.~(\ref{eq:Exotherm_varrho_tau_condit_hot}) 
and~(\ref{eq:Exotherm_varrho_tau_condit_hot_AnnDom}) are equivalent to
\begin{eqnarray}
    \frac{\varrho_M^{5/2}}{\tau^4} ~&\lesssim&~ 
    \frac{B}{\taumax\Omega_M^{1/2}}~\nonumber\\
    \frac{\varrho^2_M}{\tau^{\frac{5}{2}}} ~&\lesssim&~
\frac{B}{\taumax^{1/2}\barXi^{1/2}}~,
\end{eqnarray} 
where $B$ once again denotes the combination of parameters defined in Eq.~(\ref{eq:Bdef}).
Combining these two conditions into a single relation, we find that the overall condition 
on $\varrho_M$ and $\tau$ which we must impose in order to ensure that the effect of 
exothermic processes on the stasis dynamics may be neglected within the ``hot regime'' is
\begin{equation}
  \frac{\varrho_M^{5/2}}{\tau^4} ~\lesssim~ 
    \frac{B}{\taumax}\max\left\{\Omega_M^{-1/2},
    \frac{\taumax^{1/2}}{\barXi^{1/2}}\frac{\varrho_M^{1/2}}{\tau^{3/2}}\right\}~.
  \label{fusion_condition_full_hot}
\end{equation}

\subsection{Bose-Einstein condensation resulting in $4\to 2$ annihilation\label{sec:BEC}}

At low temperatures, a Bose-Einstein condensate (BEC) can potentially arise within 
our $\phi$-particle gas.  When this occurs, the rates for exothermic processes of 
the sort discussed in Sect.~\ref{sec:Exothermic} are significantly enhanced as a 
result of the high occupation fraction of the ground state.  Since these processes
can disrupt stasis, as discussed above, the temperature of our population of $\phi$ 
particles must exceed the critical temperature $T_c$ below which a BEC forms 
throughout the stasis epoch.

Within the regime in which our $\phi$-particle gas is reasonably weakly interacting 
and can therefore be modeled as a collection of free particles, this critical 
temperature is 
\begin{equation}
  T_c ~=~ \frac{2\pi}{m}\left[\frac{n_M}{\zeta(3/2)}\right]^{2/3}~,
\end{equation}
where $\zeta(x)$ denotes the Riemann zeta-function.  Since $\rho_M \approx m n_M$ 
for this non-relativistic population of $\phi$ particles, the corresponding 
constraint on $\tau$ is
\begin{eqnarray}
  \tau ~>~ \tau_{c} ~\equiv~ \frac{2\pi}{\zeta^{2/3}(3/2)}\varrho_M^{2/3}~.
  \label{initial_condition_bec}
\end{eqnarray}

\subsection{$X$-decay density}

In analyzing the cosmological dynamics of our stasis model, we have implicitly 
assumed that the abundance of mediator particles is negligible by the time the 
stasis epoch begins.  One way of ensuring that this this is the case is to demand 
that the lifetime of the mediator is sufficiently short that the decay rate is 
greater than the Hubble parameter ($\Gamma_X \gtrsim H$) throughout the stasis 
epoch.  Making use of Eq.~(\ref{eq:H2_rel_with_varrho}), this condition may 
be recast as an upper bound on $\varrho_M$ of the form 
\begin{equation}
  \varrho_M ~\lesssim~ \frac{3\Omega_M\Gamma_X^2}{8\pi g_G^2 m^2}~.    
\end{equation}

A conservative bound on $\varrho_M$ may be obtained by considering the
case in which $\mu/m < 0$ and our mediator particle decays exclusively
through the channel $X\to \chi\chi$.  Indeed, if $\mu/m > 0$ and the 
decay channel $X\to \phi\phi$ is kinematically accessible, this channel 
only serves to increase $\Gamma_X$.  Noting that $|\mu|/m \ll 1$ within
our parameter-space regime of interest, and thus that $m_X \approx 2m$,
we find that this conservative bound is 
\begin{equation}
  \varrho_M ~\lesssim~ \frac{3\Omega_M}{2^{15}\pi^3}\frac{g_\chi^4}{g_G^2}~.
    \label{initial_condition_Xdecay}
\end{equation}

If this condition is violated, the impact that the population of undecayed $X$ particles 
has on the cosmological dynamics depends primarily on the value of $\Xi$.  Within the 
regime in which $\Xi \ll \barXi$, the terms in Eq.~(\ref{Sweqs_expanded}) associated with 
expansion dominate over those associated with annihilation.  While the $X$ particles, which 
are non-relativistic and represent an additional contribution to the matter energy density, 
collectively have an impact of the expansion rate, the universe nevertheless continues to 
evolve toward stasis --- with $\varrho_M$ and $\tau$ both decreasing  --- until the annihilation 
term begins to have a significant impact on the cosmological dynamics.  As long as the population
of $X$ particles decays away before $\Xi$ becomes comparable to $\barXi$ and impact of 
the annihilation terms in Eq.~(\ref{Sweqs_expanded}) becomes non-negligible --- the
timescale for which we shall discuss in greater detail in Sect.~\ref{sec:Dynamics} --- 
these particles will not significantly impact the cosmological dynamics.

By contrast, within the regime in which $\Xi \gtrsim\barXi$, the stasis pump {\it does}\/ have 
a non-negligible impact on the cosmological dynamics.  Within this regime, the impact of 
annihilation in non-negligible and the presence of the $X$ particles disrupts the delicate 
interplay between the annihilation and expansion rates which ultimately gives rise to the 
stasis attractor.  That said, we note that this regime is seldom if ever reached for 
any sensible set of initial conditions.

\subsection{Thermodynamic limit}

The results we have derived for our thermal stasis model are predicated on the assumption that
the $\phi$-particle gas is in the thermodynamic limit --- \ie, that 
the number of $\phi$ particles within a given Hubble volume is large.  
For concreteness, we shall derive a bound on $\varrho_M$ by requiring that the 
number of such particles within a typical Hubble volume be larger than 
Avogadro's number $N_A$.  Since the energy density of our non-relativistic 
$\phi$-particle gas is approximately $\rho_M \approx m n_M$, this condition 
may be written as
\begin{equation}
  N_A ~<~ \frac{4\pi}{3} \frac{\rho_M}{m H^3}~.
\end{equation}
Making use of Eq.~(\ref{eq:H2_rel_with_varrho}) in order to express 
$H$ in terms of $\varrho_M$, we obtain an upper bound on this dimensionless
energy-density variable of the form
\begin{equation}
  \varrho_{M} ~<~ \frac{3 \Omega_{M}^{3}}{32 \pi N_A^{2} g_{G}^{6}}~.
  \label{eq:ThermoLimitBound}
\end{equation}

Counterintuitively, the requirement that our $\phi$-particle gas be in the 
thermodynamic limit turns out to impose an {\it upper}\/ bound on $\varrho_M$.
This is because the Hubble volume decreases as the overall energy density
of the universe increases.

\subsection{Observational constraints at late times\label{sec:BBN}}

Consistency with observation requires that the total energy density 
$\rho$ of the universe be dominated by the visible-sector radiation bath by 
the beginning of the BBN epoch.  This epoch begins when the temperature of the
visible-sector bath is approximately $T_{\rm BBN} \sim 10$~MeV and the energy 
density of the universe is
\begin{equation}
  \rho_{\rm BBN} ~=~ \frac{\pi^2}{30}g_\ast(T_{\rm BBN}) T_{\rm BBN}^4~,
\end{equation}
where $g_\ast(T_{\rm BBN}) \approx 10.75$ is the effective number of 
relativistic degrees of freedom in the visible-sector bath at this 
temperature.  By contrast, at the end of the stasis epoch, $\rho$ is jointly 
dominated in our model by dark radiation and by the $\phi$-particle gas, with 
abundances $\Omega_\gamma = 3/5$ and $\Omega=2/5$, respectively.  Between the
end of stasis and the BBN epoch, the universe must somehow transition from
this state to a state in which $\rho$ is dominated by visible-sector 
radiation and in which $\Omega_\gamma$ and $\Omega_M$ are both negligible.
Moreover, regardless of the particular manner in which this occurs, the requirement 
that $\rho$ must still exceed $\rho_{\rm BBN}$ after the transition is complete 
ultimately places a lower bound on the energy density of the universe at the 
end of the stasis epoch.

The simplest mechanism for populating the visible sector after stasis ends is
$\phi$-particle decay.  If $\phi$ couples to the fields of the visible-sector 
via highly suppressed operators which give rise to a small decay width $\Gamma_\phi$ 
for these particles into SM states, the energy density of the $\phi$-particle gas 
is effectively transferred to the visible-sector radiation bath at a timescale 
$t_\phi \sim \Gamma_\phi^{-1}$.  As long as $t_\phi$ is sufficiently large that 
this transfer of energy density occurs well after stasis begins, a stasis epoch of 
will nevertheless develop before the population of $\phi$ particles is significantly 
depleted by decays. 

That said, limits on the abundance of dark radiation at late times place additional 
constraints on $t_\phi$.  Within the regime in which $t_\phi < t_{\rm end}$, where 
$t_{\rm end}$ represents the time at which $\tau$ drops below $\taumin$, the stasis 
epoch ends prematurely at $t\sim t_\phi$ as a result of $\phi$ decay.  Thus,
at the end of stasis, the universe contains both a population of SM particles with 
abundance $\Omega_{\rm SM} \sim \barOmega_M$ and a population of dark-radiation particles 
with abundance $\Omega\gamma \sim \barOmega_\gamma$.  Since the SM particles behave 
like radiation until well after the BBN epoch begins, $\Omega_\gamma$ and 
$\Omega_{\rm SM}$ remain effectively unchanged from the end of stasis until BBN.~
As a result, a sizable abundance of dark radiation is present in the universe 
until matter-radiation equality, in conflict with observational limits.

By contrast, within the regime in which $t_\phi > t_{\rm end}$ and the energy density of 
the $\phi$-particle gas is transferred to the visible sector only after the stasis 
epoch has run its natural course, the situation is very different.  Within this regime, 
the scaling exponent in Eq.~(\ref{eq:swept_vol_rate}) effectively changes from $q =-2$ to 
$q = 0$ once $\tau$ drops below $\taumin$.  Since $\langle \sigma v \rangle$ no longer
continues to rise once this threshold has been crossed, $P_{M,\gamma}$ becomes insufficient 
to maintain stasis.  As a result, $\Omega_M$ begins rising and continues to increase until 
the universe becomes effectively matter-dominated, while $\Omega_\gamma$ decreases.  
If $t_\phi$ is sufficiently late in comparison with $t_{\rm end}$ that $\Omega_\gamma$ 
is already negligible by the time $\phi$-particle decays reheat the visible sector, the 
subsequent evolution of the universe is effectively identical to that of the standard 
cosmology.  Thus, if $\phi$-particle decay is indeed the mechanism responsible for 
populating the visible sector after stasis ends, we must require that the energy
density $\rho_{M,{\rm end}}$ of the $\phi$-particle gas at the end of stasis be 
sufficiently large that the visible-sector radiation bath is reheated to a temperature 
above $T_{\rm BBN}$, despite the decrease in $\rho_M$ that must take place 
between $t_{\rm end}$ and $t_\phi$ in order for $\Omega_\gamma$ to decrease to
a phenomenologically acceptable level.    

In deriving this bound on $\rho_{M,{\rm end}}$, we begin by noting Eq.~(\ref{eq:xi_def}) 
implies that $\rho_{M,{\rm end}} = \barXi \taumin^2 m^4$
at the moment stasis ends.  Thus, in the approximation that decays have negligible 
effect on $\rho_M$ until $t\approx t_\phi$, the energy density of the $\phi$ particle 
gas at any time $t_{\rm end} \lesssim t \lesssim t_\phi$, expressed as a function of
the scale factor $a$, is approximately  
\begin{equation}
   \rho_M ~\approx~ \barXi\,\taumin^2 m^4 \left(\frac{a}{a_{\rm end}}\right)^{-3}~,
\end{equation}
where $a_{\rm end}$ denotes the scale factor at $t_{\rm end}$. 

The relationship between $a$ and $\Omega_\gamma$ once stasis ends can be determined
Eq.~(\ref{eq:convert3}).  In the approximation that the pump terms in the evolution 
equation for $\Omega_\gamma$ can be ignored once $t \gtrsim t_{\rm end}$, we have 
\begin{equation}
  \frac{d\Omega_\gamma}{da} ~=~ 
    \frac{1}{aH}\frac{d\Omega_\gamma}{dt} ~=~ 
    \frac{1}{a}\Omega_\gamma(1-\Omega_\gamma)~.    
\end{equation}
Solving this equation for $a$ in terms of $\Omega_\gamma$, we find 
\begin{equation}
   \frac{a}{a_{\rm end}} ~=~ 
     \frac{\barOmega_\gamma(1-\Omega_{\gamma})}
       {\Omega_{\gamma}(1-\barOmega_\gamma)}~.  
\end{equation}
The relationship between $\rho_M$ and $\Omega_\gamma$ at times 
$t_{\rm end} \lesssim t \lesssim t_\phi$ is therefore
\begin{equation}
  \rho_M ~\approx~ \barXi \,\taumin^2 m^4 \left[
    \frac{\barOmega_\gamma(1-\Omega_{\gamma})}
    {\Omega_{\gamma}(1-\barOmega_\gamma)}\right]^{-3}\,.
  \label{eq:rhoM_of_Omegagamma_poststasis}
\end{equation}

As discussed above, consistency with observation demands that $t_\phi$ be
sufficiently late that the corresponding dark-radiation abundance 
$\Omega_{\gamma}(t_{\rm BBN}) < \Omega_{\gamma,\max}^{(\rm BBN)}$ lies below the  
maximum phenomenologically acceptable value $\Omega_{\gamma,\max}^{(\rm BBN)}$ at the 
time $t_{\rm BBN}$ at which the BBN epoch begins.  Since the universe is dominated 
by SM radiation at times $t_{\phi} \lesssim t \lesssim t_{\rm BBN}$, $\Omega_{\gamma}$ 
remains effectively constant throughout this time period; thus, this constraint is 
equivalent to $\Omega_{\gamma}(t_\phi) < \Omega_{\gamma,\max}^{(\rm BBN)}$.  The 
corresponding constraint on $\varrho_M$ from 
Eq.~(\ref{eq:rhoM_of_Omegagamma_poststasis}) is 
\begin{equation}
  \varrho_M ~ >~ \varrho_{M,\max}^{(\rm BBN)} ~ \equiv ~ \barXi\,\taumin^2 \left[
    \frac{\barOmega_\gamma\big(1-\Omega_{\gamma,\max}^{(\rm BBN)}\big)}
    {\Omega_{\gamma,\max}^{(\rm BBN)}(1-\barOmega_\gamma)}\right]^{-3}\,.
  \label{eq:BBN_bound_varrhoM}
\end{equation}

Bounds on the dark-radiation abundance are typically expressed in terms
of the the additional number of neutrino species $\Delta N_{\rm eff}$, the 
value of which at $t_{\rm BBN}$ is related to the dark-radiation abundance by 
\begin{equation}
  \Omega_{\gamma}(t_{\rm BBN}) ~\approx~ \frac{8}{7}\left(\frac{4}{11}\right)^{4/3} 
    \Omega_{\rm phot}(t_{\rm BBN})\,\Delta N_{\rm eff}~,     
\end{equation}
where $\Omega_{\rm phot}(t)$ is the SM photon abundance at time $t$.  Current 
observational bounds on a non-interacting dark radiation component impose a constraint 
$\Delta N_{\rm eff} < 0.16$ at $68\%$~C.L.~\cite{Chang:2025uvx}, which corresponds to 
an upper bound $\Omega_{\gamma,\max}^{(\rm BBN)} \approx 6.8\times 10^{-3}$ on 
$\Omega_\gamma$.  We adopt this as our upper bound on the dark-radiation abundance
in what follows, though we note that recent improvements in the measurement of the 
primordial $^{4}{\rm He}$ abundance~\cite{Aver:2026dxv} could reduce the value of 
$\Omega_{\gamma,\max}^{(\rm BBN)}$ by around a factor of two~\cite{Yeh:2026pil}.

We shall assume in what follows that $\phi$ decay is indeed the mechanism
responsible for populating the visible sector after stasis ends, and thus that 
Eq.~(\ref{eq:BBN_bound_varrhoM}) represents the bound on our stasis model from 
observational cosmology.  However, we note that there exist other mechanisms via 
which the universe can transition from stasis to an epoch wherein the energy density 
is dominated by SM radiation.
One such possibility is that the dark-radiation field $\chi$, which we have taken to 
be massless, in fact has a small but non-zero mass and, like $\phi$, can also decay into 
visible-sector states.  Another such possibility involves positing that an additional,
light, axion-like scalar field $\psi$ with mass $m_\psi$ whose homogeneous zero-mode is 
displaced from its potential minimum is also present in the theory.  At early times, the energy 
density associated with this zero-mode behaves as vacuum energy, but as long as the 
corresponding abundance remains negligible during stasis, the presence of this additional 
cosmological component will not affect the stasis dynamics.  However, at late times, 
after stasis, this vacuum-energy density energy can come to dominate the energy density 
of the universe, while $\Omega_M$ and $\Omega_\gamma$ to negligible levels.  
Eventually, after the Hubble parameter falls below $H \sim 2m_\psi/3$ and $\phi$ and $\psi$ 
begins to behave as massive matter rather that vacuum energy.  The subsequent decay of $\psi$ 
to visible-sector states --- provided of course that $\psi$ couples to the fields of the 
SM --- can reheat the visible sector. 

\subsection{Summary of constraints}

A summary of the considerations discussed in this section and the manner 
in which they collectively constrain the parameter space of our model is provided in 
Table~\ref{tab:constraints}.  For any given combination of 
our model parameters $g_\chi$, $g_\phi$, $g_G$, and $\mu/m$, these considerations
restrict the range of $\varrho_M^{(0)}$ and $\tau^{(0)}$ under which stasis
can arise.

\section{Evolution before the expiration date \label{sec:Dynamics}}


We have now identified the conditions under which the stasis attractor emerges 
within our thermal stasis model.  However, the emergence of the stasis attractor
within a particular region of the parameter space of this model does not necessarily 
imply that the universe will in fact achieve stasis.  As discussed in 
Sect.~\ref{sec:emergence}, the amount of time which the system takes to evolve
within the ($\Omega_M,\Xi$) plane toward the stasis fixed point along different 
trajectories can be dramatically different depending on the initial conditions
$\Omega_M^{(0)}$ and $\Xi^{(0)}$ for our cosmological system.  In this section,
we examine the impact that these initial conditions have on whether or not the 
universe in fact achieves stasis before the attractor reaches its expiration date,
and if so, how long this stasis will endure.

\subsection{Evolution under the influence of the attractor}

In assessing the impact of initial conditions on the evolution of our cosmological 
system, we shall focus for simplicity on the impact of $\Xi^{(0)}$ and fix 
$\Omega_M^{(0)}\approx 1$.  In general, there are three relevant regimes for $\Xi^{(0)}$:
\begin{itemize}
  \item $\Xi^{(0)} \approx \barXi$: the coldness parameter is initially similar to
    its stasis value.
  \item $\Xi^{(0)} \ll \barXi$: the $\phi$-particle gas is initially ``too hot'' and 
    a significant reduction in its temperature (or increase in its energy density) 
    is required before the system can achieve stasis.
  \item $\Xi^{(0)} \gg \barXi$: the $\phi$-particle gas is initially ``too cold'' and 
    a significant increase in its temperature (or reduction in its energy density) 
    is required before the system can achieve stasis.
\end{itemize}

Within the first of these regimes, the system is already close to the fixed point 
at $(\barOmega_M,\barXi)$ at $t=t^{(0)}$.  Thus, both the trajectory along which the 
system evolves toward the attractor and the number of $e$-folds which is takes for the 
system to reach stasis along that trajectory can be gleaned from 
Fig.~\ref{fig:heatmap_attractor_wait_time}.  While the precise number of $e$-folds that 
it takes the system to do this depends on the initial conditions, it is clear from the
results shown in the figure that this number is never terribly large within our regime of 
interest, wherein $\Omega_M^{(0)} \approx 1$.  

By contrast, as we shall see, within the other two regimes it can take 
$\Omega_M$ and $\Xi$
far longer to settle into their stasis values.  Indeed, in extreme cases, this process can take 
such a long time to occur that the system reaches its expiration date before $\Omega_M$ and $\Xi$ 
reach their stasis values.  Nevertheless, even in such cases, the dynamics of the stasis attractor 
can have a significant impact on the expansion history of the universe, despite the fact that 
stasis is never fully realized.  Thus, it is interesting to examine the manner in which our 
system evolves toward stasis within these other two regimes, wherein $\Xi^{(0)}$ differs 
significantly from $\barXi$.

\begin{figure*}
  \centering
  \includegraphics[width=0.55\textwidth]
    {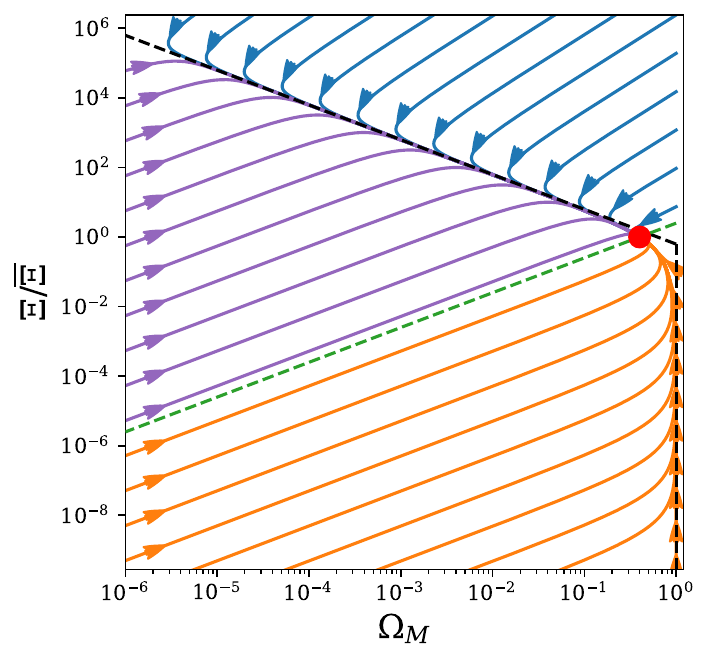}
  \caption{Trajectories (solid orange, blue, and purple curves) within the 
    $(\Omega_M,\Xi/\barXi)$ plane which illustrate the manner in which our 
    cosmological system evolves toward its stasis configuration from a given 
    initial configuration $(\Omega_M^{(0)},\Xi^{(0)}/\barXi)$ at time $t = t^{(0)}$.
    The location of the arrow along each curve indicates the state of the system 
    after the universe has undergone a single $e$-fold of expansion since $t^{(0)}$.
    For all trajectories which enter the $(\Omega_M,\Xi/\barXi)$ region shown in this 
    figure from the top or from the right, we have assumed an initial abundance 
    $\Omega_M^{(0)} = 1$.  By contrast, for all trajectories which enter this region
    from the bottom or from the left we have assumed an initial abundance 
    $\Omega_M^{(0)} = 10^{-6}$.  The red dot indicates the location of the stasis 
    fixed point.  The trajectories are shown in orange, blue, or purple to distinguish 
    their generic behaviors, as discussed in the text, while the green dashed line 
    along which $\Omega_M/\Xi = \barOmega_M/\barXi$ represents the ``fastest'' 
    trajectory discussed in Sect.~\protect\ref{sec:attractor}.~  The dashed black 
    line indicates the ``grand concourse'' along which most of our trajectories collect 
    before flowing to the fixed point, with the upper diagonal portion following the exact 
    relation $\Omega_M\Xi = (25/16)\barOmega_M\barXi$ while the lower portion is strictly 
    vertical with $\Omega_M=1$.  
\label{fig:extreme_xi_trajectories}}
\end{figure*}

As a first step in this direction, in Fig.~\ref{fig:extreme_xi_trajectories} we display 
trajectories (solid curves) within the $(\Omega_M,\Xi/\overline{\Xi})$ plane which illustrate the manner 
in which our cosmological system evolves toward its stasis configuration.  The trajectories 
which enter the region of the plane shown in the figure either from the top or from the right
correspond to the parameter choice $\Omega_M^{(0)} = 1$, whereas those which enter either 
from the bottom or from the left correspond to the parameter choice $\Omega_M^{(0)} = 10^{-6}$.
In many cases, the initial portion the of trajectory lies outside this region of the plane. 
The location of the arrow along each curve indicates the state of the system after the universe 
has undergone a single $e$-fold of expansion since $t = t^{(0)}$.  The red dot indicates the 
stasis configuration toward which the system is ultimately attracted.  The dashed 
black line indicates the contour along which $\Omega_M\Xi = \barOmega_M\barXi$.

These trajectories can be separated into three broad classes, which are indicated by the colors
of the corresponding curves.
The blue curves are representative of a class of trajectories in which $\Xi$ decreases 
monotonically while $\Omega_M$ initially decreases, eventually reaches a minimum, and then begins 
increasing again.  The trajectory thereafter approaches the black dashed line and then follows it 
to the stasis configuration, which is indicated by the red dot.  All trajectories within this 
class correspond to cases in which the $\phi$-particle gas is initially ``too cold,'' with 
$\Xi^{(0)} > \barXi$.  The location of the arrows along the trajectory indicates that the 
state of the system evolves far more rapidly for trajectories within this class during the first
$e$-fold of expansion that it does for trajectories in the other two classes. 

By contrast, the orange curves are representative of the second class of trajectories --- a 
class in which $\Omega_M$ and $\Xi$ both increase monotonically until the trajectory approaches 
the black dashed line.  Each such trajectory then effectively merges with this line and 
subsequently follows it to the stasis point.  All trajectories within this class correspond 
to cases wherein the $\phi$-particle gas is initially ``too hot,'' with $\Xi^{(0)} < \barXi$.

Finally, the purple curves are representative of the third class of trajectories --- a class 
for which $\Omega_M$ increases monotonically while $\Xi$ initially increases, reaches a maximum, 
and then decreases again.  The trajectory thereafter approaches the black dashed line and then 
follows it to the stasis point.  Some of the trajectories in this class correspond to cases in 
which the $\phi$-particle gas is initially too hot; others correspond to cases in which it is 
initially too cold.  However, trajectories within this class can only be realized in cases in
which $\Omega_M^{(0)} \ll 1$.  Thus, since we are focusing in this paper primarily on cases 
in which $\Omega_M^{(0)} \approx 1$, we shall not examine this class of trajectories in 
detail.  However, we note that the boundary between these purple curves and the orange ones 
corresponds to the ``fastest'' trajectory discussed in Sect.~\ref{sec:attractor}.  As 
the system approaches the fixed point from the left along this trajectory --- \ie, from 
the set of initial conditions in which $\Omega^{(0)} = 10^{-6}$ rather than $\Omega^{(0)} = 1$ ---
the system initially approaches the stasis fixed point in essentially the same manner as it does 
along the trajectories indicated by the orange curves.  However, it evolves toward stasis 
much more rapidly once it enters the vicinity of the fixed point.

For $\Xi^{(0)} < \barXi$ and $\barOmega_M^{(0)} = 1$, we see that $\Omega_M$ decreases 
monotonically toward $\barOmega_M$ as $\Xi$ increases monotonically toward $\barXi$.  
By contrast, for $\Xi^{(0)} > \barXi$, we see that $\Omega_M$ rapidly decreases to a 
minimum far below $\Omega_M$ and then increases toward $\barOmega_M$ as $\Xi$ monotonically 
decreases.  Moreover, we observe that once $\Omega_M$ reaches this minimum, the system evolves 
toward its stasis configuration along a trajectory for which the product $\Omega_M\Xi$ is 
approximately constant, regardless of the value of $\Xi^{(0)}$.  

In what follows, we shall 
examine the physical underpinnings of the numerical results shown in 
Fig.~\ref{fig:extreme_xi_trajectories}.  In the process, we shall establish a simple criterion
which allows us to identify regions of the ($\tau^{(0)}$, $\varrho_M^{(0)})$ plane wherein the 
system cannot reach stasis before the attractor reaches its expiration date. 

We begin by considering the manner in which the system evolves within the regime 
wherein $\Xi^{(0)} \ll \barXi$.  Within this regime, $\Xi$ must increase significantly 
before stasis is achieved.  For $q = -2$, the evolution equations for $\rho_M$ and $T$ 
appearing in Eqs.~(\ref{eq:convert3}) and~(\ref{eq:TempDiffEq}), respectively, may be
recast in the form
\begin{eqnarray}
  \frac{d\rho_M}{dt} &=& -3H\rho_M - 
    \frac{3}{5}\left(\frac{\Omega_M\Xi}{\barOmega_M\barXi}\right)^{1/2}
     H\rho_M \nonumber \\
  \frac{dT}{dt} &=& - 2HT + \frac{1}{5} 
    \left(\frac{\Omega_M\Xi}{\barOmega_M\barXi}\right)^{1/2} H T~.
  \label{eq:drhoM_dT_reformulated}
\end{eqnarray}
Changing variables from $t$ to $\calN$, we find that the evolution 
equations for the corresponding dimensionless quantities 
$\varrho_M$ and $\tau$ may be recast in the form 
\begin{eqnarray}
  \frac{d\log\varrho_M}{d\calN} &=& -3 - 
    \frac{3}{5}\left(\frac{\Omega_M\Xi}{\barOmega_M\barXi}\right)^{1/2}
    \nonumber \\
  \frac{d\log\tau}{d\calN} &=& - 2 + \frac{1}{5} 
    \left(\frac{\Omega_M\Xi}{\barOmega_M\barXi}\right)^{1/2}~.
  \label{eq:dvarrhoM_dtau_reformulated}
\end{eqnarray}

Within the regime in which $\Xi \ll \barXi$ and $\Omega_M \sim 1$, the 
second term on the right side of each of these equations is negligible in 
comparison with the first term, and we have
\begin{eqnarray}
  \frac{d\log\varrho_M}{d\calN} &\approx& -3 \nonumber\\
  \frac{d\log\tau}{d\calN} &\approx& -2~.
  \label{eq:log_evol_varrho_tau_hot}
\end{eqnarray}
It therefore follows from Eq.~(\ref{eq:xi_def}) that the corresponding 
evolution equation for $\Xi$ within this regime is 
\begin{equation}
  \frac{d\log\Xi}{d\calN} ~\approx~ 1~, 
  \label{eq:log_evol_xi_hot}
\end{equation}
the solution to which is $\Xi \approx \Xi^{(0)}e^{\calN}$.  Thus,
in the case in which $\Xi$ is initially much smaller than $\barXi$, we
find that $\Xi$ grows exponentially with $\calN$ until it 
becomes comparable to $\barXi$ and the pump terms are no longer subleading, as
shown in Fig.~\ref{fig:extreme_xi_trajectories}.

Together, the relations in Eqs.~(\ref{eq:log_evol_varrho_tau_hot}) 
and~(\ref{eq:log_evol_xi_hot}) imply that the quantity 
$\Xi^2 \tau = \varrho_M^2/\tau^3$ remains effectively constant while $\Xi$ 
remains well below $\barXi$ and the pump terms in Eq.~(\ref{eq:drhoM_dT_reformulated}) 
may therefore be neglected.  Thus, it follows that the dimensionless temperature 
$\tau^{(h1)}$ at which the coldness becomes comparable to $\barXi$ is
\begin{equation}
  \tau^{(h1)} ~=~ \left(\frac{\Xi^{(0)}}{\barXi}\right)^2\tau^{(0)} 
    ~=~ \frac{(\varrho_M^{(0)})^2}{\barXi^2 (\tau^{(0)})^3}~.
    \label{eq:taus_final_hot}
\end{equation}

In order for the system to reach stasis before the attractor reaches its expiration date,  
$\tau^{(h1)}$ must be sufficiently large in comparison with $\taumin$ that the system 
has time to evolve toward the fixed point once $\Xi$ becomes comparable to $\barXi$ 
and the pump terms in Eq.~(\ref{eq:drhoM_dT_reformulated}) become important.  Using 
Eq.~(\ref{eq:taus_final_hot}) in order to express this condition as a relation between
$\varrho_M^{(0)}$ and $\tau^{(0)}$, we have
\begin{eqnarray}
  \varrho^{(0)} ~>~ \barXi \,\taumin^{1/2} (\tau^{(0)})^{3/2}~.
  \label{eq:coolingCondition}
\end{eqnarray}
Thus, within regions of the $(\tau^{(0)},\varrho_M^{(0)})$ plane wherein this criterion is 
{\it not}\/ satisfied, the system cannot ever actually reach stasis, despite 
the influence of the attractor. 

We note that while Eq.~(\ref{eq:coolingCondition}) was derived under the assumption
$\Omega_M^{(0)} \approx 1$, the corresponding condition which obtains for any
choice of $\Omega_M^{(0)}$ and $\Xi^{(0)}$ which satisfy 
\begin{equation}
    \frac{\Omega_M^{(0)}}{\barOmega_M} ~\gg~ \frac{\Xi^{(0)}}{\barXi}
\end{equation}
will have essentially the same form.  This is the case, for example, for all of the orange 
curves shown in Fig.~\ref{fig:extreme_xi_trajectories} for which $\Omega_M^{(0)}=10^{-6}$.  Indeed, we 
observe that $\Omega_M$ and $\Xi$ initially both increase along each of these trajectories
until the trajectory approaches and then begins following the vertical dashed black line --- a 
line which constitutes one of two segments of a ``grand concourse'' onto which all of 
these trajectories are ultimately routed --- toward the stasis fixed point.  As the 
system evolves along the grand concourse it remains expansion-dominated, and indeed it is 
only at the {\it end}\/ of its evolution down the grand concourse when $\tau$ reaches 
$\tau^{(h1)}$. 

We now turn to consider the case in which $\Xi^{(0)} \gg \barXi$ and in which $\Xi$ must 
therefore decrease significantly before stasis is achieved.  Annihilation is initially 
extremely efficient in this case.  This can be due either to the thermally-averaged 
swept-volume rate $\langle\sigma v\rangle \sim 1/\tau$ or to the number density of $\phi$ 
particles being large, depending on the trajectory in question and the location along
that trajectory.  The efficiency of the annihilation process in this case is ultimately
what gives rise to the behavior exhibited by the blue curves in 
Fig.~\ref{fig:extreme_xi_trajectories}.  Along these trajectories, $\Xi$ 
decreases monotonically.  By contrast, $\Omega_M$ initially plummets to values 
well below $\barOmega_M$, but once the annihilation rate is sufficiently suppressed 
by the corresponding decrease in $\rho_M$, the matter abundance begins rising again
and the trajectory merges with the diagonal segment of the dashed black line --- 
another portion of the ``grand concourse'' onto which various trajectories in the 
$(\Omega_M,\Xi)$ plane are routed --- as the system evolves toward the stasis fixed
point.

Within this regime, the evolution equations for general values of $\Omega_M$ and $\Xi$ 
may be obtained via straightforward calculation from the evolution equations in 
Eq.~(\ref{eq:drhoM_dT_reformulated}).  In particular, we find that
\begin{eqnarray}\label{sweqs_b}
  \frac{d \log\Omega_M}{d\Ncal} &=& 1 - \Omega_M 
    - \frac{3}{5}\left(\frac{\Omega_M\Xi}{\barOmega_M\barXi}\right)^{1/2}\nonumber\\
  \frac{d\log\Xi}{d\Ncal} &=& 1  
    - \left(\frac{\Omega_M\Xi}{\barOmega_M\barXi}\right)^{1/2}~.
  \label{eq:log_evol_OmegaM_xi_gen}
\end{eqnarray}
By combining these equations, we may also obtain
an evolution equation for the product $\Omega_M\Xi$ of the form 
\begin{equation}
  \frac{d \log(\Omega_M\Xi)}{d\Ncal} ~=~ 2 -\Omega_M - \frac{8}{5}
    \left(\frac{\Omega_M\Xi}{\barOmega_M\barXi}\right)^{1/2}~.
  \label{eq:log_evol_OmegaMxiprod_gen}    
\end{equation}

The first stage in the dynamical evolution of the system, wherein $\Omega_M$ 
rapidly decreases from its initial value, ends at the time $t^{(c1)}$ at which 
$\Omega_M$ reaches a minimum.  According to Eq.~(\ref{eq:log_evol_OmegaM_xi_gen}), 
this minimum occurs when
\begin{equation} 
    \left(\frac{\Omega_M\Xi}{\barOmega_M\barXi}\right)^{1/2} ~=~
      \frac{5}{3}(1 - \Omega_M)~.
\end{equation}
Since $\Omega_M < \barOmega_M$ at this minimum, the right side of this equation
is an $\mathcal{O}(1)$ number.  Therefore, the time it takes for $\Omega_M$ to reach its 
minimum is, roughly speaking, the time it takes for $\Omega_M\Xi$ to decrease from its 
initial value $\Omega_M^{(0)}\Xi^{(0)} \gg \barOmega_M\barXi$ to a value
$\Omega_M\Xi \sim \barOmega_M\barXi$.  Until $\Omega_M\Xi$ reaches this value, the 
last term on the right side of each of the individual equations in
Eq.~(\ref{eq:log_evol_OmegaM_xi_gen}) --- the term associated with the stasis pump ---
dominates and these equations effectively reduce to
\begin{eqnarray} 
 \frac{d \log\Omega_M}{d\Ncal} &\approx& - 
   \frac{3}{5}\left(\frac{\Omega_M\Xi}{\barOmega_M\barXi}\right)^{1/2}\nonumber\\
  \frac{d\log\Xi}{d\Ncal} &\approx& - \left(\frac{\Omega_M\Xi}{\barOmega_M\barXi}\right)^{1/2}~.
  \label{eq:dWdotS_dN}
\end{eqnarray}
Together, these relations imply that $\Omega_M / \Xi^{3/5}$ is constant while 
$t^{(0)}\lesssim t \lesssim t^{(c1)}$.  Furthermore, the corresponding equation
\begin{equation}
  \frac{d \log(\Omega_M\Xi)}{d\Ncal} ~\approx~ -\frac{8}{5}
    \left(\frac{\Omega_M\Xi}{\barOmega_M\barXi}\right)^{1/2}~.
  \label{eq:dWdotS_dN_Comb}
\end{equation}
to which Eq.~(\ref{eq:log_evol_OmegaMxiprod_gen}) reduces during this time interval is 
superlinear in $\Omega_M\Xi$.  This implies that the number of $e$-folds of expansion
which the universe undergoes between $t^{(0)}$ and $t^{(c1)}$ is $\mathcal{O}(1)$ and 
not particularly sensitive to $\Omega_M^{(0)}\Xi^{(0)}$.  

Comparing the evolution equation for the product $\Omega_M\Xi$ in 
Eq.~(\ref{eq:dWdotS_dN_Comb}) to the evolution equations for $\tau$ and $\Xi$ in 
Eqs.~(\ref{eq:dvarrhoM_dtau_reformulated}) and~(\ref{eq:log_evol_OmegaM_xi_gen}),
respectively, we observe that
\begin{eqnarray}
  \frac{d\log\tau}{d\calN} &~\approx~& 
    -\frac{1}{8}\,\frac{d\log(\Omega_M\Xi)}{d\Ncal} \nonumber \\
  \frac{d\log\Xi}{d\calN} &~\approx~& 
    \frac{5}{8}\,\frac{d\log(\Omega_M\Xi)}{d\Ncal}~.
  \label{eq:dlogtau_and_dlogxi_plunge}
\end{eqnarray}
We can use these relations to obtain 
a relation between $\tau$ and $\Omega_M\Xi$ at the end of this phase of cosmological evolution.
Indeed, the first of the relations in Eq.~(\ref{eq:dlogtau_and_dlogxi_plunge}) implies that
\begin{eqnarray}
  \log\left(\frac{\tau^{(c1)}}{\tau^{(0)}}\right) &~\sim~& \frac{1}{8}
    \log\left(\frac{\Omega_M^{(0)}\Xi^{(0)}}{\barOmega_M\barXi}\right) \nonumber \\
  &~\sim~& \frac{1}{8} \log\left(\frac{\Xi^{(0)}}{\barXi}\right)~
  \label{tau_1_heating_approx}
\end{eqnarray}
at $t^{(c1)}$, where $\tau^{(c1)}$ denotes the value of $\tau$ at this time and where in 
going from the first to the second line we have assumed that $\Omega_M^{(0)}\approx 1$, 
and thus that $\Omega_M^{(0)} \sim \barOmega_M$.

A similar relationship between $\Xi$ and $\Xi^{(0)}$ may be obtained by solving the 
second equation in Eq.~(\ref{eq:dlogtau_and_dlogxi_plunge}).  Doing so, we find that
\begin{equation}
  \log\left(\frac{\Xi^{(c1)}}{\Xi^{(0)}}\right) ~\approx~ 
    -\frac{5}{8} \log\left(\frac{\Xi^{(0)}}{\barXi}\right)~,
  \label{eq:xi1_rel}
\end{equation}
where $\Xi^{(c1)}$ denoted the value of $\Xi$ at time $t^{(c1)}$.

During the second stage in the dynamical evolution of the system, after $\Omega_M$ has
reached its minimum --- a minimum which lies below $\barOmega_M$, often by several orders 
of magnitude --- $\Omega_M$ once again begins increasing toward $\barOmega_M$. 
We see from Fig.~\ref{fig:extreme_xi_trajectories} that while $\Omega_M$ is increasing in
this manner, $\Xi$ decreases in such a way that the trajectory of the system in the 
$(\Omega_M,\Xi)$ plane merges onto the diagonal segment of the ``grand concourse'' discussed 
above.  Moreover, during this stage of dynamical evolution, we find that the product 
$\Omega_M\Xi$ approaches a fixed value which is the same for every such trajectory.  
We may determine this fixed value of $\Omega_M\Xi$ by considering a trajectory for which
$\Omega_M \ll 1$ at time $t = t^{(c1)}$.  For such a small value of $\Omega_M$, 
Eq.~(\ref{eq:log_evol_OmegaMxiprod_gen}) implies that
\begin{equation}
  \left(\frac{\Omega_M\Xi}{\barOmega_M\barXi}\right)^{1/2} ~\approx~ \frac{5}{4}~.
  \label{eq:FiveFourthsEq}
\end{equation}
and thus that $\Omega_M\Xi = (25/16)\barOmega_M\barXi$.  Substituting this result into the 
evolution equations for $\tau$ and $\Xi$ in Eqs.~(\ref{eq:dvarrhoM_dtau_reformulated})
and~(\ref{eq:log_evol_OmegaM_xi_gen}), respectively, and comparing the resulting expressions, 
we observe that
\begin{equation}
  \frac{d\log\tau}{d\calN} ~\approx~ 7\,\frac{d\log\Xi}{d\calN}~.
  \label{eq:dxi_vs_dtau_rise}
\end{equation}
Solving this equation in order to obtain a relation between $\tau$ and $\Xi$ and
evaluating this expression at the time $t^{(c2)}$ at which $\Omega_M$ becomes 
comparable to $\barOmega_M$, we find that the dimensionless temperature 
$\tau^{(c2)}$ at this time is given by
\begin{equation}
  \log\left(\frac{\tau^{(c2)}}{\tau^{(c1)}}\right) ~\approx~ 
      7\log\left(\frac{\barXi}{\Xi^{(c1)}}\right) 
    ~\approx~ -\frac{21}{8}\log\left(\frac{\Xi^{(0)}}{\barXi}\right)~,
  \label{eq:taus_over_tau1}
\end{equation}
where in going from the first to the second equality we 
have used Eq.~(\ref{eq:xi1_rel}).  Combining this result with the result in 
Eq.~(\ref{tau_1_heating_approx}), we find that
\begin{eqnarray}
  \tau^{(c2)} ~=~ \left(\frac{\Xi^{(0)}}{\barXi}\right)^{-5/2} \tau^{(0)}~.
  \label{eq:taus_final_cold}
\end{eqnarray}

In order for the system to reach stasis before its expiration date, it must be the 
case that $\tau^{(c2)} > \taumin$.  Thus, using Eq.~(\ref{eq:taus_final_cold}) in order 
to express this condition as a relation between $\varrho_M^{(0)}$ and $\tau^{(0)}$, 
we find that
\begin{eqnarray}\label{heatingConstraint}
  \varrho_M^{(0)} ~<~ \frac{\barXi}{\taumin^{2/5}} (\tau^{(0)})^{12/5}~.
  \label{eq:Reach StasisCondTooHot}
\end{eqnarray}
We also note that while we have assumed in deriving this condition that 
$\Omega_M^{(0)} \approx 1$, the corresponding condition on $\tau^{(2)}$ which
obtains for $\Omega_M^{(0)} \ll 0$ is even more stringent.

Finally, the trajectories represented by the purple curves in 
Fig.~\ref{fig:extreme_xi_trajectories} correspond to initial conditions wherein 
$\Omega_M^{(0)} \ll 1$.  These trajectories are therefore outside  
our primary regime of interest
in this paper.  Nevertheless, we note that during the first stage in the dynamical
evolution of the system along these trajectories, wherein both $\Omega_M$ and $\Xi$
increase, the last term in the equation in Eq.~(\ref{eq:log_evol_OmegaM_xi_gen}) 
which governs the evolution of each of these quantities is negligible.  The evolution
of the system is therefore governed by expansion rather than the stasis pump, 
regardless of the relationship between $\Xi^{(0)}$ and $\barXi$.  As a result, the 
system evolves in essentially the same way during this stage of dynamical evolution
as it does along the trajectories represented by the orange curves shown in 
Fig.~\ref{fig:extreme_xi_trajectories}.  However, the system then merges onto the 
diagonal rather than onto the vertical segment of the grand concourse.  Once this 
occurs, the system evolves toward stasis in much the same manner as it does during 
the second stage of evolution that the system undergoes along the trajectories 
represented by the blue curves shown in the figure.

In summary, then, combining the conditions in Eqs.~(\ref{eq:coolingCondition}) 
and~(\ref{eq:Reach StasisCondTooHot}), we find that it is not possible for the system 
to reach stasis prior to the expiration date for the attractor unless
\begin{equation}
  \barXi \taumin^{2}\left(\frac{\tau^{(0)}}{\taumin}\right)^{3/2} 
    ~<~ \varrho_M^{(0)} ~<~  
    \barXi\tau_{\rm min}^2\left(\frac{\tau^{(0)}}{\taumin}\right)^{12/5}~.
  \label{eq:ExpDateConditXiOmega}
\end{equation}
These conditions --- \ie, the upper and lower bounds on $\varrho_M^{(0)}$  --- are 
also summarized in Table~\ref{tab:constraintsOnStasis}.~  

\begin{table*}[p]
    \centering
    \begin{tblr}{||Q[c,m,0.3\linewidth]|Q[c,m,0.3\linewidth]|Q[c,m,0.3\linewidth]||}
        \hline \hline
        Constraint Equation 
        & Physical Condition Imposed 
        & Consequence if Violated 
        \\ \hline \hline
          {$\displaystyle\varrho_{M} ~<~ 
          \frac{ g_{M}^{4}\Omega_M}{3 \cdot 2^{15} \pi^{3} g_{G}^{2}}\,\tau$} 
        & {General-relativistic effects on the $X$ propagator must be negligible.}
        & {Behavior unknown.} 
        \\ \hline
        {See Eqs.~(\ref{fusion_condition_full_temperate}),~(\ref{fusion_condition_full_cold}),
          and~(\ref{fusion_condition_full_hot})} 
        & {Exothermic scattering processes must have a negligible effect on the 
          $\phi$-particle gas.}
        & \SetCell[r=2]{m,0.3\linewidth}
        {$\tau$ increases rapidly while $\varrho_M$ decreases much more slowly  
           until the relevant condition is satisfied.  If $\tau$ reaches $\taumax$ 
           as a result of this process, behavior unknown.}
        \\ \cline{1-2}
          {$\displaystyle \varrho_M ~<~ \frac{\zeta(3/2)}{(2\pi)^{3/2}}\,\tau^{3/2}$}  
        & {A Bose-Einstein condensate must not form within the $\phi$-particle gas.}
        & 
        \\ \hline
          {$\displaystyle\varrho_M ~<~ 
          \frac{3\Omega_M}{2^{15}\pi^3}\frac{g_\chi^4}{g_G^2}$} 
        & {The population of relic $X$ particles must be negligible at the 
          beginning of the stasis epoch.}
        & {If $\Xi<\barXi$, then $\varrho_M$ and $\tau$ decrease such that 
            $d\ln\tau = \frac{2}{3} d\ln\varrho_M$ until this condition is 
            satisfied.  If $\Xi>\barXi$ either initially or as a result of 
            this subsequent evolution, behavior unknown.} 
        \\ \hline
          {$\displaystyle\tau ~<~ \taumax $} 
        & {The term in the denominator of $\Delta(p_X)$ quadratic in
          $|\vec{p}_{\rm CM}|$ must be subleading in comparison with
          the term linear in $|\vec{p}_{\rm CM}|$.}
        & {If $\Xi<\barXi$, then $\varrho_M$ and $\tau$ decrease such that 
          that $d\ln\tau = \frac23 d\ln\varrho_M$ until $\tau < \taumax$ or 
            until another constraint is violated.  Otherwise, behavior unknown.} 
        \\ \hline
          $\displaystyle\varrho_M ~<~ 
          \frac{3 \Omega_{M}^{3}}{32 \pi N_A^{2} g_{G}^{6}}$ 
        & {The $\phi$-particle gas must be in the thermodynamic limit.}
        & {Behavior unknown.} 
        \\ \hline
            {$\displaystyle\tau ~>~ \taumin$} 
        & {The $|\vec{p}_{\rm CM}|$-independent term in the denominator of 
          $\Delta(p_X)$ must be subleading in comparison with
          the term linear in $|\vec{p}_{\rm CM}|$.}
        & The attractor has reached its expiration date.
        \\ \hline
          $\displaystyle\varrho_M ~>~ \varrho_{M,\max}^{(\rm BBN)}$ 
        & {Stasis must end before the BBN epoch begins.}
        & The resulting cosmology conflicts with observation and is therefore excluded.
        \\ \hline \hline
    \end{tblr}
    \caption{Summary of the conditions which constrain the stasis attractor in our model. 
      The corresponding constraint equations are to be applied in the order in which they
      appear here from top to bottom.
      \label{tab:constraints}}
\bigskip
\bigskip
\bigskip
    \centering
    \begin{tblr}{||Q[c,m,0.3\linewidth]|Q[c,m,0.3\linewidth]|Q[c,m,0.3\linewidth]||}
        \hline \hline
        Constraint Equation 
        & Physical Condition Imposed 
        & Consequence if Violated 
        \\ \hline \hline
          {$\displaystyle\varrho_M ~>~ 
          \barXi \taumin^{1/2}\tau^{3/2}$}
        & {$\Xi^{(0)}$ must not be too much smaller than $\barXi$ so that the system has 
        time to ``cool down'' to $\barXi$ before $\tau$ falls below $\taumin$.}
        & The system cannot reach stasis before the expiration date.
        \\ \hline
          $\displaystyle \varrho_{M} ~<~ 
          \frac{\barXi}{\taumin^{2/5}}\tau^{12/5}$
        & {$\Xi^{(0)}$ must not be too much larger than $\barXi$ so that the system 
        has time to ``heat up'' to $\barXi$ before $\tau$ falls below $\taumin$.}
        & The system cannot reach stasis before the expiration date.
        \\ \hline \hline
    \end{tblr}
    \caption{Summary of the conditions which constrain the possibility of reaching stasis 
      in our model.  These conditions are to be considered only after all conditions in 
      Table~\ref{tab:constraints} are applied. 
      \label{tab:constraintsOnStasis}}
\end{table*}

We note that these conditions cannot be satisfied if $\tau^{(0)} < \taumin$.  Indeed,
this is a reflection of the fact that the system cannot reach stasis once the 
attractor is no longer active.  However, we also note that simply satisfying 
Eq.~(\ref{eq:ExpDateConditXiOmega}) is not a guarantee that the system
{\it will}\/ in fact reach stasis before the expiration date, since it may take the 
system a significant number of $e$-folds to come sufficiently close to the 
fixed point that the stasis criterion in Eq.~(\ref{trueDeltaDefinition}) is satisfied.

\subsection{Delivered to the doorstep of stasis: \\
Exothermic processes and 
initial conditions\label{sec:ExoInitImpact}}

There is one additional consideration for which we must account when examining 
how our cosmological system evolves from its initial configuration 
$(\Omega_M^{(0)},\Xi^{(0)})$ along its trajectory.  This is the effect that 
$4\phi\to 2\phi$ processes of the sort discussed in Sec.~\ref{sec:Exothermic} have 
on the evolution of our matter density and temperature.  Of course, when the 
condition in Eq.~(\ref{eq:PMKE_rapid_condit_T_with_strong_pump}) is satisfied 
and $4\phi\to 2\phi$ scattering has a non-negligible effect on the cosmological 
dynamics, the stasis attractor is not realized.  Nevertheless, there can be 
situations wherein $\Omega_M^{(0)}$ and $\Xi^{(0)}$ are such that the effect of 
$4\phi\to 2\phi$ scattering is initially non-negligible, but wherein this effect 
propels the system toward a region of the $(\Omega_M,\Xi)$ plane wherein the 
attractor {\it is}\/ realized and the system thereafter begins evolving toward the 
stasis fixed point. 

The qualitative effect that $4\phi \to 2\phi$ processes have on the manner in
which our system evolves within the $(\tau,\varrho_M)$ plane can be ascertained 
from the result derived in Sect.~\ref{sec:Exothermic} that $P_{M,\KE}^{(\rho)}$ always 
has a greater impact on the evolution of $\ln\tau$ than it does on the evolution of $\ln\varrho_M$.
This implies that whenever $P_{M,\KE}^{(\rho)}$ plays a dominant role in the 
evolution of our cosmological system, $\tau$ increases significantly, whereas the 
corresponding change in $\varrho_M$ is comparatively small.  This behavior will persist 
until $\tau$ becomes sufficiently large that $P_{M,\KE}^{(\rho)}$ no longer dominates.  
In this way, the expiration date for stasis is essentially reset and the system evolves 
toward stasis from a more favorable location within the $(\tau,\varrho_M)$ plane.   

In this connection, we note that regions of the $(\tau,\varrho_M)$ plane wherein the 
condition in Eq.~(\ref{initial_condition_bec}) is violated and the $\phi$-particle 
gas forms a BEC are regions wherein $4\phi\to 2\phi$ annihilation processes are likely 
to play a significant role in the evolution of our cosmological system.  Indeed, the 
amplitude for such processes in Eq.~(\ref{eq:general_M42}) is such that the corresponding 
annihilation rate increases as the momenta of the incoming $\phi$ particles decrease.  Within
the regime in which the $\phi$-particle gas forms a BEC, a significant fraction of 
these particles are in the ground state and thus have exceedingly low momenta.  If
$\rho_M$ is sufficiently large that the formation of the BEC occurs while this gas 
is within the  ``hot'' or ``temperate'' regimes, despite its low temperature,  
the $4\phi \to 2\phi$ scattering rate can be dramatically enhanced.  Within the 
``temperate'' regime the enhancement factor is $\mathcal{O}(\tau^2/\taumin^2)$, while 
within the ``hot'' regime it is even larger.  Indeed, as we shall see, the effect of 
$4\phi\to 2\phi$ processes within the region of the $(\tau,\varrho_M)$ plane wherein 
the $\phi$-particle gas forms a BEC in fact renders it possible for the universe either 
to achieve stasis or to at least spend a significant number of $e$-folds of expansion 
under the influence of the stasis attractor before the expiration date is reached.


\section{Results \label{sec:Results}}


In Sects.~\ref{sec:constraints} and~\ref{sec:Dynamics}, we established a set of
exclusion contours within the $(\tau,\varrho_M)$ plane which correspond to the
consistency conditions and constraints applicable to our thermal stasis model 
and derived a set of approximate, analytic expressions which describe the 
trajectories along which the state of our system evolves under the influence of 
the stasis attractor.  In this section, we present our numerical results for 
these trajectories within the region of the $(\tau,\varrho_M)$ plane allowed 
by these exclusion contours and examine the timescales involved in the evolution 
toward stasis from different initial conditions.  As we shall see, a 
significant number of $e$-folds of stasis can be realized within the context of 
this model.  However, we shall also see that the duration of the stasis epoch is
quite sensitive to the initial conditions for the system.

\subsection{Dynamical evolution and initial conditions\label{sec:ResultsEvol}}

\begin{figure*}
  \centering
    \includegraphics[keepaspectratio, width=1 \textwidth]
      {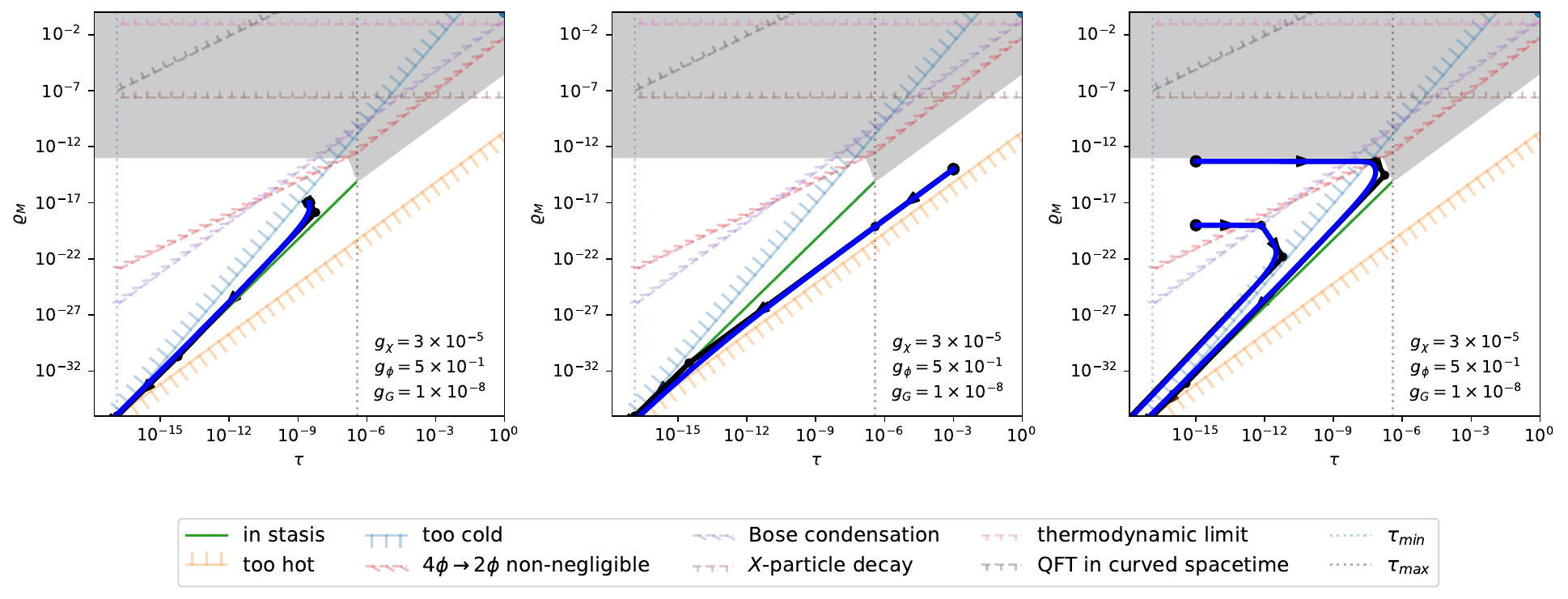}
\caption{ 
Representative trajectories for our system, similar to those in 
Fig.~\ref{fig:extreme_xi_trajectories} but now plotted within the $(\tau,\varrho_M)$ 
plane (thick blue lines) and separated into different panels for clarity.
The trajectory shown in the left panel corresponds to one of the blue trajectories in 
Fig.~\ref{fig:extreme_xi_trajectories}, while the trajectory shown in the middle panel 
corresponds to one of the orange trajectories.  By contrast, the two trajectories shown 
in the right panel are initially dominated by $4\phi\to 2\phi$ processes;  eventually 
these processes become too slow to matter, after which these trajectories resemble the 
blue trajectories in Fig.~\ref{fig:extreme_xi_trajectories}.  The larger black dot(s) 
in each panel indicate the initial conditions adopted for the system.  The black 
trajectories represent piecewise approximations to the true trajectories --- approximations 
evaluated according to the analytic approach we developed in Sect.~\ref{sec:Dynamics}.~  The 
smaller black dots represent the transition points between the pieces of these piecewise
paths.  Also shown in each panel are the various constraint contours discussed in 
Sects.~\ref{sec:constraints} and~\ref{sec:Dynamics} and summarized in 
Tables~\ref{tab:constraints} and~\ref{tab:constraintsOnStasis}.~  Note that the BBN 
constraint does not exclude any portion of the $(\tau,\varrho_M)$ plane shown.  
Within each panel we have assumed the same values for the couplings 
$g_\chi = 3.0\times 10^{-5}$, $g_\phi = 5.0\times 10^{-1}$, $g_G = 1.0 \times 10^{-8}$, 
and $\mu$ taken to satisfy Eq.~(\ref{mumaximumeff}), while the gray regions indicate 
where the dynamical evolution of the 
system involves additional effects beyond the scope of our analysis.  Finally, the green 
solid line appearing in each panel represents the relationship $\varrho_M=\barXi \tau^2$ 
between $\tau$ and $\varrho_M$ which holds once the stasis epoch begins.  Note that the 
results shown in this figure do not account for the manner in which the stasis pump 
changes when $\tau \sim \taumin$ and the $\phi$-particle gas begins to depart from thermal 
equilibrium. 
    \label{fig:trajectories}}    
\end{figure*}

We begin by examining the manner in which our cosmological system 
evolves within the $(\tau,\varrho_M)$ plane under the influence of the stasis 
attractor, subject to the constraints and model-consistency conditions summarized 
in Tables~\ref{tab:constraints} and~\ref{tab:constraintsOnStasis}.~  
In Fig.~\ref{fig:trajectories}, we display a number 
of trajectories illustrating how the system evolves from several
different initial conditions within this plane (indicated by the black dot or 
dots in each panel).  The blue curves represent our full numerical 
results, while the black curves represent the piecewise approximations to these 
curves using the analytic expressions derived in Sect.~\ref{sec:Dynamics}.~
The results shown in all panels correspond to the parameter choices 
$g_\chi = 3.0\times 10^{-5}$, $g_\phi = 5.0\times 10^{-1}$, 
$g_G = 1.0 \times 10^{-8}$, and $\mu= 0$.  We neglect the effect on these 
trajectories of modifications to the form of the stasis pump which arise 
when $\tau \sim \taumin$ and the $\phi$-particle gas begins to depart from thermal 
equilibrium.  The exclusion contours associated 
with the constraints itemized in Tables~\ref{tab:constraints} 
and~\ref{tab:constraintsOnStasis} are shown in each panel.  The short segments 
extending from each contour indicate the region which is excluded by the 
constraint.  The vertical dotted lines indicate the corresponding values of $\tau_{\rm min}$
and $\tau_{\rm max}$.  The solid green contour, along which we have $\rho_M = \barXi \tau^2$,
indicates the relationship between $\tau$ and $\varrho_M$ which holds while 
the system is in stasis.  Within the solid gray region of each panel, the 
dynamical evolution of the system involves additional effects beyond the scope 
of our analysis.

Before we discuss the trajectories along which our system evolves within the 
$(\tau,\varrho_M)$ plane, we begin by highlighting several important observations 
regarding the constraint contours themselves.  First, we observe that the constraint
in Table~\ref{tab:constraints} associated with general-relativistic effects on the $X$ 
propagator --- the constraint which corresponds to the gray contour in 
Fig.~\ref{fig:trajectories} --- is vastly 
subleading in comparison with other constraints on our model.  Thus, our results are 
insensitive to the precise form of the coefficient $b_X^{(1)}(p^2)$ defined in 
Sect.~\ref{subsec:GravQFT}.~  Second, we note that no exclusion contour 
associated with the bound in Table~\ref{tab:constraints} imposed by BBN 
constraints appears in Fig.~\ref{fig:trajectories}.  Indeed, we find that the 
corresponding bound within the $(\tau,\varrho_M)$ plane is sufficiently weak that 
no portion of that plane shown in the figure is excluded by it.  Finally, we note that 
the relationship between the blue and orange contours plays a crucial role within
the context of this model.  In particular, it is within the wedge-shaped region 
in each panel of Fig.~\ref{fig:trajectories} which lies between these two contours 
and between the $\tau_{\rm min}$ and $\tau_{\rm max}$ lines that the stasis attractor 
is realized and the two consistency conditions in Table~\ref{tab:constraintsOnStasis} 
are both satisfied.  Thus, for initial conditions such that $(\tau^{(0)},\varrho_M^{(0)})$ 
lies within this ``stasis wedge,'' the system is capable of reaching stasis unless other 
consistency conditions are violated.

The trajectory shown in the left panel of Fig.~\ref{fig:trajectories} is 
representative of the class of trajectories for which $(\tau^{(0)},\varrho_M^{(0)})$ 
lies within the stasis wedge and $\Xi^{(0)} \gg \barXi$.
This class of trajectories within the $(\tau,\varrho_M)$ plane corresponds to the 
class of trajectories within the $(\Omega_M,\Xi/\barXi)$ plane represented by the blue 
curves in Fig.~\ref{fig:extreme_xi_trajectories}.  For trajectories within this class,
the second term on the right side of each of the evolution equations in 
Eq.~(\ref{eq:log_evol_varrho_tau_hot}) initially dominates.  These evolution
equations therefore together imply that the trajectory along which the system initially 
evolves is one along which the quantity $\varrho_M \tau^{3}$ remains approximately 
constant as $\tau$ increases.  This first portion of the trajectory within 
the $(\tau,\varrho_M)$ plane corresponds to the portion of the trajectory within the 
$(\Omega_M,\Xi)$ plane where $\Omega_M$ and $\Xi$ are both decreasing and the system has not 
yet merged onto the grand concourse.  The quantity $\Omega_M \Xi$ decreases while the system 
evolves along this first portion of the trajectory, as discussed in Sect.~\ref{sec:Dynamics}, 
and once this quantity decreases to the point where it becomes comparable to $\barOmega_M\barXi$,
the behavior of the system changes.  Thereafter, $\Omega_M \Xi$ remains approximately constant 
at the value implied by Eq.~(\ref{eq:FiveFourthsEq}).  Thus, as the system evolves along 
this second portion of the trajectory the equations in Eq.~(\ref{eq:log_evol_varrho_tau_hot}) 
together imply that the quantity $\varrho_M\tau^{-15/7}$ remains constant as $\tau$ decreases.  
This second portion of the trajectory within the $(\tau,\varrho_M)$ plane corresponds to the 
portion of the trajectory in the $(\Omega_M,\Xi)$
plane where the system proceeds along the 
grand concourse toward the stasis fixed point.

By contrast, the trajectory shown in the right panel of Fig.~\ref{fig:trajectories} 
is representative of the class of trajectories for which $(\tau^{(0)},\varrho_M^{(0)})$ 
likewise lies within the stasis wedge, but for which $\Xi^{(0)} \ll \barXi$.
This class of trajectories within the $(\tau,\varrho_M)$ plane corresponds to the class of 
trajectories within the $(\Omega_M,\Xi)$ plane represented by the orange curves with 
$\Omega^{(0)}\sim 1$ in Fig.~\ref{fig:extreme_xi_trajectories}.  For trajectories
within this class, the evolution equations for $\varrho_M$ and $\tau$ in 
Eq.~(\ref{eq:log_evol_varrho_tau_hot}) together imply that the trajectory along which 
the system evolves toward the stasis contour is one along which the quantity 
$\varrho_M \tau^{-3/2}$ remains approximately constant as $\tau$ decreases.  

The upper trajectory shown in the right panel of Fig.~\ref{fig:trajectories} 
represents yet another set of initial conditions from which our cosmological system can
reach stasis.  This is a set of initial conditions wherein $(\tau^{(0)},\varrho_M^{(0)})$ 
lies outside the stasis wedge, but within a region of the $(\tau,\varrho_M)$ plane 
wherein the energy-density pump $P^{(\rho)}_{\rm M, KE}$ associated with 
$4\phi\to 2\phi$ scattering has a significant effect on the evolution of 
the system.  As discussed in Sect.~\ref{sec:ExoInitImpact}, $4\phi\to 2\phi$ 
scattering causes $\tau$ to increase rapidly, but has a comparatively small effect
on $\varrho_M$.  As a result, the system is propelled from its initial location within 
the $(\tau,\varrho_M)$ plane to the edge of the red constraint contour, beyond which
the impact of $4\phi \to 2\phi$ processes on the evolution of the system becomes 
negligible.  Since the point at which the trajectory crosses this contour is within
the stasis wedge, the system then evolves toward stasis.  

By contrast, the lower trajectory shown in the right panel in Fig.~\ref{fig:trajectories} 
represents a set of initial conditions for which the system does not reach stasis.
For this set of initial conditions, the system is propelled from its initial 
location within $(\tau,\varrho_M)$ plane to the edge of the red constraint contour,
just as it is for the upper trajectory.  However, since the point at which the 
lower trajectory crosses this contour lies outside the stasis wedge, the system 
does not have sufficient time to reach stasis from this point in the $(\tau,\varrho_M)$
plane before $\tau$ becomes comparable to $\tau_{\rm min}$.

In summary, the results shown in Fig.~\ref{fig:trajectories} indicate that for reasonable 
values of the parameters $g_\chi$, $g_\phi$, $g_G$, and $\mu$ there indeed exists a broad range
of initial conditions for the dynamical variables which characterize our thermal stasis 
model for which the stasis attractor is active and the system has sufficient time to 
achieve stasis before that attractor reaches its expiration date.  Furthermore, there also 
exists a broad range of initial conditions for these variables for which the attractor
is not initially active, but for which the system is propelled by the action of exothermic
processes into a region of the $(\tau,\varrho_M)$ plane within which it is active.

\subsection{The approach to and duration of stasis\label{sec:stasisDurationFind}}

\begin{figure}
  \centering
    \includegraphics[keepaspectratio, width=0.48\textwidth]
      {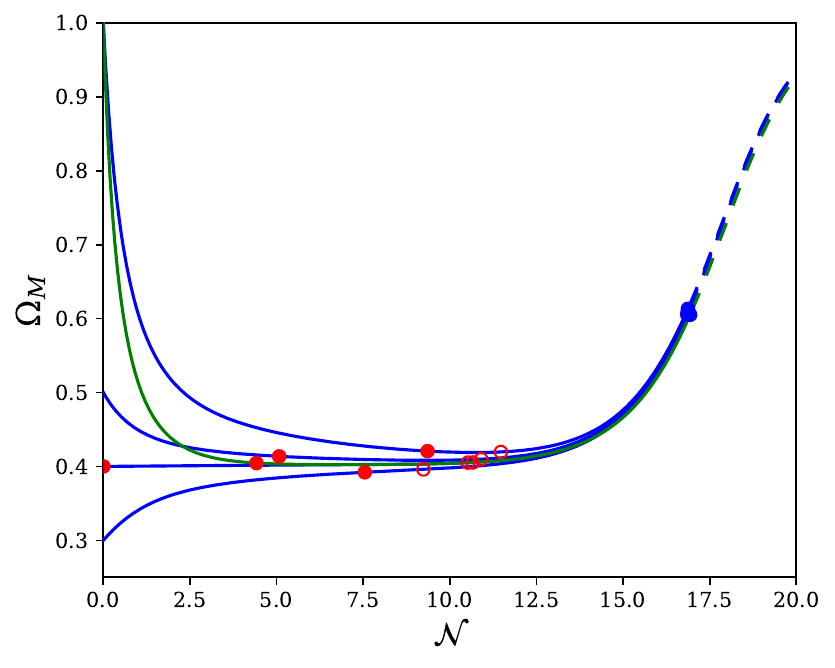}
  \caption{The evolution of $\Omega_M$ as a function of $\mathcal{N}$ for several different 
    choices of initial conditions.  All curves shown correspond to the parameter choices
    $g_\gamma = 1.25\times 10^{-5}$, $g_M = 1.00$, and $g_G = 9.21 \times 10^{-10}$, with
    $\mu$ taken to satisfy Eq.~(\ref{mumaximumeff}).
    For each of the trajectories shown, we have taken $\tau^{(0)} = \taumax/10$.  The initial 
    conditions for the green curve are such that the system evolves toward the stasis fixed point
    along the fast trajectory specified by Eq.~(\ref{fastest}).  The initial conditions for the 
    various blue curves are discussed in the text.  The solid red dot and hollow red circle along
    each curve indicate the times at which the system first starts to and finally ceases to 
    satisfy the stasis criterion in Eq.~(\ref{trueDeltaDefinition}), respectively, while the blue 
    dot shows the time at which $\tau = \taumin$, beyond which $q$ is effectively no longer equal 
    to $-2$. 
    \label{fig:stasisplot}}
\end{figure}

Finally, we examine how the initial conditions for the dynamical variables
which govern our thermal stasis model affect the duration of the stasis epoch in
cases wherein the system in fact reaches stasis before the expiration date.
To this end, in Fig.~\ref{fig:stasisplot} we illustrate the manner in which 
$\Omega_M$ evolves as a function of $\mathcal{N}$ under the influence of the
stasis attractor for a variety of initial conditions for $\Omega_M^{(0)}$ and 
$\Xi^{(0)}$.  All curves shown correspond to the choice of parameters which we find yields 
the longest stasis epoch possible within the context of this thermal stasis model.  
In particular, we find that the duration of the stasis epoch is maximized for the choices
$g_\gamma = 1.25\times 10^{-5}$, $g_M = 1.00$, $g_G = 9.21 \times 10^{-10}$, 
and a value of $\mu$ which satisfies the condition in Eq.~(\ref{mumaximumeff}) and 
therefore has essentially no impact on the cosmological dynamics.  For all curves shown,
we have taken the $\tau^{(0)} = \tau_{\rm max}/10$, where the factor of $1/10$ has been
included in order both to mitigate effects which arise due to the modification of the 
structure of the stasis pump when $\tau \sim \taumax$ and to ensure that the effect of 
$4\phi\to 2\phi$ processes on the dynamics can safely be neglected.  The solid red dot along 
each curve indicates the value of $\mathcal{N}$ at which the stasis criterion in 
Eq.~(\ref{trueDeltaDefinition}) is first satisfied, while the hollow red circle indicates the 
value of $\mathcal{N}$ at which this criterion ceases to be satisfied as $\tau$ approaches 
$\taumin$.  Finally, the solid blue dot indicates the value of $\mathcal{N}$ at which 
$\tau = \taumin$.  We note that many of the hollow red circles shown in the figure overlap, 
as do many of the blue dots.    

In constructing each of the $\Omega_M$ curves in Fig.~\ref{fig:stasisplot}, we have accounted 
for the full dependence of the scattering cross-section on temperature for $\tau$ around or 
below $\taumin$.  However, we have not accounted for the departure from thermal equilibrium 
which occurs around $\tau \sim \tau_{\rm therm} \approx \pi\tau_{\rm min}$ for the parameter 
choices we have adopted here.  The solid portion of each curve indicates the range of 
$\mathcal{N}$ within which $\tau > \tau_{\rm therm}$ and the $\phi$-particle gas is in 
thermal equilibrium.  By contrast, the dashed portion of each curve represents an extrapolation 
of our results into the range of $\mathcal{N}$ within which $\tau < \tau_{\rm therm}$.  Within 
this latter range of $\mathcal{N}$, the departure from thermal equilibrium in general modifies the
form of $P_{M,\gamma}^{(\rho)}$ and thus modifies the manner in which $\Omega_M$ 
evolves with $\mathcal{N}$.  However, we expect this modification not to be particularly severe
across the range of $\mathcal{N}$ shown in the figure, and thus we expect that the dashed portion 
of the curve remains a reasonable approximation to the true behavior of $\Omega_M$.

The green curve 
in Fig.~\ref{fig:stasisplot}
corresponds to the fast trajectory, with the initial value of the variable $s$
introduced in Eq.~(\ref{fastest}) taken to be $s^{(0)} = 1/\barOmega_M$, such that 
$(\Omega_M^{(0)},\Xi^{(0)}) = (1,\barXi/\barOmega_M)$.  We find that this choice of 
initial conditions yields a stasis epoch with a duration of $\mathcal{N}_s = 6.3$~$e$-folds.
The various blue curves correspond to different choices of 
$\Omega_M^{(0)}$ with $\Xi^{(0)} = \barXi$.  These include
\begin{itemize}
  \item $\Omega_M^{(0)} = \barOmega_M$: For this choice of $\Omega_M^{(0)}$, the system
    is already in stasis at $t=t^{(0)}$ and remains in stasis until the criterion in 
    Eq.~(\ref{trueDeltaDefinition}) is no longer satisfied.  This choice
    of $\Omega_M^{(0)}$ of course leads to the longest stasis epoch --- an epoch lasting 
    $\mathcal{N}_s = 10.5$~$e$-folds.
  \item $\Omega_M^{(0)} = \barOmega_M\pm 0.1$: For these two choices of $\Omega_M^{(0)}$,
    the initial conditions are such that the system is close to stasis but not yet in 
    stasis.  We find that for $\Omega_M^{(0)} = \barOmega_M + 0.1$, the stasis criterion
    in Eq.~(\ref{trueDeltaDefinition}) is satisfied for 
    $\mathcal{N}_s = 5.8$~$e$-folds, whereas for $\Omega_M^{(0)} = \barOmega_M - 0.1$ 
    it is only satisfied for $\mathcal{N}_s = 1.7$~$e$-folds.
  \item $\Omega_M^{(0)} = 1$: This choice of $\Omega_M^{(0)}$ is representative of a more 
    general class of initial conditions wherein the system begins further away from stasis.
    For this particular choice of initial conditions, we find that the system only satisfies 
    the stasis criterion for $\mathcal{N}_s = 2.1$~$e$-folds.
\end{itemize}

The results shown in Fig.~\ref{fig:stasisplot} indicate that a significant number of $e$-folds
of stasis can be achieved within our thermal stasis model.  However, we also find that due to the 
expiration date for the attractor, the value of $\mathcal{N}_s$ is highly sensitive to the initial 
conditions for the system.  Indeed, a significant number of $e$-folds is typically only obtained 
if the system initially already reasonably close to stasis at $t = t^{(0)}$ or else 
if the initial conditions are such that the system evolves along the fast trajectory described 
by Eq.~(\ref{fastest}).

\FloatBarrier
\section{Conclusions\label{sec:conclusions}}


In this paper, we have presented a self-consistent model realization of the thermal 
stasis mechanism introduced in Ref.~\cite{Barber:2024vui} and have used this
model in order to investigate the dynamics associated with the stasis attractor.
We have shown that there exist trajectories associated with a particular Jacobian
eigenvalue for the equations of motion for $\Omega_M$ and $\Xi$ near the attractive 
fixed point along which our cosmological system evolves quite rapidly toward stasis.  
By contrast, along other trajectories the system evolves toward stasis much more slowly.  

This distinction is crucial because the stasis attractor in this thermal stasis model 
has an ``expiration date'' determined by the temperature of the non-relativistic 
particle gas.  Indeed, while we find that this model can
give rise to a stasis epoch lasting a significant number of $e$-folds, we also find
that the duration of stasis is comparatively short unless the initial conditions 
are such that the system is already very close to stasis or the system evolves toward 
stasis along the fast trajectory associated with the more negative Jacobian eigenvalue. 
Moreover, we also find in this model that there is a broad range of initial conditions
for which the stasis attractor is realized but the system never reaches stasis. 

One effect which we have not considered in this paper is the impact that the 
cosmological dynamics associated with our stasis model could have on the growth of 
density perturbations.  This could be important because such density perturbations can 
affect the $\phi$-particle annihilation rates which give rise to our cosmological pumps.  
During any cosmological epoch wherein the growth of matter-density perturbations 
is significantly enhanced --- as it is, for example, during an early matter-dominated 
era --- a cosmologically significant fraction of the matter abundance can end 
up being bound into halos prior to the end of that epoch if the epoch lasts more than a few 
$e$-folds~\cite{Blanco:2019eij,Barenboim:2021swl,Ganjoo:2024mie}.  The growth of such perturbations 
is likewise enhanced, though to a slightly lesser degree, during an epoch of matter/radiation stasis 
in tower-based stasis scenarios~\cite{Dienes:2025tox}.  It is therefore possible that perturbations
in the density of any cosmological components which behave like massive matter also experience enhanced 
growth within the context of our thermal stasis model as well.  Such components include both
the $\phi$-particle gas itself and any other spectator matter components which might be present 
during the stasis epoch --- \ie, matter components whose energy densities are sufficiently small 
throughout the period of interest that they  do not appreciably impact the stasis dynamics.  
This enhanced perturbation growth could occur not only during the stasis epoch itself, but 
also while the system is evolving toward stasis under the influence of the 
attractor --- even if the system never reaches stasis before $\tau$ falls below $\tau_{\rm min}$.

That said, the manner in which perturbations evolve both prior to and during the stasis epoch 
within the context of our thermal stasis model is more complicated than in tower-based realizations 
of stasis, primarily because the pump term $P_{M,\gamma}^{(\rho)} \propto \rho_M^2$ associated 
with $\phi$-particle annihilation is non-linear in $\rho_M$ and decreases with $T$.
If a non-negligible fraction of the $\phi$ particles in the universe were to become bound in 
self-gravitating structures during the period wherein the stasis attractor is active,
both the energy density and velocity  distribution of the bound $\phi$ particles would 
be modified relative to the energy density and velocity distribution of the $\phi$ particles 
associated with the homogeneous background.  The structure of the stasis pump 
$P_{M,\gamma}^{(\rho)}$ would therefore also be modified.  Were this modification significant, 
it could alter the constraints on our model.  We leave the analysis of these effects for future 
work.


\begin{acknowledgments}
We thank S.~Gralla for discussions.  The research activities of 
JB and KRD are supported in part by the U.S.\ Department of Energy under Grant 
DE-FG02-13ER41976 / DE-SC0009913; the research activities of KRD are also 
supported in part by the U.S.\ National Science Foundation through its employee 
IR/D program.  The research activities of BT are supported in part by the 
U.S.\ National Science Foundation under Grant PHY-2310622. 
Parts of this work were performed at the Aspen Center for Theoretical Physics, 
which is supported by the U.S.\ National Science Foundation under Grant PHY-2210452.
The opinions and conclusions expressed herein are those of the authors, and do not 
represent any funding agencies. 
\end{acknowledgments}

\appendix


\section{Consistency conditions for quartic couplings\label{sec:Quartics}}


The quartic interaction terms which appear in the second line of Eq.~(\ref{eq:ScalarPot})
can in principle have an impact on the physics which gives rise to stasis in 
our model.  
In principle, this could invalidate portions of the analysis in Sect.~\ref{sec:Results}.~ 
In this appendix, we evaluate the conditions under which the 
contributions from these terms can be neglected and demonstrate that there are large 
regions of our model-parameter space within which $\lambda_\phi$, $\lambda_\chi$, and 
$\lambda_{\phi\chi}$ have essentially no impact on the analysis.  Thus the analysis we 
perform within Sect.~\ref{sec:Results} remains valid.

\subsection{Stabilizing the potential}

The couplings $\lambda_\phi$ and $\lambda_\chi$ play an important role in 
stabilizing the scalar potential $U$ in our model.  Indeed, in order to ensure that 
$\langle \phi\rangle = \langle \chi\rangle =0$ at the global minimum of $U$, 
it is sufficient to impose the conservative bounds
\begin{eqnarray}
  \lambda_\phi ~\geq~ \frac{6 g_{\phi}^{2}}{\mu^{2}/m^{2} + 4}~,~~~~~ 
  \lambda_\chi ~\geq~ \frac{6 g_{\chi}^{2}}{\mu^{2}/m^{2} + 4}~
  \label{eq:PotentialStabilCondits}
\end{eqnarray}
on these couplings.  We emphasize, however, that $\lambda_{\phi\chi}$ is not 
subject to a similar constraint. 

\subsection{Comparison to \texorpdfstring{$s$}{s}-channel variants for 
\texorpdfstring{$\lambda_{\phi\chi}$}{lamphichi} and 
\texorpdfstring{$\lambda_{\phi}$}{lamphi}}

The interaction terms in Eq.~(\ref{eq:ScalarPot}) can affect the cosmological
dynamics in other ways as well.  For example, in the presence of such interaction 
terms, the amplitude for the annihilation process $\phi\phi\to \chi\chi$ --- 
a process which plays a pivotal role in the emergence of the stasis attractor --- 
receives contributions not only from the $s$-channel process depicted in 
Fig.~\ref{fig:pump_diagram}, but also from a contact interaction between the initial- 
and final-state particles with coupling strength $\lambda_{\phi\chi}$.  Since these
two contributions do not scale with $|\vec p_\CM|$ in the same way, the latter
contribution must be negligible in comparison with the former one in order for
$\sigma v$ to scale with $|\vec p_\CM|$ in the desired manner.

Comparing the magnitudes of these two contributions to the matrix element, 
we find that within the regime in which $T$ lies within the range specified 
in Eq.~(\ref{Trange}) and $\Delta(p_X)$ is therefore well approximated by 
Eq.~(\ref{eq:DeltaInAttractorLand}), we find that the contact interaction 
may be neglected when
\begin{equation}
  \lambda_{\phi\chi} ~\ll~ \frac{32\pi g_\chi m}{g_\phi |\vec p_\CM|}~.
\end{equation}
This constraint is the most stringent when $T \sim T_{\rm max}$ lies at the 
upper end of this range and $|\vec p_\CM|\sim \sqrt{m T_{\rm max}}$ is
maximized.  Thus, the contact interaction may be neglected throughout the 
period during which the stasis attractor is realized, provided that
\begin{equation}
    \lambda_{\phi\chi} ~\ll~ \frac{4096 \pi^{2} g_{\chi}}{g_{\phi}^{3}}~.
  \label{eq:lambdaphichibound}
\end{equation}

By contrast, there is no tree-level contribution to the annihilation process 
$\phi\phi\to \chi\chi$ involving $\lambda_{\phi}$.  However, there do exist 
tree-level contributions involving this coupling to both the elastic scattering 
process $\phi\phi\to \phi\phi$ which serves to maintain kinetic equilibrium 
among the $\phi$ particles and the exothermic scattering process 
$4\phi\to 2\phi$ --- contributions in which each $X$ propagator is replaced by
a contact interaction between the the four $\phi$ particles.  Within the regime
in which which $\Delta(p_X)$ is well approximated by 
Eq.~(\ref{eq:DeltaInAttractorLand}), requiring that this contact-interaction 
contribution be negligible imposes an upper bound on $\lambda_\phi$ analogous 
to the upper bound on $\lambda_{\phi\chi}$ in Eq.~(\ref{eq:lambdaphichibound}):
\begin{equation}
  \lambda_\phi ~\ll~ \frac{32\pi m}{|\vec p_\CM|}~.   
\end{equation}
Once again taking $|\vec p_\CM|\sim \sqrt{m T_{\rm max}}$, we find that 
this contribution may be neglected throughout the period during which the stasis 
attractor is realized, provided that
\begin{equation}
  \lambda_{\phi} ~\ll~ \frac{4096 \pi^{2}}{g_{\phi}^{2}}~.
  \label{eq:lambdaphibound}
\end{equation}
This bound on $\lambda_\phi$ is compatible with the potential-stabilization 
condition in Eq.~(\ref{eq:PotentialStabilCondits}).

\subsection{Radiation self-scattering} \label{sec:rad_rad_scattering}

We have assumed that the effect of self-interactions among the $\chi$ particles 
which collectively constitute the radiation in our thermal stasis scenario can be 
neglected.  While such self-interactions do not have a direct impact on the 
stasis dynamics --- \ie, on the equations of motion for $\rho_M$ and $T$ --- they
do have an impact on the phase-space distribution $f_\chi(p)$ of the $\chi$ particles.
Thus, since $f_\chi(p)$ has an impact on the rate at which kinetic energy is transferred 
from radiation to the $\phi$-particle gas via processes such as $\phi\chi\to \phi\chi$, 
self-interactions among the $\chi$ particles can in principle affect the evolution of $T$
indirectly.  Nevertheless, in this Appendix, we demonstrate that there exists a regime 
consistent with the applicable model-consistency conditions and constraints wherein this 
effect is negligible. 

The leading contributions to the amplitude for $\chi\chi\to\chi\chi$ scattering are that 
associated with the four-point interaction in the scalar potential, which is proportional
to $\lambda_\chi$, and those associated with $s$-{}, $t$-{}, and $u$-channel diagrams 
involving a virtual $X$ particle, all of which are proportional to $g_\chi^2$.  The 
swept-volume rate for this process is
\begin{eqnarray}
  (\sigma v)_{\chi\chi\to\chi\chi} 
    &\,=\,&  \frac{1}{64\pi \abs{\vec{p}_1}\abs{\vec{p}_2}} 
    \nonumber\\
    & & \times
    \Bigg|\lambda_\chi
    - \!\!\!\sum_{q^2 = \{s, t, u\}}\frac{g_\chi^2 m^2 }
      {q^2 - m_X^2 + \Pi_X(q^2)}\Bigg|^2\,, \nonumber \\
  \label{eq:sigmavfullchichiscat}
\end{eqnarray}
where $\Pi_X(q^2)$ is the one-loop radiative correction to the $X$ propagator
and where $s$, $t$, and $u$ denote the usual Mandelstam variables.  While interference 
between the contribution from the four-point interaction and the other contributions
has an impact on $(\sigma v)_{\chi\chi \to \chi\chi}$, this effect does not in the 
absence of fine-tuning impact $(\sigma v)_{\chi\chi \to \chi\chi}$ at the 
order-of-magnitude level.  Thus, we establish an approximate bound on $\lambda_\chi$
and by taking $g_\chi\to 0$ in Eq.~(\ref{eq:sigmavfullchichiscat}) and vice versa.

The swept-volume rate $(\sigma v)_{\chi\chi \to \chi\chi}^{(\lambda)}$ that we obtain
by taking $g_\chi\to 0$ in Eq.~(\ref{eq:sigmavfullchichiscat}) is simply
\begin{equation}
  (\sigma v)_{\chi\chi\to\chi\chi}^{(\lambda)} ~=~  
    \frac{\lambda_\chi^2}{64\pi \abs{\vec{p}_1}\abs{\vec{p}_2}} ~.
\label{eq:SweptVolChiChiLambda}
\end{equation}
By contrast, the swept-volume rate $(\sigma v)_{\chi\chi\to\chi\chi}^{(g)}$ that 
we obtain by taking $\lambda_\chi \to 0$ is the more complicated expression
\begin{eqnarray}
  (\sigma v)_{\chi\chi\to\chi\chi}^{(g)} 
    &\,\approx\,&  \frac{g_\chi^2 m^2 }{64\pi \abs{\vec{p}_1}\abs{\vec{p}_2}} 
    \nonumber\\
    & & \times
    \Bigg|\sum_{q^2 = \{s, t, u\}}\frac{1}
      {q^2 - 4m^2 + \Pi_X(q^2)}\Bigg|^2\,, \nonumber \\
  \label{eq:sigmavgchichiscat}
\end{eqnarray}
where we have used the fact that $m_X \approx 2m$.

Within our regime of interest --- the regime in which $\chi\chi\to \chi\chi$ 
scattering does not have a significant impact on $f_\chi(p_\chi)$ --- we can
obtain an order-of-magnitude estimate for $(\sigma v)_{\chi\chi\to\chi\chi}^{(g)} $ 
by noting certain qualitative properties that $f_\chi(p_\chi)$ has within this regime.
Since our $\phi$-particle gas is non-relativistic, the magnitude of the momentum of 
any $\chi$ particle produced by $\phi\phi\to\chi\chi$ annihilation is initially 
$|\vec{p}_\chi| \approx m$ in the background frame, but rapidly decreases below $m$ 
as a result of cosmological redshifting.  As a result, 
$f_\chi(p_\chi) \approx 0$ for $|\vec{p}_\chi| > m$ and the vast majority of $\chi$ 
particles have $|\vec{p}_\chi| \ll m$.  For the $t$-{} and $u$-channel contributions
to the sum in Eq.~(\ref{eq:sigmavgchichiscat}), $|q^2| < 4m^2$ for $|\vec{p}_1| < m$ 
and $|\vec{p}_2| < m$.  Thus, within our regime of interest, the $q^2$ and $\Pi_X(q^2)$ 
terms in these contributions do not affect $(\sigma v)_{\chi\chi\to\chi\chi}^{(g)}$ at 
the order-of-magnitude level.  Moreover, since very few $\chi$ particles have momenta
$|p_\chi| \approx m$ within this regime, the $s$-channel resonance is unimportant and 
the the $q^2$ and $\Pi_X(q^2)$ terms in the $s$-channel contribution to the sum in 
Eq.~(\ref{eq:sigmavgchichiscat}) can likewise be neglected.  Thus, at the 
order-of-magnitude level, we can approximate
\begin{equation}
  (\sigma v)_{\chi\chi\to\chi\chi}^{(g)} ~\sim~  
    \frac{9g_\chi^4 }{1024\pi \abs{\vec{p}_1}\abs{\vec{p}_2}}~.
\label{eq:SweptVolChiChi}
\end{equation}

In order to ensure that $\chi\chi\to \chi \chi$ scattering has a negligible 
effect on $f_\chi(p)$, it is sufficient to require that the scattering rate
\begin{equation}
    \Gamma_{\chi\chi\to\chi\chi} ~=~ 
      \frac{\rho_\gamma}{m} (\sigma v)_{\chi\chi\to\chi\chi}
  \label{eq:GammaChiChiGenDef}
\end{equation}
associated with this process is negligible in comparison with the 
expansion rate --- \ie, that $\Gamma_{\chi\ch\to\chi\chi} \ll H$.  Indeed, we
may interpret $\Gamma_{\chi\chi\to\chi\chi}$ as the rate at which $\chi$ particles 
with initial momentum magnitudes $|\vec{p}_1|$ and $|\vec{p}_2|$ redistribute their 
momentum and energy.  In order to derive approximate bounds on $\lambda_\chi$ and $g_\chi$,
we can define scattering rates 
\begin{eqnarray}
  \Gamma_{\chi\chi\to\chi\chi}^{(\lambda)} &~\sim~&  
    \frac{\lambda_\chi^2\rho_\gamma}{64\pi m \abs{\vec{p}_1}\abs{\vec{p}_2}} 
    \nonumber \\
  \Gamma_{\chi\chi\to\chi\chi}^{(g)} &~\sim~&  
    \frac{9g_\chi^4 \rho_\gamma}{1024\pi m\abs{\vec{p}_1}\abs{\vec{p}_2}}
\label{eq:ScatRateChiChi}
\end{eqnarray}
by replacing $(\sigma v)_{\chi\chi\to\chi\chi}$ in Eq.~(\ref{eq:GammaChiChiGenDef})
with $(\sigma v)_{\chi\chi\to\chi\chi}^{(\lambda)}$ and 
$(\sigma v)_{\chi\chi\to\chi\chi}^{(g)}$, respectively.

At times when the universe is either in or very close to stasis, the expansion
rate $H$ is very similar to the annihilation rate $\Gamma_{\phi\phi\to\chi\chi}$ 
associated with the stasis pump.  It follows from Eq.~(\ref{quadratically}) that
\begin{equation}
  \Gamma_{\phi\phi\to\chi\chi} ~=~ 
    \frac{P^{(\rho)}_{M,\gamma}}{\rho_M} ~=~ 
      \frac{ \rho_M}{m} \, C \,\left(\frac{T}{m}\right)^{q/2} \, A(q)~.
\end{equation}
Thus, at such times, requiring that 
$\Gamma_{\chi\chi\to\chi\chi}^{(\lambda)} \ll \Gamma_{\phi\phi\to\chi\chi}$ and
$\Gamma_{\chi\chi\to\chi\chi}^{(g)} \ll \Gamma_{\phi\phi\to\chi\chi}$ yields the
constraints
\begin{eqnarray}
  \lambda_\chi^2 &~\ll~& \frac{2^{11}\pi^2g_\chi^2}{g_\phi^2} 
    \left(\frac{\abs{\vec{p}_1}\abs{\vec{p}_2}}{mT}\right)
    \frac{\rho_M}{\rho_\gamma} 
    \nonumber \\
  g_\chi^2 &~\ll~& \frac{2^{15}\pi^2}{9g_\phi^2}
    \left(\frac{\abs{\vec{p}_1}\abs{\vec{p}_2}}{mT}\right)
    \frac{\rho_M}{\rho_\gamma}~,
\label{eq:ScatRateChiChiBounds}
\end{eqnarray}
respectively, where we have taken $q=-2$ and used the expression for $C$ in 
Eq.~(\ref{formula_for_C}).

Although scattering processes such as $\phi\chi\to\phi\chi$ occur for $\chi$ 
particles with any momentum $|\vec{p}_\chi|>0$, it is only those $\chi$ particles with 
momenta $|p_\chi| \sim \mathcal{O}(m)$ or larger which have a significant impact on the 
kinetic energy of the $\phi$-particle gas.  It is also only those $\chi$ particles with
$|\vec{p}_\chi| \sim \mathcal{O}(m)$ or larger which are capable of producing $\phi$ particles 
via the process $\chi\chi\to\phi\phi$, which represents the inverse of the process 
associated with the stasis pump.  Thus, since the population of $\chi$ particles produced 
directly by $\phi\phi\to\chi\chi$ scattering with $|\vec{p}_\chi|$ significantly above $m$ 
is negligible, we may estimate the bounds on $\lambda_\chi$ and $g_\chi$ by taking 
$|\vec{p}_1|\sim |\vec{p}_2| \sim m$ in Eq.~(\ref{eq:ScatRateChiChiBounds}).  Moreover,
$\rho_\gamma \sim \rho_\phi$ and $T < T_{\rm max}$ at times when the stasis attractor is 
realized and the universe is either in or very close to stasis.  Thus, through use of 
Eq.~(\ref{eq:T_bounds_propagator}), we may express these bounds as 
\begin{eqnarray}
  \lambda_\chi &~\ll~& \frac{2^{25/2}\pi^2 g_\chi}{g_\phi^3}
    \left(\frac{T_{\rm max}}{T}\right)^{1/2}
    \nonumber \\
  g_\chi &~\ll~& \frac{2^{29/2}\pi^2}{3g_\phi^3}
    \left(\frac{T_{\rm max}}{T}\right)^{1/2}~.
\label{eq:ScatRateChiChiBoundsFinal}
\end{eqnarray}
These constraints can easily be satisfied within our parameter-space regime of interest.


\section{Correction to $\barOmega_M$ from $\phi\chi\to\phi\chi$ scattering
\label{sec:compton}}


Another process which necessarily occurs within our model is the elastic scattering 
process $\phi\chi \to \phi\chi$.  Within the regime in which $\lambda_{\phi\chi}$
satisfies the condition in Eq.~(\ref{eq:lambdaphichibound}), the dominant contribution 
to the cross-section for this process is the contribution from the $t$-channel 
analogue of the $s$-channel diagram depicted in Fig.~\ref{fig:pump_diagram}.  Since 
this process facilitates the transfer of kinetic energy between $\phi$ 
and $\chi$ particles, it can in principle alter the manner in which $T$ and $\rho_M$ 
evolve and thereby disrupt the stasis attractor.
In this Appendix, we demonstrate that $\phi\chi\to\phi\chi$ scattering does not
disrupt stasis entirely, but merely alters the value of $\barOmega_M$.
Moreover, we demonstrate that within our parameter-space region of interest, the 
shift in $\barOmega_M$ is negligible.

We begin by considering the effect that $\phi\chi\to\phi\chi$ scattering has on the  
cosmological dynamics.  The process $\phi\chi\to\phi\chi$ scattering does not alter the 
number density of the $\phi$ particles and therefore does not modify the equation of 
motion for $\Omega_M$ directly.  However, this process does transfer kinetic energy from 
radiation to matter and thus gives rise a additional pump term $P_{\rm \phi\chi}$ in the 
equation of motion for $T$ --- a pump whose dependence on $\rho_M$ and $T$ differs from that 
of the the annihilation pump $P_{\KE,\g}$.  In the presence of this additional 
pump, Eq.~(\ref{eq:TempDiffEq}) is modified to
\begin{eqnarray}
\frac{dT}{dt} ~&=&~
    -2 H T  - 
    \frac{2m}{3\Omega_M} 
    \left( P_{\KE,\gamma} 
    - \frac{\Omega_\KE}{\Omega_M}
    \,P_{M,\gamma} \right)\nonumber\\
    &&~+ \frac{2m}{3\Omega_M} P_{\rm \phi\chi}~.
\label{dTdtequation_compton}
\end{eqnarray}

Without loss of generality, we can express the corresponding energy pump 
$P^{(\rho)}_{\phi\chi}$ in terms of the stasis pump $P^{(\rho)}_{M,\g}$ as 
follows
\begin{equation}
    P^{(\rho)}_{\phi\chi} ~=~ P^{(\rho)}_{M,\g} 
      \frac{\rho_\gamma}{\rho_M} \frac{T}{2m} \xi~,
  \label{eq:PPhiChivsPMGammaScaling}
\end{equation}
where $\xi$ is a dimensionless scaling factor.  In our model, this scaling factor is 
time-independent.  During stasis, $P^{(\rho)}_{M,\g} = \Omega_\gamma H \rho_M$,
and thus we have
\begin{equation}
  P^{(\rho)}_{\rm \phi\chi} ~=~  
    H T \frac{\rho_M \Omega_\gamma^2 }{\Omega_M}  \frac{\xi}{2m} ~.
\end{equation}
The corresponding abundance pump is therefore
\begin{equation}
    P_{\rm \phi\chi} ~=~  H T \W_\g^2  \frac{\xi}{2m} ~.
\end{equation}
We may therefore write the equation of motion for $T$ as 
\begin{eqnarray}
  \frac{dT}{dt} &~=~&
      -\left[2-\frac{(1-\Omega_M)^2}{3\Omega_M}  \xi\right] H T  \nonumber\\
    & &~ -\frac{2m}{3\Omega_M} \left( P_{\KE,\gamma} - \frac{\Omega_\KE}{\Omega_M}
      \,P_{M,\gamma} \right)~.
  \label{dTdtequation_compton_2}
\end{eqnarray}

This result implies that the overall effect of $\phi\chi\to\phi\chi$ scattering on the 
cosmological dynamics is to modify the coefficient of the first term on the right side 
of Eq.~(\ref{eq:TempDiffEq}).  However, this modification does not disrupt the stasis 
attractor; rather, it simply modifies the expressions for $\barOmega_M$ and $\barXi$ 
relative to those given in Eq.~(\ref{eq:stasisvalues}).  In particular, one finds that
\begin{eqnarray}
  \barOmega_M &=& \frac{(1+q^2/6)+(2q+3)+2q\xi/3}{2(1+q^2/6)+2q\xi/3} 
    \nonumber \\
    & &\times \left[1 + \sqrt{1-\frac{8 q\xi (6 + 2 q\xi + q^2)}{(24 +12q+4q\xi+q^2)^2}}\,
    \right],~~~~
\label{stasisvalues_WM_compton_alt}
\end{eqnarray}
while $\barXi$ is still given by the expression in Eq.~(\ref{eq:stasisvalues}), but 
with $\barOmega_M$ replaced by this shifted value.

We now calculate $P^{(\rho)}_{\rm \phi\chi}$ and thus the scaling factor $\xi$ 
for our model.  In general, this kinetic-energy pump is given by
\begin{equation}
   P^{(\rho)}_{\rm \phi\chi} ~=~ n_M n_\gamma 
   \Big\langle  \, (\sigma v)_{\phi\chi\to\phi\chi} \Delta{\rm KE} 
   \Big\rangle~,
   \label{eq:PumpComptonGenerality}
\end{equation}
where $n_M$ and $n_\gamma$ denote the number densities of $\phi$ and $\chi$ particles,
respectively, where $\Delta{\rm KE}$ denotes the change in kinetic energy of the $\phi$ 
particle during the scattering process, and where $\langle X \rangle$ denotes the average of
the quantity $X$ over the momentum distribution of both initial-state particles --- \ie,
\begin{equation}
  \langle X\rangle ~\equiv~ \frac{1}{(2\pi)^6n_\gamma n_\phi}\int d^3p_\phi f_\phi(p_\phi) 
    \int d^3p_\chi f_\phi(p_\chi) \, X~.
\end{equation}
Within our regime of interest, wherein the $\phi$-particle gas is non-relativistic, 
$(\sigma v)_{\phi\chi\to\phi\chi}$ and $\Delta{\rm KE}$ are both approximately 
independent of the energy and momentum of the incoming $\phi$ particle, while 
$n_M \approx \rho_M/m$.  Thus, within this regime, the integral over $d^3p_\phi$ is 
trivial and Eq.~(\ref{eq:PumpComptonGenerality}) reduces to
\begin{eqnarray}
   P^{(\rho)}_{\rm \phi\chi} &=&~ \frac{\rho_M}{2\pi^2 m} \int_0^m d|\vec{p}_\chi|\, 
     |\vec{p}_\chi|^2 f_\chi(p_\chi)  \nonumber\\
&&~\times\int d\W \frac{d(\sigma v)_{\phi\chi\to\phi\chi}}{d\W} \Delta{\rm KE} ~,
\label{eq:ComptonPumpGen}
\end{eqnarray}
where $d\Omega$ denotes the solid angle element in the cosmological background frame 
into which the final-state $\chi$ particle scatters. 

In order to evaluate Eq.~(\ref{eq:ComptonPumpGen}), we begin by noting that the 
initial- and final-state $\chi$-particle momenta in the background 
frame, which we respectively denote as $\vec{p}_\chi$ and $\vec{p}_{\chi,f}$, are
related for this Compton-scattering-like process by
\begin{equation}
    \frac{1}{|\vec{p}_{\chi,f}|} ~=~ 
      \frac{1}{|\vec{p}_\chi|} + (1-\cos\theta) \frac{1}{m}~.
\end{equation}
where $\theta$ is the angle between $\vec{p}_{\chi,f}$ and $\vec{p}_\chi$.  It therefore
follows that the kinetic energy transferred to the $\phi$ particle as a result of the 
scattering is
\begin{equation}
  \Delta {\rm KE} ~=~ |\vec{p}_{\chi}| - |\vec{p}_{\chi,f}| ~=~ 
    \frac{|\vec{p}_{\chi}|^2 (1-\cos\t)}{m + |\vec{p}_{\chi}| (1 - \cos\t)}~.
\end{equation}

Next, we note that he leading contributions to $(\sigma v)_{\phi\chi\to\phi\chi}$ arise
at tree-level due to four-point interaction in the scalar potential involving the coupling
$\lambda_{\phi\chi}$ and due to a $t$-channel Feynman diagram involving the exchange of 
a virtual $X$ particle.  The amplitude for this process is 
\begin{equation}
  i\mathcal{M}_{\phi\chi\to\phi\chi} ~=~ -i\lambda_{\phi\chi}
    -\frac{ig_\phi g_\chi m^2}{(p_\chi - p_{\chi,f})^2-4m^2}~,    
\end{equation}
where we have used the fact that $m_X \approx 2m$ within our regime of interest.  In
Appendix~\ref{sec:Quartics}, we demonstrated that $\lambda_{\phi\chi}$ may be taken 
to arbitrarily small such that its impact on the dynamics of our model is negligible.
Hence, for simplicity we shall ignore the contribution to $\mathcal{M}_{\phi\chi\to\phi\chi}$ 
from the four-point interaction and focus on the contribution from the $t$-channel process.
In the background frame, we have
\begin{eqnarray}
  (p_\chi - p_{\chi,f})^2 &~=~& -2 |\vec{p}_\chi| |\vec{p}_{\chi,f}| (1-\cos\t) 
    \nonumber \\ 
    &~=~& -2 m \Delta {\rm KE}~.
\end{eqnarray}
Thus, after some algebra, one finds that the differential swept-volume rate for 
$\phi\chi\to\phi\chi$ scattering in this frame can be written as
\begin{eqnarray}
  \frac{d(\sigma v)_{\phi\chi\to\phi\chi}}{d\Omega} &~=~& 
    \frac{g_\phi^2g_\chi^2}{256\pi^2} \frac{m}
    {(m+2|\vec{p}_\chi|)(\Delta{\rm KE}+2m)^2} 
    \nonumber \\ & & \times ~
    \frac{1-\beta_{\rm CM}^2}{(1-\beta_{\rm CM}\cos\theta)^2}~,
  \label{eq:sigmaphichibeforeinteg}
\end{eqnarray}
where $\beta_{\rm CM} ~\equiv~ |\vec{p}_\chi|/(m+|\vec{p}_\chi|)$ is the velocity 
of the center-of-mass frame relative to the background frame.

In order to proceed further, we must derive an expression for $f_\chi(p_\chi)$.
Our primary regime of interest is that within which $\lambda_\chi$ and $g_\chi$
satisfy the bounds in Eq.~(\ref{eq:ScatRateChiChiBoundsFinal}) and 
$\chi\chi\to\chi\chi$ scattering has a negligible effect on $f_\chi(p_\chi)$ for 
momenta $|\vec{p}_\chi|\sim \mathcal{O}(m)$.  Given this, we shall begin by evaluating 
$P_{\phi\chi}^{(\rho)}$ in the limit in which $\lambda_\chi$ and $g_\chi$ are
sufficiently small that they have no appreciable impact on $f_\chi(p_\chi)$ for any 
value of $|\vec{p}_\chi|$.  On the basis of these results, we shall then argue that 
even in cases in which $\lambda_\chi$ and/or $g_\chi$ are sufficiently large that 
$\chi\chi\to\chi\chi$ scattering has a non-trivial impact on $f_\chi(p_\chi)$ for 
momenta $|\vec{p}_\chi| \ll m$, the impact on $P_{\phi\chi}^{(\rho)}$ will be
relatively unimportant.  In addition, since our primary aim is to determine the 
impact that $\phi\chi\to\phi\chi$ scattering has on $\barOmega_M$, we shall also 
continue to focus in what follows on the regime wherein the universe is either already 
in or else very close to stasis. 

We begin be noting that within this regime, $\rho_\gamma$ is approximately proportional 
to the critical density.  Thus, the relationship between the radiation energy densities 
at any two times $t$ and $t'$ is
\begin{equation}
  \rho_\gamma(t) ~\approx~ \rho_\gamma(t') \left[\frac{a(t)}{a(t')}\right]^{-3(1+\overline{w})}
  ~=~ \rho_\gamma(t') \left[\frac{a(t)}{a(t')}\right]^{-4+\Omega_M}\!.
  \label{eq:HowRhoGammaScales}
\end{equation}
Since the $\chi$ particles are massless, with energies $E_\chi \approx |\vec{p}_\chi|$, the 
energy density of radiation at any time $t$ can be expressed as  
\begin{equation}
  \rho_\gamma(t) ~=~ \frac{1}{2\pi^2} \int_0^\infty d|\vec{p}_\chi| |\vec{p}_\chi|^3
  f_\chi(p_\chi,t)~.
  \label{eq:RhoGammaAndPSD}
\end{equation}
A similar relation also holds at any other time $t'$.  It therefore follows from 
Eq.~(\ref{eq:HowRhoGammaScales}) that when the universe is either already in or 
else very close to stasis, we have
\begin{eqnarray}
&&  \int_0^\infty d|\vec{p}_\chi| |\vec{p}_\chi|^3
  f_\chi(p_\chi,t) ~=~ \nonumber \\
&& ~~~~~~~~~~\left[\frac{a(t)}{a(t')}\right]^{-4+\Omega_M}
  \int_0^\infty d|\pvec{p}_\chi'| |\pvec{p}_\chi'|^3
  f_\chi(p_\chi',t')~,~~~~~~~~
\end{eqnarray}
where we have defined $|\pvec{p}_\chi'| \equiv |\vec{p}_\chi(t')|$.  In order to 
derive the form of the differential quantity $f_\chi(p_\chi,t)$ from this 
integral relation, we note that since all $\chi$ particles produced 
via $\phi\phi\to \chi\chi$ scattering initially have $|\vec{p}_\chi| \sim m$, as
as discussed in Appendix~\ref{sec:rad_rad_scattering}.  However, once a population 
of $\chi$ particles is produced with this initial value of $|\vec{p}_\chi|$, their 
contribution to the overall energy density subsequently scales like that of radiation.  
Since $|\vec{p}_\chi|$ scales with time according to the relation
\begin{equation}
  |\vec{p}_\chi(t)| ~=~ |\pvec{p}_\chi'|\frac{a(t')}{a(t)}~,   
\end{equation}
it therefore follows that $f_\chi(p_\chi,t) = f_\chi(p_\chi',t')$ at times $t > t'$ for a 
population of $\chi$ particles which have momentum $|\pvec{p}'_\chi| < m$ at time $t'$.
It therefore follows from inspection that the phase-space distribution for the 
$\chi$ particles must take the form
\begin{equation}
  f_\chi(p_\chi,t) ~=~ \frac{2\pi^2\rho_\gamma \Omega_M}{m^4} 
    \left(\frac{|\vec{p}_\chi|}{m}\right)^{\Omega_M - 4}\Theta(m-|\vec{p}_\chi|)~,
\end{equation}
where the overall normalization factor is determined by the condition in 
Eq.~(\ref{eq:RhoGammaAndPSD}).

Substituting this result and the result for the differential swept-volume rate in 
Eq.~(\ref{eq:sigmaphichibeforeinteg}) into Eq.~(\ref{eq:ComptonPumpGen}), we find after
some algebra that $P^{(\rho)}_{\phi\chi}$, which can be expressed in terms of an 
integral over the dimensionless ratio $x \equiv |\vec{p}_\chi|/m$, takes the form 
\begin{equation}
  P^{(\rho)}_{\phi\chi}
    ~=~ \frac{g_\chi^2 g_\phi^2}{32\pi m^3} \rho_M \rho_\g  
    \int_0^1 dx\, \mathcal{J}(x)~,
  \label{eq:PChiGammaInTermsofJint}
\end{equation}
where we have defined
\begin{equation}
  \mathcal{J}(x) \,\equiv\, 
    \frac{\barOmega_M}{4} x^{\barOmega_M-4}\Bigg[ 
    \log\left(\frac{x^2+2x+1}{2x+1}\right) -\frac{x^2}{(x+1)^2}\Bigg]~.
  \label{eq:integral_BT_version}
\end{equation}

We are now equipped to evaluate the scaling factor $\xi$ in 
Eq.~(\ref{eq:PPhiChivsPMGammaScaling}) and thereby determine the extent to which 
$\phi\chi\to\phi\chi$ scattering shifts the value of $\barOmega_M$.  
Substituting our ``temperate''-regime expression for 
$P^{(\rho)}_{M,\gamma}$ in Eq.~(\ref{eq:PMgamma_cold}) and our result for 
$P^{(\rho)}_{\phi\chi}$ in Eq.~(\ref{eq:PChiGammaInTermsofJint}) into 
Eq.~(\ref{eq:PPhiChivsPMGammaScaling}) and solving for $\xi$, we find that
\begin{equation}
  \xi ~=~ \frac{2m P^{(\rho)}_{\phi\chi}}{T P^{(\rho)}_{M,\gamma}}
      \left(\frac{\rho_M}{\rho_\g}\right)
    ~=~\frac{g_\phi^4}{512\pi^2} \int_0^1 dx\,\mathcal{J}(x)~.
\end{equation}
For $q=-2$, which corresponds to a matter abundance $\barOmega = 2/5$ during stasis, 
the integral in this expression evaluates to
\begin{equation}
   \int_0^1 dx\, \mathcal{J}(x) ~\approx~ 8.41\times 10^{-3}~. 
\end{equation}
Substituting this result into Eq.~(\ref{stasisvalues_WM_compton_alt}) and expanding the
resulting expression as a power series in $\xi$, we find that for $q = -2$, the matter 
abundance during stasis is modified as a consequence 
of $\phi\chi\to\phi\chi$ scattering to   
\begin{eqnarray}
  \barOmega_M &~=~& \frac{2}{5}\left[1 + \frac{9}{10}\xi  +\mathcal{O}(\xi^2)\right] 
    \nonumber \\ &~\approx~& 
    \frac{2}{5}\bigg[1 + \left(1.5\times 10^{-6}\right) g_\phi^4 
      \bigg]~.  
  \label{eq:ModifiedOmegaBar}
\end{eqnarray}
Since the correction to $\barOmega_M$ is extremely small, we may conclude that the impact
of $\phi\chi\to\phi\chi$ scattering on the stasis dynamics is negligible when the system is 
either already in or else very close to stasis.

\begin{figure}[t]
  \centering
  \includegraphics[width=0.43\textwidth]{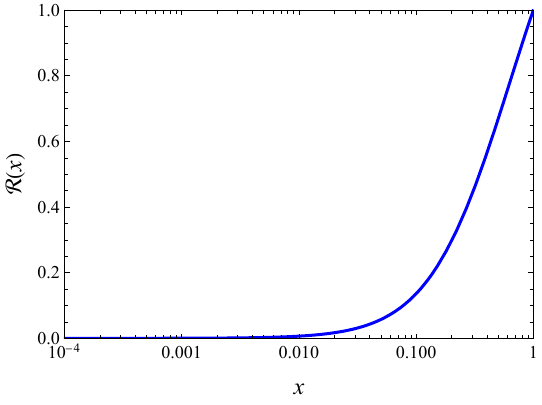}
  \caption{The quantity $\mathcal{R}(x)$ defined in Eq.~(\protect\ref{eq:ScriptRDef}), 
    plotted as a function of $x$.  We observe that $\mathcal{R}(x)$ remains negligible 
    for $x \ll 1$.
\label{fig:BT_int}}
\end{figure}

In deriving the result in Eq.~(\ref{eq:ModifiedOmegaBar}), we have assumed that 
$\lambda_\chi$ and $g_\chi$ 
are sufficiently small that $\chi\chi\to\chi\chi$ scattering has a negligible 
effect on $f_\chi(p_\chi)$.  In light of Eq.~(\ref{eq:PChiGammaInTermsofJint}),  
we now consider how relaxing this assumption might affect $P^{(\rho)}_{\phi\chi}$.
Since we must nevertheless require that $\lambda_\chi$ and $g_\chi$ satisfy the 
conditions in Eq.~(\ref{eq:ScatRateChiChiBoundsFinal}), relaxing this assumption 
can only affect the shape of $f_\chi(p_\chi)$ at $|\vec{p}_\chi| \ll m$.  In 
order to assess the impact that modifying $f_\chi(p_\chi)$ in this manner could
have on $P^{(\rho)}_{\phi\chi}$, in Fig.~\ref{fig:BT_int} we plot the normalized
integral
\medskip
\noindent
\begin{equation}
  \mathcal{R}(x) ~\equiv~ \frac{\int_0^x dx'\, \mathcal{J}(x')}
    {\int_0^1 dx'\, \mathcal{J}(x')}
\label{eq:ScriptRDef}
\end{equation}
for $\barOmega_M = 2/5$.  We observe that $\mathcal{R}(x)$ is negligible for 
$x \ll 1$, implying that the contribution to $P^{(\rho)}_{\phi\chi}$ from 
scattering events involving $\chi$ particles with $|\vec{p}_\chi| \ll m$ is 
likewise negligible.  We may therefore infer that modifications of $f_\chi(p_\chi)$
due to the redistribution of energy and momentum among $\chi$ particles with 
$|\vec{p}_\chi| \ll m$ have little impact on $P^{(\rho)}_{\phi\chi}$.  Thus, 
provided that $\lambda_{\phi\chi}$ is small and that $\lambda_\chi$ and $g_\chi$ 
satisfy the conditions in Eq.~(\ref{eq:ScatRateChiChiBoundsFinal}), the energy-density 
pump associated with $\phi\chi\to\phi\chi$ scattering should be given --- at least 
to a good approximation --- by Eq.~(\ref{eq:PChiGammaInTermsofJint}).

\bibliography{TheLiterature2}

\end{document}